\documentclass[a4paper,11pt]{article}
\pdfoutput=1 

\usepackage{jheppub} 

\usepackage[T1]{fontenc} 

\usepackage[dvipsnames]{xcolor}

\title{Spinning Conformal Correlators \\ from Neural Networks}

\author[a]{Manas Dogra}
\author[a,b]{James Halverson}
\author[a,b,c]{Joydeep Naskar}

\affiliation[a]{Department of Physics, Northeastern University, Boston, MA, 02115, USA}
\affiliation[b]{The NSF AI Institute for Artificial Intelligence and Fundamental Interactions, Cambridge, MA, U.S.A.}
\affiliation[c]{Beijing Institute of Mathematical Sciences and Applications, Huaibei Town, Huairou District, Beijing 101408, China}

\emailAdd{dogra.ma@northeastern.edu}
\emailAdd{j.halverson@northeastern.edu}
\emailAdd{joydeepnaskar@bimsa.cn}

\abstract{
We construct spinning conformal fields from neural networks and the embedding formalism, computing their two-, three- and four-point functions in examples, building on scalar conformal field techniques introduced in \cite{Halverson:2024axc}. For a particular ensemble of i.i.d. neurons we recover the 4d Maxwell CFT in the infinite-width limit.}

\begin{document} 
\maketitle
\raggedbottom

\section{Introduction}
Neural networks and quantum field theories, despite arising in distinct domains, exhibit a deep structural correspondence. This connection traces back to a foundational result \cite{Neal:1996blnn} by Neal  in the 1990s, which proved that infinite-width neural networks converge to Gaussian processes. Over the past decade, this asymptotic behavior has been generalized far beyond simple networks \cite{Novak:2018cnnGP,GarrigaAlonso:2018dcnnGP,Hron:2020infiniteAttention, Williams:1996infiniteNetworks,Matthews:2018gpWideDNN, Schoenholz:2017correspondence, Jacot:2018ntk}. For a review see \cite{Halverson:2026pmb,Yang:2019tensorProgramsI,Yang:2020tensorProgramsII,Halverson:2024hax}. From the physics perspective, the mathematical equivalence between free field theories and Gaussian processes has recently provided an alternate framework for field theory calculations \cite{Halverson:2020trp}. This led to the first neural network construction of Euclidean free scalar theory \cite{Halverson:2021aot}, followed by a series of other interesting works including a NN description of symmetries \cite{Maiti:2021fpy}, architectural dependence on correlators at finite $N$ \cite{Zhang:2026tss}, quantum mechanics \cite{Ferko:2025ogz} and several field theories including $\phi^4$ scalar theories \cite{Demirtas:2023fir,Sen:2025vzl}, fermionic and supersymmetric models \cite{Frank:2025zuk}, Liouville theory \cite{Ferko:2026axm}, anomalies \cite{Ferko:2026kkm}, topological field theory \cite{Ferko:2026ken, Ferko:2026ukw} and even bosonic worldsheet string theory \cite{Frank:2026bui}.
NN/FT has also been used to investigate conformal defects \cite{Capuozzo:2025ozt}, Virasoro symmetry \cite{Robinson:2025ybg}.

Scalar conformal fields were constructed directly from neural networks at initialization in \cite{Halverson:2024axc} by a subset of us. This  approach leverages the embedding formalism, which utilizes the fact that the global conformal group on $\mathbb{R}^d$ is isomorphic to the Lorentz group $SO(d+1,1)$ acting linearly on the ambient Minkowski space $\mathbb{R}^{d+1,1}$. Conformal fields in $d$ dimensions can then be constructed by restricting rotationally invariant ensembles of homogeneous neural networks in $(d+2)$ dimensions, followed by a Wick rotation, and restriction of the correlators to the projective null cone. More concretely, implementing this construction of CFTs requires Lorentz-invariant theory of homogeneous fields on the embedding space $\mathbb{R}^{d+1,1}$ defined via
\begin{equation}
    Z[J] = \langle e^{\int d^{d+2}X~J(X)\Phi(X)} \rangle, \label{partfunc}
\end{equation}
where we will endow the field $\Phi(X)$ with statistics. In our construction, we replace the field $\Phi(X)$ as a parameterized family of functions, $\Phi_\Theta(X)$ where $\Theta$ is called the ``neural network parameter'', denoting the collection of all weights and biases. The specific functional form of $\Phi_\Theta(X)$ is referred to as the network's architecture. Since the parameters in $\Theta$ are probabilistic, the function $\Phi_\Theta(X)$ is also probabilistic.
 
 This paradigm is part of a broader program called neural network-field theory (NN/FT) correspondence. The scalar partition function is defined as
\begin{equation}
    Z[J] = \int D\Theta \, P(\Theta) \, e^{\int d^{d+2}X~J(X)\Phi_\Theta(X)}. 
\end{equation}
where $P(\Theta)$ is a probability distribution on the parameters of the NN field $\Phi_{\Theta}(X)$. Correlators can be computed by appropriate insertions of $\Phi_{\Theta}(X)$ functionals.
The proposal of NN/FT correspondence is that this statistical partition function will give us the same correlators as an ordinary field theory, where the interesting dynamics of the action is now encoded in the neural network architecture. Of course, we do not know fully yet, how this mapping precisely works. For all we know for now, is that it exists and we can produce (many of) the symmetries of the theory correctly. In other words, this is an alternative definition of Euclidean field theory. Indeed, it was proved in \cite{Ferko:2026axm} that under appropriate assumptions every Euclidean QFT admits a NN representation with a countable infinity of parameters using measure theoretic techniques. This is analogous to the universal approximation theorem which states that every continuous function can be approximated by some NN up to arbitrary precision. In particular for a scalar primary $\Phi(X)$, the NN field ansatz proposed in \cite{Halverson:2024axc} is 

\begin{equation}
    \Phi(X)=(\Theta \cdot X)^{-\Delta}
\end{equation}
which yielded the correct CFT correlators and the $4$-point functions yielded conformal block decomposition, at least for non-unitary examples with $\Delta<0$.

In this work, we extend the work of \cite{Halverson:2024axc} to include integer-spin fields and compute spinning correlators. By introducing reference polarization vectors $Z$ to go index-free \cite{Costa:2011mg}, we propose specific homogeneous neural network architectures capable of representing spinning primary fields. We explicitly evaluate their two-, three-, and four-point correlation functions, demonstrating that appropriate permutations of the network parameters yield different valid CFT tensor structures. A related work has appeared in \cite{CarmoTerin:2025lyx} where the author has constructed gauge fields. We will apply the techniques for the case of free Maxwell CFT in $d=4$.

The paper is organized as follows: In Section \ref{sec:review} we review the embedding formalism in CFT, especially concerning spinning correlators following \cite{Rychkov:2016cft,Costa:2011mg}. In section \ref{sec:main} we construct neural network architectures and compute the spinning correlators. In section \ref{Maxwellsection} we discuss applications of the formalism to the $d=4$ Maxwell theory. In section \ref{sec:discussion}, we conclude our discussions with some potential applications and future work.

\section{Review of the Embedding Formalism}\label{sec:review}

It is well known that the conformal group for Euclidean $CFT_{d}$ with $d\ge3$ is $SO(d+1,1)$ which also happens to be the Lorentz group for $\mathbb{R}^{d+1,1}$. This identification allows the conformal transformations to act linearly on $\mathbb{R}^{d+1,1}$ which is called the \emph{embedding space} in this context. We can parameterize this ambient space with coordinates $X = (X^+, X^-, X^\mu)$ (where $\mu = 1, ... , d$). This space is equipped with a metric defined by,
\begin{equation}
    X^2 = -X^+X^- + \sum_{\mu=1}^d (X^\mu)^2
\end{equation}
The coordinates transform via linear matrix multiplication, $X \mapsto gX$, for $g \in SO(d+1,1)$. Of course this manifest linearity comes at the cost of redundancies and hence to recover the original $d$-dimensional theory, we must impose the following two constraints:
\begin{enumerate}
\item \textbf{Restriction to the Null Cone:} One dimension is removed by restricting to the null cone $X^2 = 0$ which is manifestly preserved by all $SO(d+1,1)$ transformations. This constraint reduces the effective dimensionality of the space from $d+2$ to $d+1$.
\item \textbf{Projectivization:} Since the null cone condition $X^2=0$ is unaffected by a rescaling $X \to \lambda X$, all points along a single ray from the origin are physically equivalent. We therefore identify them:
\begin{equation}
    X \sim \lambda X, \quad \text{for } \lambda \in \mathbb{R}_{\ge0}.
\end{equation}
\end{enumerate}
\noindent The resulting space known as the projective null cone (PNC), is geometrically the space of all light-like rays in $\mathbb{R}^{d+1,1}$. The action of the conformal group is simplified to the linear action of $SO(d+1,1)$ that acts on these rays. Similar themes have played an important role in viewing $(3+1)$-D Poincaré transformations as linear transformations in $\mathbb{R}^{1,4}$ \cite{Tung:1985iqd}, superspace in supersymmetric field theories \cite{Salam:1974yz} and also the projective twistor space $\mathbb{PT}$ in twistor theory \cite{Penrose:1967wn}. 

Now we will review the embedding formalism for spinning conformal fields, closely following \cite{Costa:2011mg}. Conformal primaries $\varphi_{\mu_1 \mu_2... \mu_l}$ of spin $l$ on $\mathbb{R}^d$ can be lifted to tensor operators $\Phi_{A_1 A_2 ... A_l}$ on the null cone in $\mathbb{R}^{d+1,1}$ which transform homogeneously with degree $-\Delta$ where $\Delta$ is the scaling dimension of $\varphi_{\mu_1 \mu_2... \mu_l}$ i.e., 
\begin{equation}
    \Phi_{A_1 A_2 ... A_l} (\lambda X)=\lambda^{-\Delta} \Phi_{A_1 A_2 ... A_l}(X) 
\end{equation}
where $\lambda \in \mathbb{R}_{>0}$. 
We also need to choose a covariant gauge condition (transversality) 
\begin{equation}\label{transversality_intro}
    X^A \Phi_{A A_2... A_l}=0
\end{equation}
to kill the redundancies introduced in recasting the $\phi_{\mu_1...\mu_l}$ in terms of embedding fields $\Phi_{A_1...A_l}$. 
Explicitly the fields on $\mathbb{R}^d$ can be obtained by evaluating it on a section of the PNC (typically the Poincar\'e section (PS) $X^+=1$) and then projecting back to $\mathbb{R}^d$:
\begin{equation}
\varphi_{\mu_1...\mu_l}=\frac{\partial X^{A_1}}{\partial x^{\mu_1} }...\frac{\partial X^{A_l}}{\partial x^{\mu_l}} \Phi_{A_1 ... A_l} \vert_{PS}
\end{equation}

We will restrict to symmetric and traceless tensors for simplicity. To reduce notational clutter and go index-free \cite{Costa:2011mg}, one can contract the tensor indices by a set $\{Z_{A_k}\}_{k=1}^l$ of reference null ``polarization'' vectors in $\mathbb{R}^{d+1,1}$ so as to obtain the so called \emph{embedding polynomials} in $Z$ of degree $l$: 
\begin{equation}
   \Phi(X,Z)=\Phi_{A_1...A_l}Z^{A_1} ... Z^{A_l} 
\end{equation}
This is a simplifying tool related ultimately to  the polarization of an algebraic form \cite{Procesi:2007liegroups}. This has also proven to be useful in the calculation of $\mathcal{N}=4$ Super Yang-Mills 1/2-BPS correlators where the analogue of the condition $Z^2=0$ is necessary to make the operators in the correlator BPS \cite{Komatsu:2017buu}.
The polarization may be shifted by any multiple of $X$ without affecting the physical field: on the null cone $Z \to Z + \alpha X$ preserves both $Z^2=0$ and $X\cdot Z=0$. This is the index-free realization of \eqref{transversality_intro}. In other words, transversality is the condition that $\Phi(X,Z)$ depend on $Z$ only through its equivalence class $Z \sim Z + \alpha X$. The architecture of Section \ref{sec:main} realizes this invariance exactly. Parity invariant $n$-point correlators for primaries of arbitrary spins $l_i$ can be written as \cite{Costa:2011mg}
\begin{equation}
    G(\{ X_i,Z_i\}) = \prod_{i<j}^n X_{ij}^{-\alpha_{ij}} \sum_k f_k(u_a) Q^{(k)}(\{X_i, Z_i\}) \label{CPPRcorrelator}
\end{equation}
Here, the lowercase indices label different points in the embedding space and the uppercase indices label different components of a particular tensor. Also, $X_{ij}:=-2X_i \cdot X_j$ and the $\cdot$ has been taken with respect to the natural $\mathbb{R}^{d+1,1}$ inner product. The exponents are
\begin{equation}
\alpha_{ij} = \frac{\tau_i + \tau_j}{n-2} - \frac{1}{(n-1)(n-2)} \sum_{k=1}^n \tau_k
\label{alphas}
\end{equation}
and $\tau_i=\Delta_i+l_i$ is the ``conformal spin'' \cite{Caron-Huot:2017vep} and $\Delta_i$ and $l_i$ are the scaling dimensions and spins of the $i^{th}$ primary. The $f_k$'s are theory-specific functions of the $u_a$'s, which in turn represent the $n(n-3)/2$ conformal cross ratios
\begin{equation}
    u = \frac{X_{12} X_{34}}{X_{13} X_{24}}\quad\quad\quad v = \frac{X_{14} X_{23}}{X_{13} X_{24}}.\label{crossratios}
\end{equation}
Finally, $ Q^{(k)}(\{X_A, Z_A\})$ is the tensor structure contracted with the appropriate factors of $Z_A$ and take the following form \cite{Costa:2011mg}
\begin{equation}
    \left( \prod_{i=1}^n \prod_{j \neq i, i+1}^n V_{i,(i+1)j}^{m_{ij}} \right) \prod_{i<j}^n H_{ij}^{n_{ij}}
\end{equation}
such that 
\begin{equation}
    \quad \sum_{j \neq i, i+1}^n m_{ij} + \sum_{j \neq i}^n n_{ij} = l_i,
\end{equation}
where $H_{ij}$ and $V_{i,jk}$ are building blocks of the tensor structures of the correlators.

In particular $H_{ij}$ is the tensor structure that appears in $2$-point correlator of two spin-1 primaries and $V_{i,jk}$ is the one that appears in a $3$-point correlator of a spin-1 and two scalar primaries where the spin-1 operator is at $X_i$. Explicitly, they can be expressed as:
\begin{equation}
H_{ij} = -2[(Z_i \cdot Z_j)(X_i \cdot X_j) - (Z_i \cdot X_j)(Z_j \cdot X_i)] \label{Hij}
\end{equation}
\begin{equation}
V_{i,jk} = \frac{(Z_i \cdot X_j)(X_i \cdot X_k) - (Z_i \cdot X_k)(X_i \cdot X_j)}{(X_j \cdot X_k)}.
\label{Vijk}
\end{equation}
For fixed $i$ the three structures $V_{i,jk}$ are not independent so any two of them may be used as a basis. Below we choose, for each $i$, whichever pair is most convenient, rather than the $V_{i,(i+1)j}$ of \eqref{CPPRcorrelator}.
For example, the embedding polynomial of the 4-point correlator $\langle A_M(X_1)A_N(X_2) \Phi(X_3) \Phi(X_4) \rangle$ contains the tensor structures $H_{12}, V_{1,24}V_{2,34} , V_{1,24}V_{2,31}, V_{1,23}V_{2,34}$ and $ V_{1,23}V_{2,31}$ where $A_M$ and $\Phi$ are  spin-1 and spin-0 primaries respectively on the embedding space. Just like the original fields, the correlators on the physical spacetime can be obtained by restricting to the Poincar\'e section. Since the expressions for the correlators frequently involve factors of $X_1 \cdot X_2$, $X_1 \cdot Z_2$ and $X_2 \cdot Z_1$ it is helpful to note their projections to $\mathbb{R}^d$ \cite{Costa:2011mg} 
\begin{equation}\label{projections}
    X_1 \cdot X_2 \to -\frac{1}{2}x_{12}^2, \quad Z_A \cdot Z_B \to z_A \cdot z_B, \quad Z_A \cdot X_2 \to- z_A \cdot x_{12}, \quad X_1 \cdot Z_B \to z_B \cdot x_{12}
\end{equation}
These can be easily derived by choosing a representative point on the Poincar\'e section. For example, Using $X_1 = (1, x_1^2, x_1)$ and $X_2 = (1, x_2^2, x_2)$ we find 
\begin{align}
    X_1 \cdot X_2 &= -\frac{1}{2}(X_1^+ X_2^- + X_1^- X_2^+) + X_1^\mu X_2^\mu \notag \\
    &= -\frac{1}{2}(x_2^2 + x_1^2) + x_1 . x_2 \notag \\
    &=  -\frac{1}{2}(x_1 - x_2)^2 = -\frac{1}{2}x_{12}^2
\end{align}
Similarly choosing $Z_2 = (0, 2x_2 . z_2, z_2)$ one finds 
$$\begin{aligned}
X_1 \cdot Z_2 &= -\frac{1}{2}(X_1^+ Z_2^- + X_1^- Z_2^+) + X_1^\mu Z_2^\mu \\
 &= -\frac{1}{2}\big(2 x_2 . z_2 + 0\big) + x_1 . z_2 \\
 &= z_2 . (x_1 - x_2) = z_2 . x_{12}
\end{aligned}$$
where the $.$ is the inner product on $\mathbb{R}^d$ and $x_{ij}=(x_i-x_j)$ in contrast to $X_{ij}:=-2 X_i \cdot X_j$. 
\section{Spinning Conformal Correlators}\label{sec:main}

In this section we present our construction of spinning conformal fields, exemplified  by computing correlation functions for a variety of spins and conformal dimensions.

\subsection{Principles for Spinning NN-CFT}
As in the scalar case \cite{Halverson:2024axc} the following must hold: 
\begin{enumerate}
    \item \textbf{Homogeneity in $X$:} The field must transform homogeneously under scalings of $X$, which is achieved by a suitable choice of the neural network architecture.
    \item \textbf{Lorentz-invariance:} The theory must be Lorentz-invariant in $(d+2)$ dimensions, which is ensured by an appropriately chosen parameter density $P(\Theta)$ and choosing a $SO(d+2)$-symmetric Euclidean theory first and later Wick rotating the correlators to Lorentzian signature, followed by a restriction to the PNC.
    \item \textbf{Finiteness:} The resulting correlators must be well-defined and finite, avoiding co-incident point divergence integration.
\end{enumerate}
Additionally, to extend to the spinning case, we impose the following conditions in accordance with the discussion in section \ref{sec:review}:
\begin{enumerate}
    \setcounter{enumi}{3}
    \item{ \textbf{Homogeneity in $Z$}}: The embedding space polynomials must be a homogeneous polynomial in $Z$ of degree $l$
    \item{\textbf{Transversality:}} The field $\mathcal{F}$ must satisfy the transversality condition $X \cdot \mathcal{F}=0$ so that the projection to physical space is unique.
\end{enumerate}
Together, these define our prescription for constructing spinning conformal correlators.

In studying correlation functions of operators with different spin we will sometimes utilize the same parameters in the architectures associated to different primaries. This is an essential mechanism to generate non-trivial correlations between those operators, building on results from \cite{Halverson:2024axc}, where such parameter overlaps were essential for partial OPE matching.

\subsection{Concrete Realization}
We exemplify the construction with a concrete architecture for a primary $\mathcal{F}$ of $SO(d)$ spin $J$ and scaling dimension $\Delta$:
\begin{align}
    \mathcal{F}_{\Theta,\eta;\tilde{\Theta}}(X,Z)&=(\tilde{\Theta}\cdot X)^{-\Delta-J}
\Big((\Theta\cdot X)(\eta\cdot Z)-(\eta\cdot X)(\Theta\cdot Z)\Big)^J \nonumber \\
\Theta, \tilde\Theta, \eta &\sim \mathcal{N}(0,1)\label{ansatz},
\end{align}
where the $\sim$ means that the neural network parameters are drawn i.i.d. from a Gaussian with zero mean and unit covariance, in which case $\langle \Theta^M\rangle=0, \langle \Theta^M \Theta^N \rangle=\delta^{M N}$, and likewise for the other parameters \footnote{In principle we can include a non-trivial constant in the $2$-point function of the parameters which can be later utilized for fixing the normalization of the fields and hence the correlators; but this won't affect most of our discussions except for certain parts of section \ref{Maxwellsection}.}. 

We will check the construction for various spins $J$ and conformal dimensions $\Delta$.
As a first check, we find that for $J=0$ the above ansatz reduces to
\begin{equation}\label{scalaransatzfirst}
    \mathcal{F}(X,Z)=(\tilde{\Theta}\cdot X)^{-\Delta}
\end{equation}
which was introduced in \cite{Halverson:2024axc}. Therefore, the scalar correlators can be deduced using the methods of that paper. However, our ansatz will also allow us to study $J\neq 0$. For brevity, we will hereafter drop the $\Theta, \eta,\tilde{\Theta}$ subscript from the NN fields and use $\Phi(X)$ to denote the neural network field rather than the original QFT field whenever notational confusion is unlikely. Henceforth, we denote scalar fields by $\Phi(X)$, spin $J=1$ fields by
\begin{equation}
A(X,Z)=(\tilde{\Theta}\cdot X)^{-\Delta-1}
\left((\Theta\cdot X)(\eta\cdot Z)-(\eta\cdot X)(\Theta\cdot Z)\right). \label{spin1ansatz}
\end{equation}
and 
for $J=2$ we use 
\begin{equation}
    T(X,Z)=(\tilde{\Theta} \cdot X)^{-\Delta-2}\left((\Theta\cdot X)(\eta\cdot Z)-(\eta\cdot X)(\Theta\cdot Z)\right)^2.
\end{equation}
In each case we check whether the two-, three-, and four-point functions have the required conformal structure, with non-trivial functional dependence arising at four points.  We note on the passing that the bracket in \eqref{ansatz} is not merely transverse but exactly invariant under $Z \to Z + \alpha X$:
\begin{align}
(\Theta\cdot X)\big(\eta\cdot Z + \alpha\,\eta\cdot X\big) - (\eta\cdot X)\big(\Theta\cdot Z + \alpha\,\Theta\cdot X\big)
&= (\Theta\cdot X)(\eta\cdot Z) - (\eta\cdot X)(\Theta\cdot Z) \notag \\
&\quad + \alpha\big[(\Theta\cdot X)(\eta\cdot X) - (\eta\cdot X)(\Theta\cdot X)\big],
\end{align}
and the last bracket vanishes identically. The field \eqref{ansatz} therefore depends on $Z$ only through the equivalence class $Z \sim Z+\alpha X$, and so defines a section on the projective null cone rather than just one transverse representative of it.

\subsubsection*{Two-point functions}\label{2ptfn}
We briefly review the NN method of obtaining the scalar $2$-point correlator $\langle \Phi(X_1) \Phi(X_2)\rangle$ here following \cite{Halverson:2024axc}. From the above discussion the NN architecture for a scalar field at $X_1$ for scaling dimension $\Delta=-1$ (for simplicity) is
\begin{equation}\label{scalarr}
    \Phi(X_1)=\Theta \cdot X_1
\end{equation}
So the $2$-point correlator is 
\begin{align}
    \langle \Phi(X_1) \Phi(X_2)\rangle&=\langle (\Theta \cdot X_1) (\Theta \cdot X_2)\rangle \notag\\
    &=\langle \Theta^{M} \Theta^{N}\rangle X_{1M} X_{2N} \notag \\
    &=\delta^{MN}X_{1M}X_{2N} \notag \\
    &=X_1 \cdot X_2
\end{align}
The case of other negative $\Delta$ would involve higher-point Wick contractions and can be dealt with similarly. Methods of dealing with positive $\Delta$ are discussed in \ref{scalar2pt}. If we had used two different fields $\Phi(X_1)$ as in \eqref{scalarr} and $\tilde{\Phi}(X_2)=\tilde{\Theta} \cdot X_2$ where $\tilde{\Theta}$ is an independent parameter, then the $2$-point function would vanish due to $\langle \Theta^{\mu} \rangle=\langle \tilde{\Theta}^{\mu} \rangle=0$ despite being of the same scaling dimension: this can be interpreted as vanishing of the $2$-point normalization constant. Finally the $2$-point correlator $\langle \tilde{\Phi}(X_1) \tilde{\Phi}(X_2)\rangle$ is also $X_1 \cdot X_2$.

For the $2$-point correlator of two spin-1 primaries using the proposed ansatz, we find (cf. Appendix \ref{2ptappendix})
\begin{align}
\left\langle A(X_1,Z_1) A(X_2,Z_2)\right\rangle
&=
2 (X_1\!\cdot\! X_2)^{-\Delta-1}
\Big[
(X_1\!\cdot\! X_2)(Z_1\!\cdot\! Z_2)
-(X_1\!\cdot\! Z_2)(X_2\!\cdot\! Z_1)
\Big]
\label{eq:AA_final}
\end{align}
which is indeed the correct $2$-point function for spin-1 primaries. The term within the brackets is proportional to the tensor structure $H_{12}$ and the factor sitting outside is to ensure correct overall scaling. Indeed under $X_i \to \lambda_i X_i$ for $i=1,2$ the correlator $\langle A(X_1,Z_1) A(X_2,Z_2)\rangle$ scales as $\lambda_i^{-\Delta}$ as a CFT correlator should.

For a spin-2 traceless symmetric primary operator $T(X,Z)$ the two point correlator is 
\begin{equation}
    \langle T(X_1,Z_1) T(X_2,Z_2) \rangle=
    3(X_1\!\cdot\! X_2)^{-\Delta-2} H_{12}^2 
\end{equation}
as can be shown by detailed computations in \ref{TTApendix}. Since the ansatz for traceless symmetric 2nd rank tensors is just the vector ansatz \eqref{spin1ansatz} ``squared'', then in $d=2$ if the vector is also a Kac-Moody current of dimension $\Delta_A=1$, then $T$ manifestly is reminiscent of a stress tensor constructed out of Sugawara construction \cite{DiFrancesco:1997nk} \footnote{Normal ordering is trivial because the coincident point expectation value $\langle A(X,Z)^2\rangle=2(X^2Z^2-(X\cdot Z)^2)$ vanishes on the PNC, so that no subtraction is required.}.  

\subsubsection*{Three-point functions}
The three point function of three same scalar fields $\langle \Phi(X_1) \Phi(X_2) \Phi(X_3)\rangle$ with scaling dimension $\Delta=-1$ vanish. This is not unusual: a concrete example of this is the scalar primary $\Phi(X):=F_{\mu \nu}^2$ in free Maxwell theory in $d=4$ \cite{El-Showk:2011xbs}. To get the non-zero functional form one can compute the three-point function $\langle \Phi^2(X_1) \Phi^2(X_2) \Phi^2(X_3) \rangle$ which turns out to be non-zero and consistent with the usual CFT results as shown in \cite{Halverson:2024axc}.

Similarly we can compute the three point function $\langle \Phi(X_1) \tilde{\Phi}(X_2) A(X_3)\rangle$
where as before we choose
\begin{align}
\Phi(X_1) &= (\Theta \cdot X_1)^{-\Delta_1} \notag \\
\tilde{\Phi}(X_2) &= (\eta \cdot X_2)^{-\Delta_2} \notag \\
A(X_3,Z_3) &= (\tilde{\Theta} \cdot X_3)^{-\Delta_3-1} \left[ (\Theta \cdot X_3)(\eta \cdot Z_3) - (\eta \cdot X_3)(\Theta \cdot Z_3) \right] \label{PhiPhiAansatz}
\end{align}
This gives $\langle \Phi(X_1) \tilde{\Phi}(X_2) A(X_3)\rangle=0$ because $\tilde{\Theta}$ is an independent local parameter appearing only in $A(X_3)$, and hence must self-contract to yield terms like $X_3 \cdot X_3=X_3^2=0$ for $\Delta_3 <-1$. This is consistent with the known results because we are implicitly omitting the $X_i$ independent scaling in the $\Theta$ correlations. The vanishing of this three point correlator is interpreted as this scaling to be zero in this case. This is similar to the case of the 3-point function $ \langle \Phi(X_1) \Phi(X_2) \Phi (X_3) \rangle$ as discussed above. As a non-trivial check we can also see that $\Delta_1=\Delta_2=\Delta_3=-1$ gives us the non-zero result (cf. Appendix $\ref{PhiPhiAappendix}$)
\begin{align}
\langle \Phi(X_1) \tilde{\Phi}(X_2) A(X_3) \rangle 
&=
(X_1\!\cdot\! X_3)(X_2\!\cdot\! Z_3) - (X_1\!\cdot\! Z_3)(X_2\!\cdot\! X_3).
\label{eq:PhiPhiA_expanded}
\end{align}
This is indeed the correct result, as expected from the general formula 
\begin{align}
\langle \Phi(X_1) \tilde\Phi(X_2) A(X_3) \rangle &= (X_1 \cdot X_2)^{-\frac{1}{2}(\Delta_1 + \Delta_2 - \Delta_3 - 1)} \notag \\
&\quad \times (X_1 \cdot X_3)^{-\frac{1}{2}(\Delta_1 - \Delta_2 + \Delta_3 + 1)} \notag \\
&\quad \times (X_2 \cdot X_3)^{-\frac{1}{2}(-\Delta_1 + \Delta_2 + \Delta_3 + 1)} \notag \\
&\quad \times \frac{ (X_1 \cdot X_3)(X_2 \cdot Z_3) - (X_1 \cdot Z_3)(X_2 \cdot X_3) }{X_1 \cdot X_2}
\end{align}
which is obtained from \eqref{CPPRcorrelator} in the special case of $l=(0,0,1)$ and $n=3$. 

As a final check we also compute the three point correlator of three different vector primaries $\langle A_1(X_1,Z_1) A_2(X_2,Z_2) A_3(X_3,Z_3)\rangle$
using
\begin{align}
A_1(X_1,Z_1) &= (\Theta \cdot X_1)(\eta \cdot Z_1) - (\eta \cdot X_1)(\Theta \cdot Z_1) \notag \\
A_2(X_2,Z_2) &= (\tilde{\Theta} \cdot X_2)(\Theta \cdot Z_2) - (\Theta \cdot X_2)(\tilde{\Theta} \cdot Z_2) \notag \\
A_3(X_3,Z_3) &= (\eta \cdot X_3)(\tilde{\Theta} \cdot Z_3) - (\tilde{\Theta} \cdot X_3)(\eta \cdot Z_3)
\label{AAAansatz}\end{align}
As before the self 2-point correlators yields expected results from a CFT and the cross-correlators vanish. Also note that taking any two of these spin-1 fields would reproduce the spin-1 part of the architecture \eqref{AAPhiPhiansatz2} used to find the $4$-point correlator $\langle A_1(X_1) A_2(X_2) \Phi(X_3) \Phi(X_4) \rangle$.
The three point function $\langle A_1(X_1) A_2(X_2) A_3(X_3) \rangle$ is
\begin{align}
\langle A_1(X_1) A_2(X_2) A_3(X_3) \rangle 
&=
(X_1\!\cdot\! Z_2)(Z_1\!\cdot\! X_3)(X_2\!\cdot\! Z_3) 
- (X_1\!\cdot\! Z_2)(Z_1\!\cdot\! Z_3)(X_2\!\cdot\! X_3) 
\nonumber\\
&\qquad
- (X_1\!\cdot\! X_2)(Z_1\!\cdot\! X_3)(Z_2\!\cdot\! Z_3) 
+ (X_1\!\cdot\! X_2)(Z_1\!\cdot\! Z_3)(Z_2\!\cdot\! X_3) 
\nonumber\\
&\qquad
- (X_1\!\cdot\! X_3)(Z_1\!\cdot\! Z_2)(X_2\!\cdot\! Z_3) 
+ (X_1\!\cdot\! Z_3)(Z_1\!\cdot\! Z_2)(X_2\!\cdot\! X_3) 
\nonumber\\
&\qquad
+ (X_1\!\cdot\! X_3)(Z_1\!\cdot\! X_2)(Z_2\!\cdot\! Z_3) 
- (X_1\!\cdot\! Z_3)(Z_1\!\cdot\! X_2)(Z_2\!\cdot\! X_3).\label{AAAexpanded}
\end{align}
It is interesting to note that all other parameter permutations either yield zero or exactly this result up to a sign. The same isn't true for the $4$-point functions as will be evident from \eqref{target} and other similar equations in Appendix \ref{AAPhiPhiAll}.

The result in \eqref{AAAexpanded} can be expanded into the $4$ possible tensor structures for $\langle A_1(X_1) A_2(X_2) A_3(X_3) \rangle$ i.e. $H_{12}V_{3,12}, H_{13}V_{2,13}, H_{23}V_{1,23}$ and $V_{1,23}V_{2,13}V_{3,12}$ 
\begin{equation} \label{AAAtensorexpand}
    \langle A_1(X_1) A_2(X_2) A_3(X_3) \rangle = -\frac{1}{2} H_{12} V_{3,12} + \frac{1}{2} H_{13} V_{2,13} - \frac{1}{2} H_{23} V_{1,23} + V_{1,23} V_{2,13} V_{3,12}    
\end{equation}
as is verified in Appendix \ref{expansion3pt}. Unlike the $4$-point function $\langle A(X_1) A(X_2) \Phi(X_3) \Phi(X_4)\rangle$ the coefficients are not functions of $u$ and $v$ but pure numbers as expected from a CFT $3$-point function. The fact that none of them are required to be zero is a necessary condition for a fairly generic CFT that is devoid of obvious symmetries.

\subsubsection*{Four-point functions}
Next we move on to evaluate the $4$-point function $\langle A(X_1) A(X_2) \Phi(X_3) \Phi(X_4) \rangle$ restricting to $\Delta=-1$ for all the fields and using the ansatz:
\begin{align}
A(X_1,Z_1)&=(\Theta \cdot X_1) (\eta \cdot Z_1)-(\eta \cdot X_1) (\Theta \cdot Z_1) \notag\\
A(X_2,Z_2)&=(\Theta \cdot X_2) (\eta \cdot Z_2)-(\eta \cdot X_2) (\Theta \cdot Z_2) \notag \\
\Phi(X_3)&=\tilde{\Theta} \cdot X_3 \notag\\
\Phi(X_4)&=\tilde{\Theta} \cdot X_4 \label{GFFansatz}.
\end{align}
The four point function is (cf. Appendix \ref{AAPhiPhiGFF}) 
\begin{align}
\langle A(X_1) A(X_2) \Phi(X_3) \Phi(X_4) \rangle 
&=
2(X_1\!\cdot\! X_2)(Z_1\!\cdot\! Z_2)(X_3\!\cdot\! X_4) - 2(X_1\!\cdot\! Z_2)(Z_1\!\cdot\! X_2)(X_3\!\cdot\! X_4).
\label{eq:AAPhiPhi_expanded}
\end{align}
We recognize the above as $-(X_3 \cdot X_4)H_{12}$ where the $X_3 \cdot X_4$ arises from the $2$-point function $\langle \Phi(X_3)  \Phi(X_4) \rangle$  and the $H_{12}$ arises from the $2$-point function $\langle A(X_1) A(X_2) \rangle$ i.e. the $4$-point function factorizes as $\langle A(X_1) A(X_2) \Phi(X_3) \Phi(X_4) \rangle= \langle A(X_1) A(X_2) \rangle \langle \Phi(X_3) \Phi(X_4)\rangle$. But one should not think from this particular correlator that the architecture here describes the Generalized Free Field (GFF) sector of a CFT with spin-$0$ and spin-$1$ primaries, because as we will see shortly the correlator $\langle A(X_1) A(X_2) A(X_3) A(X_4)\rangle$ does not factorize, so this is not a GFF. This is expected because the architecture for spin-$1$ (and $\Delta=-1$) is  bilinear in the Gaussian parameters $\Theta$ and $\eta$. A different architecture  inspired by the large-$N$ methods of \cite{Halverson:2024axc} which produces the GFF sector is discussed in Appendix \ref{sec:exact_gff}.

Recall from end of section \ref{sec:review} that there are $4$ more tensor structures that could appear in such a $4$-point function. By choosing appropriate neural network parameters, we can get a variety of other correlators. For example if we use
\begin{align}
A_1(X_1,Z_1)&=(\eta_1 \cdot Z_1)(\Theta_1 \cdot X_1) - (\eta_1 \cdot X_1)(\Theta_1 \cdot Z_1) \notag \\
A_2(X_2,Z_2)&=(\Theta_1 \cdot Z_2)(\Theta_2 \cdot X_2) - (\Theta_1 \cdot X_2)(\Theta_2 \cdot Z_2) \notag \\
\Phi_3(X_3)&=\Theta_2 \cdot X_3 \notag \\
\Phi_4(X_4)&=\eta_1 \cdot X_4 \label{AAPhiPhiansatz2}
\end{align}
we find (cf. Appendix \ref{AAPhiPhiInteracting})
\begin{align}
&\langle A_1(X_1,Z_1) A_2(X_2,Z_2) \Phi_3(X_3) \Phi_4(X_4) \rangle \notag \\
&= (Z_1 \cdot X_4)(X_1 \cdot Z_2)(X_2 \cdot X_3) - (Z_1 \cdot X_4)(X_1 \cdot X_2)(Z_2 \cdot X_3) \notag \\
&\quad - (X_1 \cdot X_4)(Z_1 \cdot Z_2)(X_2 \cdot X_3) + (X_1 \cdot X_4)(Z_1 \cdot X_2)(Z_2 \cdot X_3)\label{obtained}
\end{align}
To see that this indeed produces some of the $VV$-type structures we note that \eqref{obtained} can be written as a linear combination of $H_{12}$ and $V_{1,24}V_{2,13}$ with the coefficients being functions of cross-ratios $u$ and $v$ defined in \eqref{crossratios}, as expected from \eqref{CPPRcorrelator}. To wit,
\begin{equation}
\langle A_1(X_1,Z_1) A_2(X_2,Z_2) \Phi_3(X_3) \Phi_4(X_4) \rangle = P \left[ -\frac{1}{4} u^{-2/3} v^{5/6} H_{12} + \frac{1}{2} u^{-2/3} v^{-1/6} V_{1,24}V_{2,13} \right] \label{target} 
\end{equation}
where $P$ is the overall factor $\prod_{i<j}^4 X_{ij}^{-\alpha_{ij}}$ with $\alpha_{ij}$ as in \eqref{alphas}. The explicit expression for $P$ is in \eqref{prefactor4} of Appendix \ref{expansion4pt} where the expansion \eqref{target} has also been verified. We can also express the above result in a form similar to the more familiar form for $4$-point functions for scalars \cite{Costa:2011mg,DiFrancesco:1997nk}: 
\begin{equation}
    G_4(\{X_i,Z_i\}_{i=1}^4) = \frac{\left(\frac{X_{24}}{X_{14}}\right)^{\frac{\tau_1-\tau_2}{2}} \left(\frac{X_{14}}{X_{13}}\right)^{\frac{\tau_3-\tau_4}{2}}}{(X_{12})^{\frac{\tau_1+\tau_2}{2}} (X_{34})^{\frac{\tau_3+\tau_4}{2}}} \sum_k f_k(u, w) Q^{(k)} \label{altform}
\end{equation} where the $Q^{(k)}$s are $H_{12},W_1 W_2, \bar{W_1}\bar{W_2}$ and $\bar{W}_2 W_1-\bar{W}_1 W_2$ and $w=v/u$. The definitions of the $W_i$s are \cite{Costa:2011mg}
\begin{align*}
W_1 &:= V_{1,23} + V_{1,24} , & \bar{W}_1 &:= V_{1,23} - V_{1,24} \\
W_2 &:= V_{2,13} + V_{2,14} , & \bar{W}_2 &:= V_{2,13} - V_{2,14}
\end{align*}
The overall factor in \eqref{altform} evaluates to be $X_{34}$ for our case and \eqref{target} gives us
\begin{align}
        &\langle A_1(X_1,Z_1) A_2(X_2,Z_2) \Phi_3(X_3) \Phi_4(X_4) \rangle \notag \\&=X_{34} \left[ -\frac{w}{4} H_{12} + \frac{1}{8u} W_1 W_2 - \frac{1}{8u} \bar{W}_1 \bar{W}_2 + \frac{1}{8u} (W_1 \bar{W}_2 - \bar{W}_1 W_2) \right]
\end{align}
The other permutations of the parameters in \eqref{AAPhiPhiansatz2} yields linear combinations of different tensor structures. A complete list of replacements and the obtained tensor structures is listed in Appendix \ref{AAPhiPhiAll}.
One can also check the two-point functions of the fields in \eqref{AAPhiPhiansatz2}: the correlators of the same field trivially like $\langle A_1(X_1,Z_1)A_1(X_2,Z_2) \rangle$ reduce to the correlators evaluated earlier, and the cross correlators like $\langle A_1(X_1,Z_1) A_2(X_2,Z_2) \rangle$ are zero because all the parameters have vanishing $1$-point functions.

Now let us look at the four-point correlator $\langle A(X_1) A(X_2) A(X_3) A(X_4)\rangle$ given by\footnote{See Appendix \ref{AAAAresults} for details and also the $\langle A_1(X_1,Z_1) A_1(X_2,Z_2) A_2(X_3,Z_3) A_2(X_4,Z_4)$ case.},

\begin{equation}\label{eq:4AAAA}
\begin{aligned}
&\langle A(X_1) A(X_2) A(X_3) A(X_4)\rangle\\
&= \left(\frac{1+u+2v}{2v}\right) H_{14}H_{23}
+ \left(\frac{u+v+2}{2}\right) H_{13}H_{24}
+ \left(\frac{1+2u+v}{2u}\right) H_{12}H_{34} \\[6pt]
&\quad - \left(\frac{1+v}{u}\right) H_{12}V_{3,14}V_{4,13}
+ H_{12}V_{3,14}V_{4,12}
+ H_{12}V_{3,12}V_{4,13}
+ H_{13}V_{2,14}V_{4,13} \\
&\quad - (u+v) H_{13}V_{2,14}V_{4,12}
+ H_{14}V_{2,14}V_{3,12}
+ H_{13}V_{2,13}V_{4,12}
+ H_{14}V_{2,13}V_{3,14} \\
&\quad - \left(\frac{1+u}{v}\right) H_{14}V_{2,13}V_{3,12}
+ \left(\frac{1}{v}\right) H_{23}V_{1,24}V_{4,13}
+ \left(\frac{1}{v}\right) H_{23}V_{1,24}V_{4,12}
\\
&\quad - H_{24}V_{1,24}V_{3,12} - \left(\frac{1}{u}\right) H_{34}V_{1,24}V_{2,13}
- H_{23}V_{1,23}V_{4,12}
+ v\, H_{24}V_{1,23}V_{3,14} \\
&\quad + v\, H_{24}V_{1,23}V_{3,12}- \left(\frac{v}{u}\right) H_{34}V_{1,23}V_{2,14} \\[6pt]
&\quad + 2\, V_{1,24}V_{2,14}V_{3,12}V_{4,12}
+ \left(\frac{2}{u}\right) V_{1,24}V_{2,13}V_{3,14}V_{4,13}
- \left(\frac{2}{v}\right) V_{1,24}V_{2,13}V_{3,12}V_{4,13} \\
&\quad - \left(\frac{2}{v}\right) V_{1,24}V_{2,13}V_{3,12}V_{4,12}
+ \left(\frac{2v}{u}\right) V_{1,23}V_{2,14}V_{3,14}V_{4,13}\\
&\quad - 2v\, V_{1,23}V_{2,14}V_{3,12}V_{4,12}
+ 2\, V_{1,23}V_{2,13}V_{3,12}V_{4,12} - 2v\, V_{1,23}V_{2,14}V_{3,14}V_{4,12}
\end{aligned}
\end{equation}
We do not attempt an explicit computation of four-point functions of spin-2 fields in this section, which we expect to be more complicated than \eqref{eq:4AAAA}. However, there is immense simplification in the case of Maxwell CFT in section \ref{Maxwellsection} due to its Gaussian nature, and we will compute the exact four-point correlators of rank-2 antisymmetric fields.

The expansion coefficients $f_k(u,v)$ can be found in principle by writing a generic linear combination of all the tensor structures and comparing the correlator obtained from the NN ansatz $G$ to the expansion in terms of the tensor structures $\sum_k f_k Q^k$ where the index $k$ runs over different tensor structures expected from a particular correlator and we have absorbed the prefactor $P$ into $Q^k$ for notational clarity.
Since the vector space spanned by the tensor structures does not come with a natural inner product, we cannot use orthonormality of the basis to extract the $f_k$ in any straightforward way.
One way would be to completely expand both sides to write them in terms of $X_i\cdot Z_j$ and $Z_i \cdot Z_j$ and compare each term of both sides to extract the $f_k$s but this is tedious because the number of tensor structures as well as the number of terms in the expansion grow exponentially in $n$ \cite{Costa:2011mg}.
To make this more efficient, one can substitute several different choices for all the polarization vectors in the NN ansatz so as to simplify both sides of $\sum_k f_k Q^k=G$. This would yield a linear system of equations $G_{\{l\}}=\sum_k f_k Q^k_{\{l\}}$ where $l$ indices different choices of $Z_i$ i.e. $l=1,2, ..., N$ with $N \ge \text{number of tensor structures}$. For judiciously chosen $Z_i$s the system is non-singular and can be solved for $f_k$. For example one choice is $Z_i=X_j$ with $i \ne j$: The case of $i=j$ would kill all the terms in the tensor structures $Q^k$ as can be seen from the building blocks \eqref{Hij} and \eqref{Vijk}  and make the system singular. However, the above substitution with $i \ne j$ will yield too many possible choices for the polarization vectors and might also render the system of equations singular. For example, for the $4$-point correlator $\langle A_1(X_1) A_2(X_2) \Phi_3(X_3) \Phi_4(X_4) \rangle$ we can set $Z_1= X_2, X_3, \text{or~} X_4$ and $Z_2= X_1, X_3, \text{or~} X_4$ giving us $9$ choices for the $2$ polarization vectors $Z_1, Z_2$ but since there are $5$ tensor structures for this correlator, we can minimally use only $5$ out of them. So we can build $\binom{9}{5}=126$ linear systems to find the coefficients $f_k(u,v)$. Out of them $45$ are singular and $81$ of them yield the same result \eqref{obtained} which was also verified directly in Appendix \ref{expansion4pt}. Another way the $f_k$s could be found is by fixing the $X$s instead: in particular we can set some of the $X$ to $0,1, \infty$ to extract the $f_k$. These architectures almost always yield a linear combination of different tensor structures with non-zero coefficients. The vanishing of some of the coefficients translates to absence of certain primaries in the OPE channels. A straightforward conformal block decomposition computation should yield the operators that appear in the intermediate OPE channels and their conformal data for the architecture we used and lead to a concrete identification of the CFT we found. One of the ways to obtain the missing tensor structures is to modify the statistics accordingly by either breaking Gaussianity or breaking the assumption of independent and identically distributed NN parameters \cite{Demirtas:2023fir}. An alternative would be to modify the architecture by itself but still endow it with with Gaussian statistics. These methods could potentially be used to reverse-engineer a desired CFT i.e. given particular forms of $f(u,v)$, one can tune the statistics of the NN parameters to match the CFT correlators. Explicit examples of this are left for future work.

\section{4d Maxwell CFT as NN-CFT}\label{Maxwellsection}

In this section, we reproduce the local gauge invariant sector of free Maxwell theory in $d=4$ by using NN/FT techniques similar to those in the previous sections \footnote{The embedding space here projects to the Euclidean space, but one can easily analytically continue the results to those of Lorentzian Maxwell at least for the local gauge invariant sector \cite{Gross1975}.}.  Free Maxwell theory in $d=4$ is Gaussian, and a Gaussian theory is
determined completely by its two-point function. This dictates the
strategy of the construction:
\begin{enumerate}
\item First in Section \ref{maxwelllikesection} we construct theories whose field strength two-point function
is exactly that of $d=4$ Maxwell theory upon restriction to the
Poincar\'e section, in the normalization of \cite{Dolan:2000ut}. These
theories are however not exactly Maxwell, since the connected piece of
their four-point correlators is non-zero: we call them ``Maxwell-like''.
\item Then, in Section \ref{largeNmain}, we take a large-$N$ ensemble of
such theories. The non-Gaussianities are suppressed at large $N$, while
the two-point function is unchanged.
\item At large $N$ the ensemble is Gaussian in field strength with vanishing one-point
function and two-point function matching that of Maxwell, so all of its
correlators are determined by Wick's
theorem. The Bianchi identity is not an additional requirement either:
our field strength is defined as the curvature of a vector potential, so
it holds identically rather than only under $\langle \cdot \rangle$. In
other words, every correlator which is polynomial in field strength and its derivatives agrees with that of free
Maxwell theory as defined by the action 
\begin{equation}
    S=\frac{1}{4}\int d^4x F_{\mu \nu} F^{\mu \nu}
\end{equation}
Hence, we obtain the local sector of $d=4$ Maxwell CFT.
\end{enumerate}

\subsection{Maxwell-like theories}\label{maxwelllikesection}

In this section we construct a finite-width NN-CFT whose two-point function coincides with that of 4d Maxwell theory.

We start by taking inspiration from the previous sections: Naively, one might think that we could directly use the architecture \eqref{spin1ansatz} to model free Maxwell theory, because it is known to be a CFT in $d=4$ \cite{Bateman:1910mvi,El-Showk:2011xbs,Cunningham:1910pxu}. However, recall that the vector potential $A_\mu(x)$ in pure Maxwell theory in $(3+1)$ dimensions is not a conformal primary because it transforms anomalously under conformal transformations. Moreover, the scaling dimensions of the Maxwell field $\Delta=1$ is below the unitarity bound for vectors $\Delta \ge d-1$ and hence is not a part of the unitary CFT spectrum \cite{El-Showk:2011xbs}. However, one can construct primaries in this CFT: the field strength $F_{\mu \nu}$ which is an anti-symmetric tensor primary and scalar primary $\Phi:=\frac{1}{4}F_{\mu\nu}^2$. Consequently, the results of \cite{Costa:2011mg} do not directly apply to the embedding space uplifts $A_M(X)$ and $F_{MN}(X)$: namely we expect new tensor structures which are not in \eqref{CPPRcorrelator} at least for field strength correlators. In Maxwell theory, the 2-point correlator $\langle A(X_1,Z_1)A(X_2,Z_2) \rangle$ is expected to have gauge-dependent terms which we do not aim to reproduce using our techniques. In fact, the vector potential $2$-point function $\langle A_\mu(x) A_\mu(0)\rangle$ in any gauge does not match the one obtained from $\langle A(X_1,Z_1) A(X_2,Z_2) \rangle$ of section \ref{sec:main} after projection to the physical spacetime. This is because the vector primary $2$-point correlator has the tensor structure of an involution unlike Maxwell theory which has the tensor structure of a projector. To wit, the Maxwell vector potential two-point function is 
\begin{equation}\label{eq:gaugepropagator}
    \langle A^\text{Maxwell}_\mu(x) A^\text{Maxwell}_\nu(0) \rangle = \frac{1}{4\pi^2 x^2} \left( \frac{1+\xi}{2} \delta_{\mu\nu} + (1-\xi) \frac{x_\mu x_\nu}{x^2} \right)
\end{equation}
whereas the CFT vector primary two-point function which was recovered using NNCFT techniques is 
\begin{equation}
    \langle A_\mu(x) A_\nu(0) \rangle = \frac{1}{(x^2)^\Delta} I_{\mu\nu}(x)
\end{equation}
where
\begin{equation}
    I_{\mu \nu}(x) = \delta_{\mu \nu} - 2\frac{x_\mu x_\nu}{x^2}
\end{equation}
There exists no $\xi$ for which the former reduces to the latter: for example, the ratio of the two terms for the vector primary case is $-2$; equating with the same for the Maxwell case we get 
$$\frac{1-\xi}{\frac{1+\xi}{2}} = -2$$ which does not admit any consistent solution for $\xi$.

Nonetheless, we directly compute the field strength correlator using  \eqref{spin1ansatz} and compare it with the same as obtained in the Maxwell CFT. The construction (after a few modifications explained below) will give us the correct $2$-point correlators for field strength but not so for $F_{\mu \nu}^2$ or other higher-point correlators. Hence, we call these ``Maxwell-like'' theories in what follows. Since the genuine Maxwell theory is free, we use the trick used to construct $d=4$ free boson in \cite{Halverson:2024axc} by taking an ensemble of $N$ neural networks, and finally taking the $N\rightarrow\infty$ limit, thereby reproducing exactly the Maxwell CFT in $d=4$. This is discussed in \ref{largeNmain}.

We start with \eqref{spin1ansatz} rewritten here for convenience,
\begin{equation}
A(X,Z)=(\tilde{\Theta}\cdot X)^{-\Delta-1}
\left((\Theta\cdot X)(\eta\cdot Z)-(\eta\cdot X)(\Theta\cdot Z)\right). \label{spin1ansatzMaxwell}
\end{equation}
Although we are primarily interested in the case $\Delta=1$, we will keep $\Delta$ arbitrary as of now and impose $\Delta=1$ later.
Since the projection to the physical space $\mathbb{R}^d$ is compatible with the tensorial form of the field strength in embedding space i.e., the projection is $SO(d+1,1)$ covariant, it takes the usual form \footnote{We will use the same symbol $A$ and $F$ for both the fields in $d$ dimensions and also its embedding space uplift. No confusion should arise, as the distinction will be clear from the indices and the arguments wherever relevant.}
\begin{equation}
    F_{MN}=\partial_M A_N-\partial_N A_M \label{FMN}
\end{equation}
To construct the field strength embedding space polynomial we use
\begin{equation}
    F(X,Z_A,Z_B)=F_{MN}(X)Z_{A}^MZ_{B}^N \label{Fdef}
\end{equation}
Note the use of two auxiliary polarization vectors $Z_A$ and $Z_B$ at the same point $X$. This is a departure from the rules in \eqref{CPPRcorrelator} that locally assigns a single polarization vector to a point in $\mathbb{R}^{d+2}$: if we would have used a single polarization vector (as in the discussion of $TT$ correlators in \ref{TTApendix}) then the embedding polynomial would have vanished due to antisymmetry of $F_{MN}$ (which follows from antisymmetry of the field strength $F_{\mu \nu}(x)$ at $x \in \mathbb{R}^d$). Using \eqref{FMN}, \eqref{Fdef} simplifies to 
\begin{equation}
    F(X,Z_1,Z_2)= \left(Z_1 \cdot \frac{\partial}{\partial X}\right) A(X,Z_2) - \left(Z_2 \cdot \frac{\partial}{\partial X}\right) A(X,Z_1)
\end{equation}
In particular using \eqref{spin1ansatzMaxwell} we have 
\begin{align}
    F(X,Z_1,Z_2)&= -(\Delta+1) (\tilde{\Theta} \cdot X)^{-\Delta-2} \Big\{ (\tilde{\Theta} \cdot Z_1) \big[ (\Theta \cdot X)(\eta \cdot Z_2) - (\eta \cdot X)(\Theta \cdot Z_2) \big] \notag \\
    &\qquad \qquad \qquad \qquad \quad - (\tilde{\Theta} \cdot Z_2) \big[ (\Theta \cdot X)(\eta \cdot Z_1) - (\eta \cdot X)(\Theta \cdot Z_1) \big] \Big\} \notag \\
    &\quad + 2 (\tilde{\Theta} \cdot X)^{-\Delta-1} \big[ (\Theta \cdot Z_1)(\eta \cdot Z_2) - (\eta \cdot Z_1)(\Theta \cdot Z_2) \big] \label{FansatzgenericDelta}
\end{align}

The field strength \eqref{Fdef} descends to the projective null cone only at $\Delta=1$. Under $Z_1 \to Z_1 + \alpha X$ the bracket $(\Theta\cdot X)(\eta\cdot Z_1)-(\eta\cdot X)(\Theta\cdot Z_1)$ is invariant, while $(\tilde{\Theta}\cdot Z_1) \to (\tilde{\Theta}\cdot Z_1) + \alpha(\tilde{\Theta}\cdot X)$ and similar for $\Theta \cdot Z$. Therefore from \eqref{FansatzgenericDelta} we get under $\delta Z_1=\alpha X$
\begin{align}
\delta F(X,Z_1,Z_2) &= \Big[-(\Delta+1)\,\alpha\,(\tilde{\Theta}\cdot X)^{-\Delta-1} + 2\alpha\,(\tilde{\Theta}\cdot X)^{-\Delta-1}\Big]\big[(\Theta\cdot X)(\eta\cdot Z_2)-(\eta\cdot X)(\Theta\cdot Z_2)\big] \notag \\
&= \alpha(1-\Delta)\,(\tilde{\Theta}\cdot X)^{-\Delta-1}\big[(\Theta\cdot X)(\eta\cdot Z_2)-(\eta\cdot X)(\Theta\cdot Z_2)\big] \notag \\
&= \alpha\,(1-\Delta)\,A(X,Z_2), \label{eq:Fshift}
\end{align}
with $A$ as in \eqref{spin1ansatzMaxwell}. The redundancy is thus exact precisely when $\Delta=1$; away from that value $F$ shifts by the vector field itself. This is a restatement of the result proved in Appendix \ref{cftresults}, that an antisymmetric level-one descendant of a vector primary is primary only for $\Delta=1$. 

The field strength $1$-point function vanishes because $\langle \Theta \rangle= \langle \eta\rangle=0$. The field strength $2$-point correlator $\langle F(X_1, Z_1, Z_2) F(X_2, Z_3, Z_4) \rangle$ computation can be done in two different ways which lead to the same result. One of them is by taking derivatives of the $\langle A(X_1,Z_1)A(X_2,Z_2)\rangle$ correlator similar to how one finds descendant correlators from primary correlators using usual CFT techniques. Namely, we use the following 

\begin{align}
F(X_1, Z_1, Z_2) &= \left(Z_1 \cdot \frac{\partial}{\partial X_1}\right) A(X_1, Z_2) - \left(Z_2 \cdot \frac{\partial}{\partial X_1}\right) A(X_1, Z_1) \\
F(X_2, Z_3, Z_4) &= \left(Z_3 \cdot \frac{\partial}{\partial X_2}\right) A(X_2, Z_4) - \left(Z_4 \cdot \frac{\partial}{\partial X_2}\right) A(X_2, Z_3)
\end{align}
so that the two point correlator $\langle F(X_1,Z_1,Z_2) F(X_2,Z_3,Z_4)\rangle$ becomes
\begin{align}
\langle F(X_1, Z_1, Z_2) F(X_2, Z_3, Z_4) \rangle &= \Big\langle \left[ \left(Z_1 \cdot \frac{\partial}{\partial X_1}\right) A(X_1, Z_2) - \left(Z_2 \cdot \frac{\partial}{\partial X_1}\right) A(X_1, Z_1) \right] \notag \\
&\qquad \times \left[ \left(Z_3 \cdot \frac{\partial}{\partial X_2}\right) A(X_2, Z_4) - \left(Z_4 \cdot \frac{\partial}{\partial X_2}\right) A(X_2, Z_3) \right] \Big\rangle \notag \\
&= \left(Z_1 \cdot \frac{\partial}{\partial X_1}\right) \left(Z_3 \cdot \frac{\partial}{\partial X_2}\right) \langle A(X_1, Z_2) A(X_2, Z_4) \rangle \notag \\
&\quad - \left(Z_1 \cdot \frac{\partial}{\partial X_1}\right) \left(Z_4 \cdot \frac{\partial}{\partial X_2}\right) \langle A(X_1, Z_2) A(X_2, Z_3) \rangle \notag \\
&\quad - \left(Z_2 \cdot \frac{\partial}{\partial X_1}\right) \left(Z_3 \cdot \frac{\partial}{\partial X_2}\right) \langle A(X_1, Z_1) A(X_2, Z_4) \rangle \notag \\
&\quad + \left(Z_2 \cdot \frac{\partial}{\partial X_1}\right) \left(Z_4 \cdot \frac{\partial}{\partial X_2}\right) \langle A(X_1, Z_1) A(X_2, Z_3) \rangle \label{FFtoAA}
\end{align}
Now using \eqref{eq:AA_final} we get (See Appendix \ref{FFfromAA})
\begin{align}
\langle F(X_1, Z_1, Z_2) F(X_2, Z_3, Z_4) \rangle &= -4(\Delta-1) (X_1 \cdot X_2)^{-\Delta-1} \big[ (Z_1 \cdot Z_3)(Z_2 \cdot Z_4) - (Z_1 \cdot Z_4)(Z_2 \cdot Z_3) \big] \notag \\
&\quad + 2(\Delta^2-1)(X_1 \cdot X_2)^{-\Delta-2} K_{\text{diff}} \label{eq:full_correlator_generic}
\end{align}
where $K_{\text{diff}}$ is 
\begin{align}
K_{\text{diff}} &:= (Z_1 \cdot Z_3)(X_1 \cdot Z_4)(X_2 \cdot Z_2) + (Z_2 \cdot Z_4)(X_1 \cdot Z_3)(X_2 \cdot Z_1) \notag \\
&\quad - (Z_1 \cdot Z_4)(X_1 \cdot Z_3)(X_2 \cdot Z_2) - (Z_2 \cdot Z_3)(X_1 \cdot Z_4)(X_2 \cdot Z_1). \label{KDiff}
\end{align}
For comparison to the result expected by taking an uplift (See Appendix \ref{DirectMaxwell}) of the Maxwell theory computation using CFT/perturbative QFT techniques, we can express the above in terms of the new tensor structures (cf. Appendix \ref{match})
\begin{align}
    H(Z_A, Z_B ; X_1, X_2) := -2 \big[ (Z_A \cdot Z_B)(X_1 \cdot X_2) - (Z_A \cdot X_2)(Z_B \cdot X_1) \big].\label{newtensorstructure}
\end{align}
as 
\begin{align}
\langle F(X_1, Z_1, Z_2) F(X_2, Z_3, Z_4) \rangle
&= \frac{1-\Delta}{(X_1 \cdot X_2)^{\Delta+3}} \Big[ H(Z_1, Z_3 ; X_1, X_2) H(Z_2, Z_4 ; X_1, X_2)
\notag \\
& \quad - H(Z_1, Z_4 ; X_1, X_2 \big) H(Z_2, Z_3 ; X_1, X_2) \Big] 
+ \frac{2(\Delta-1)^2}{(X_1 \cdot X_2)^{\Delta+2}} K_{\text{diff}}
\label{eq:closed_form}
\end{align}
For the case of interest $\Delta=1$, the above vanishes. 
This is expected because the $2$-point function of level-$1$ $2$-form descendant of vector conformal \emph{primaries} vanishes when $\Delta=1$ (cf. Appendix \ref{cftresults}). However, we can still recover the Maxwell field strength correlators using this architecture. Our construction manifestly forces the field strength  $F(X_k,Z_{i_1},Z_{i_2})$ to be a descendant of $A(X,Z)$ besides also being a primary: it is a descendant because it is obtained by taking derivatives of the conformal primary $A(X,Z)$ as defined in \eqref{spin1ansatzMaxwell}, and it is also a primary because $\Delta=1$ (That primality of an antisymmetric descendant of a conformal primary implies $\Delta=1$ is proved in Appendix \ref{cftresults}). But unlike our case, in pure Maxwell theory, $F$ is \emph{not} a descendant of some primary \cite{El-Showk:2011xbs}. The way we now impose this in our construction is by making the primary $A(X,Z)$ singular, so that it is decoupled from the theory. We implement this by rescaling and taking the limit 
\begin{equation}
    \tilde{A}(X_1,Z_1)=\lim_{\Delta\to 1}\frac{A(X_1,Z_1)}{\sqrt{1-\Delta}} \label{rescaled}
\end{equation}
Then ignoring subtleties related to the branches appearing from the square root\footnote{We can choose the cut to lie along some complex direction in the $\Delta$ space in which scaling dimensions of physical interest do not lie, which would give a phase factor that can be reabsorbed into the definition of $\tilde{A}(X,Z)$ and is not relevant for our purposes} the field strength correlator of these rescaled fields gives us 
the correct form of Maxwell field strength $2$-pt correlator:
\begin{align}
&\langle \tilde{F}(X_1, Z_1, Z_2) \tilde{F}(X_2, Z_3, Z_4) \rangle \notag \\
&= \frac{H(Z_1, Z_3 ; X_1, X_2) H(Z_2, Z_4 ; X_1, X_2)-H(Z_1, Z_4 ; X_1, X_2) H(Z_2, Z_3 ; X_1, X_2) }{(X_1 \cdot X_2)^4}\label{MaxwellFF}
\end{align}
which after projection to $\mathbb{R}^4$ becomes (cf. Appendix \ref{DirectMaxwell})
\begin{align}
\langle F_{\mu\nu}(x_1) F_{\lambda\sigma}(x_2) \rangle \propto \frac{1}{(x_{12}^2)^2} \big( I_{\mu\lambda}I_{\nu\sigma} - I_{\mu\sigma}I_{\nu\lambda} \big),
\end{align}
where $x_{12} = (x_1 - x_2)$ with $x_i \in \mathbb{R}^4$ and the inversion tensor is defined as 
\begin{equation}
    I_{\mu\nu} := \delta_{\mu\nu} - 2\frac{x_{12\mu} x_{12\nu}}{x_{12}^2}.
\end{equation}
The overall constant is not fixed at this stage, because \eqref{eq:AA_final} was written with the $\tilde{\Theta}$ average of the two prefactors $(\tilde{\Theta}\cdot X_i)^{-\Delta-1}$ retained only up to proportionality. It is fixed in Section \ref{largeNmain}, where the field normalization is chosen so that the projection reproduces the Dolan--Osborn value.

We should note that the limit in \eqref{rescaled} is understood to be taken after evaluating the correlator: this ensures that the $1$-point function of $\tilde{F}$ still vanishes, while the $\langle A(X_1,Z_1) A(X_2,Z_2)\rangle$ diverges. An alternative way to obtain the field strength two-point correlators is discussed in \ref{MaxwellfromnonprimaryA}.

One could argue that there is almost no new use of the NN technique here except for the fact that $\langle A(X_1,Z_1)A(X_2,Z_2) \rangle$ correlators can be found using NN techniques as in section \ref{2ptfn}. In that case, the above serves as a consistency check on the following method of computation which uses NN techniques more directly: Multiply expressions of the form \eqref{FansatzgenericDelta} and compute correlators of the resulting expressions w.r.t. the NN parameters. This involves calculating correlators which involve factors of the form $(\Theta \cdot X)^{-\alpha}$ with $\alpha \ge0$ which are technically more laborious than the case of $\alpha<0$ where Wick's theorem applies. However, using Schwinger parameterization techniques inspired from \cite{Capuozzo:2025ozt}, one can evaluate the field strength $2$-point correlator as done for the case of interest $\Delta=1$ in Appendix \ref{FFfromFF} using the field strength expression \eqref{FansatzgenericDelta} and obviously the answer matches \eqref{eq:closed_form} and after rescaling and taking the limit $\Delta \to 1$ it also matches \eqref{MaxwellFF}.

Higher-point correlators can be computed in a straightforward manner and one can show that the connected components are typically not zero as in a genuine Maxwell theory.
A moment of thought makes us realize that these deviations from pure Maxwell theory are really due to the non-Gaussianities arising from the architecture: although the NN parameters $\Theta, \eta$ are chosen from a Gaussian distribution, the architecture \eqref{spin1ansatz} is not linear in these parameters and hence this makes the theory non-Gaussian. One way of achieving Gaussianity and hence getting exact Maxwell is discussed next.
\subsection{Maxwell CFT at Large-$N$}\label{largeNmain}
In this section we promote the single-channel field
\eqref{spin1ansatzMaxwell} to a large ensemble of $N$ i.i.d.\ copies
and take the neural-network Gaussian-process (NNGP) limit $N \to \infty$ following \cite{Halverson:2024axc}. As detailed in Appendix \ref{sec:largeN}, this ensemble construction ``averages'' over multiple independent network channels. Each individual channel is
non-Gaussian, but the normalized sum over channels converges to a Gaussian
field by the central-limit theorem, while the two-point function is
unchanged.
If we fix the field normalization of $A(X,Z)$ (cf. Appendix \ref{match}), then
all the correlators of $F$ and $\Phi$ follow with the correct numerical coefficient after the rescaling, now that the sum over channels is Gaussian. Namely, take $N$ i.i.d.\ copies of the (normalized) single-channel field, labelled by the
channel index $c = 1,\dots,N$,
\begin{equation}\label{eq:normalizedAmaintext}
A^{(c)}(X,Z)=\frac{1}{2\sqrt{2} \pi} \lim_{\Delta \to 1} \frac{1}{\sqrt{1-\Delta}}(\tilde{\Theta}^{(c)}\cdot X)^{-\Delta-1}
\left(({\Theta}^{(c)}\cdot X)(\eta^{(c)}\cdot Z)-(\eta^{(c)}\cdot X)({\Theta}^{(c)}\cdot Z)\right).
\end{equation}
where each channel carries its own independent Gaussian parameters and no sum over $c$ is to be assumed 
\begin{equation}\label{eq:channelstatsmaintext}
    \langle \Theta^{(c)}_M \rangle = \langle \eta^{(c)}_M \rangle = 0, \qquad
    \langle \Theta^{(c)}_M \Theta^{(c')}_N \rangle = \delta^{cc'}\delta_{MN}, \qquad
    \langle \eta^{(c)}_M \eta^{(c')}_N \rangle = \delta^{cc'}\delta_{MN}.
\end{equation}
and it is implied that the limit is being taken after evaluating the correlators, and after taking $N\to\infty$ in the field $A_M^{(N)}(X)$ and its field strength defined as the 
channel ``averages''
\begin{equation}\label{eq:ensemblefieldsmaintext}
    A_M^{(N)}(X) = \frac{1}{\sqrt{N}}\sum_{c=1}^{N} A_M^{(c)}(X), 
    \qquad
    F^{(N)}(X,Z_A,Z_B) = \frac{1}{\sqrt{N}}\sum_{c=1}^{N} F^{(c)}(X,Z_A,Z_B),
\end{equation}
The field-strength two-point correlator
is diagonal in the channel index, since by \eqref{eq:channelstatsmaintext} distinct channels are
uncorrelated:
\begin{align}\label{eq:MaxwellFFchannelmaintext}
    \langle F^{(N)}(X_1, Z_1, Z_2)\, F^{(N)}(X_2, Z_3, Z_4) \rangle 
    &= \frac{1}{N}\sum_{c,c'=1}^{N}
       \langle F^{(c)}(X_1, Z_1, Z_2)\, F^{(c')}(X_2, Z_3, Z_4) \rangle \notag \\
    &= \frac{1}{N}\sum_{c,c'=1}^{N} \delta^{cc'}\,
       \langle F^{(c)}(X_1, Z_1, Z_2)\, F^{(c)}(X_2, Z_3, Z_4) \rangle \notag \\
    &= \frac{1}{N}\sum_{c=1}^{N}
       \langle F^{(c)}(X_1, Z_1, Z_2)\, F^{(c)}(X_2, Z_3, Z_4) \rangle .
\end{align}
Each channel contributes the same single-channel value, so the sum has $N$ identical terms. The factor of $N$ cancels the $1/N$ infront of it and the
ensemble correlator after taking $\Delta \to 1$ becomes,
\begin{align}\label{eq:MaxwellFFmaintext}
    \lim_{N\to\infty}\langle F^{(N)}(X_1, Z_1, Z_2)\, &F^{(N)}(X_2, Z_3, Z_4) \rangle \notag 
   \\ &= \frac{H(Z_1, Z_3 ; X_1, X_2) H(Z_2, Z_4 ; X_1, X_2)-H(Z_1, Z_4 ; X_1, X_2) H(Z_2, Z_3 ; X_1, X_2)}{16 \pi^2(X_1 \cdot X_2)^4}.
\end{align}
Restricting to the Poincar\'e section via \eqref{projections}, this projects exactly to the
standard $d=4$ Maxwell field-strength correlator
\begin{equation}\label{eq:MaxwellFFphysmaintext}
    \langle F_{\mu\nu}(x_1) F_{\sigma\rho}(x_2) \rangle =
    \frac{1}{\pi^2 r_{12}^4}
    \left( I_{\mu\sigma}(x_{12}) I_{\nu\rho}(x_{12}) - I_{\mu\rho}(x_{12}) I_{\nu\sigma}(x_{12}) \right),
\end{equation}
where $r_{12}^2 = x_{12}^2$ and $I_{\mu\nu}(x) = \delta_{\mu\nu} - 2\,x_\mu x_\nu / x^2$ is the
conformal inversion tensor, reproducing the Dolan--Osborn normalized result \cite{Dolan:2000ut}.
The four-point field-strength correlator factorizes into products of two-point
functions:
\begin{align}
    \langle F^{(N)}(X_1)\, F^{(N)}(X_2)\, F^{(N)}(X_3)\, F^{(N)}(X_4) \rangle
    &= \frac{1}{N}\, \langle F(X_1)\, F(X_2)\, F(X_3)\, F(X_4) \rangle
    \notag \\
    &\quad + \left(1 - \frac{1}{N}\right) \Big(
        \langle F(X_1) F(X_2) \rangle \langle F(X_3) F(X_4) \rangle
    \notag \\
    &\qquad\qquad\;\;
        + \langle F(X_1) F(X_3) \rangle \langle F(X_2) F(X_4) \rangle
    \notag \\
    &\qquad\qquad\;\;
        + \langle F(X_1) F(X_4) \rangle \langle F(X_2) F(X_3) \rangle
    \Big),
\end{align}
where we have temporarily suppressed the polarization vectors for notational simplicity, i.e.
$F(X_i) := F(X_i, Z_{a_i}, Z_{b_i})$. In the large-$N$ limit, the $1/N$-dependent
terms drop out and we obtain
\begin{align}
    \lim_{N \to \infty}
    \langle F^{(N)}(X_1)\, F^{(N)}(X_2)\, F^{(N)}(X_3)\, F^{(N)}(X_4) \rangle
    &= \langle F(X_1) F(X_2) \rangle \langle F(X_3) F(X_4) \rangle
    \notag \\
    &\quad + \langle F(X_1) F(X_3) \rangle \langle F(X_2) F(X_4) \rangle
    \notag \\
    &\quad + \langle F(X_1) F(X_4) \rangle \langle F(X_2) F(X_3) \rangle .
\end{align}
The two limits $N \to \infty$ and $\Delta \to 1$ do not commute, and the order above is the required one: $N\to\infty$ at fixed $\Delta\neq1$, and only afterwards $\Delta\to1$. This is because each rescaled field carries a factor $(1-\Delta)^{-1/2}$ from \eqref{rescaled}, while the single-channel connected correlators are finite and generically non-vanishing at $\Delta=1$, so the connected part of the $2n$-point function of $F^{(N)}$ scales as $N^{1-n}(1-\Delta)^{-n}$. At fixed $N$ the $\Delta\to1$ limit is therefore singular, and the $1/N$ suppression never acts. Taking $N\to\infty$ first removes every connected contribution at generic $\Delta$, after which the rescaled two-point function is finite and the $\Delta\to1$ limit is smooth.

Similarly, using this architecture all the higher order correlators of Maxwell theory can be determined essentially by Wick's theorem. Therefore, we obtain the local sector of Maxwell theory generated by the field strength. 

As additional consistency checks, we compute the correlators of the composite scalar primary $\Phi := \tfrac{1}{4}F_{MN} F^{MN}$. For the two-point function, the normal-ordered disconnected contractions dominate at large $N$,
\begin{equation}\label{eq:PhiPhiWickmaintext}
    \langle \Phi(X_1)\Phi(X_2)\rangle 
    = \frac{1}{8}\,\langle F_{M_1N_1}(X_1)F_{M_2N_2}(X_2)\rangle\,
    \langle F^{M_1N_1}(X_1)F^{M_2N_2}(X_2)\rangle + \mathcal{O}\!\left(\tfrac{1}{N}\right),
\end{equation}
and carrying out the tensor contractions of \eqref{eq:MaxwellFFmaintext} (cf. Appendix
\ref{sec:largeN}) gives a result in $d=4$ dimensions,
\begin{equation}\label{eq:PhiPhimaintext}
    \langle \Phi(X_1)\Phi(X_2)\rangle = \frac{4(4-1)}{64\pi^4 (X_1\cdot X_2)^4}
    \;\xrightarrow{\;\text{PS}\;}\;
    \langle \Phi(x_1)\Phi(x_2)\rangle = \frac{3}{\pi^4\, x_{12}^8}.
\end{equation}
which is precisely the free Maxwell value \cite{Dolan:2000ut}. This should be contrasted with the single-channel computation of Appendix \ref{sec:phiphifromprimaryF}, where the non-Gaussian artifacts spoiled the coefficient by a
factor of $5/8$. The same logic fixes the three-point function. Since $\Phi$ is quadratic in the Gaussian field
$F$, Wick's theorem leaves only the single connected loop of three $\langle FF\rangle$
contractions,
\begin{equation}\label{eq:PhiPhiPhiWickmaintext}
    \langle \Phi(X_1)\Phi(X_2)\Phi(X_3)\rangle 
    = \frac{1}{8}\,\langle F_{M_1N_1}(X_1)F_{M_2N_2}(X_2)\rangle
    \langle F^{M_2N_2}(X_2)F_{M_3N_3}(X_3)\rangle
    \langle F^{M_3N_3}(X_3)F^{M_1N_1}(X_1)\rangle.
\end{equation}
Evaluating the trace over the six inversion-tensor building blocks (cf. Appendix
\ref{sec:largeN}) gives
\begin{equation}\label{eq:PhiPhiPhimaintext}
    \langle \Phi(X_1)\Phi(X_2)\Phi(X_3)\rangle \propto 
    \frac{(d-1)(d-4)}{(X_1\cdot X_2)^2 (X_2\cdot X_3)^2 (X_1\cdot X_3)^2},
\end{equation}
where the proportionality constant is $d$-independent. The factor of $(d-4)$ makes the three-point function vanish identically in $d=4$, exactly as required for free Maxwell theory where
$C_{\Phi\Phi\Phi}=0$ \cite{Dolan:2000ut,El-Showk:2011xbs}.

\section{Discussions}\label{sec:discussion}

In this work, we extended the neural network-field theory construction of scalar conformal fields to spinning operators using the embedding-space formalism. We introduced architectures with the required homogeneity and transversality properties and showed explicitly that their two-, three-, and four-point functions reproduce the tensor structures required by conformal symmetry. An important feature of the construction is that correlations between network parameters provide a direct mechanism for controlling which tensor structures and consequently OPE data appear. Thus, neural-network architecture and parameter statistics together encode nontrivial information about the resulting conformal field theory \footnote{An algebraic perspective on these NN-CFTs will be discussed in an upcoming work \cite{future:JN-SL}}. 

As a concrete application, we constructed the local gauge-invariant sector of four-dimensional free Maxwell theory. A single network channel can reproduce the Maxwell field-strength two-point function but remains non-Gaussian and therefore does not reproduce the higher-point functions of the free theory. By instead considering a ensemble of $N$ independent channels and taking $N \to \infty$ connected non-Gaussian contributions are suppressed while the field-strength two-point function is retained. This is Maxwell CFT.

These results illustrate that for spinning theories, the sharing and statistics of parameters determine which allowed conformal tensor structures are realized. Understanding this map systematically, in particular, determining how desired conformal data constrain network architecture and parameter distributions - could provide a route toward constructing or reverse engineering more general CFTs: since $n$-point CFT correlators for $n > 4$ are determined in terms of conformal data available from low-point correlators, this becomes essentially a constrained factorization problem for the lower point correlators where we systematically derive one of the possible NN architectures for a given CFT without having to guess them. Extending NN/FT constructions to incorporate these structures, and developing a systematic relation between network data and CFT OPE data, are natural directions for future work.

There are some limitations to our approach.  We have restricted our attention to the local gauge-invariant sector and have not addressed the nonlocal  observables such as Wilson lines, ’t Hooft lines, flux operators, etc. Gauge symmetry related aspects of the Maxwell theory like the transformation law of $A_\mu(x)$ under gauge transformations was also not explored in this work because we decoupled the gauge potential from our theory. Finally, extending to $d \ne 4$ was out of scope because free Maxwell theory is not a CFT in $d=3,5$. However, see the upcoming work \cite{FerkoHackettHalverson2026} for an alternate NN/FT construction of Maxwell theory which directly addresses some of these limitations.

\section*{Acknowledgments}
We thank Christian Ferko, Samuel Leutheusser and Benjamin Suzzoni for discussions. M.D. and J.N. are partially supported by the Graduate Assistantship at Northeastern University. J.N. is currently supported by the Beijing Institute of Mathematical Sciences and Applications (BIMSA). J.H. is supported by NSF grant PHY-220990. This research was supported in part by grant NSF PHY-2309135 to the Kavli Institute for Theoretical Physics (KITP) and Cooperative Agreement PHY-2019786 (The NSF AI Institute for Artificial Intelligence and Fundamental Interactions). J.N. would like to thank the KITP for hospitality and accommodation during the Graduate Fellowship program.

\appendix
\addtocontents{toc}{\protect\setcounter{tocdepth}{1}}
\section{Calculation Details}
This appendix details the computations of the correlators discussed in the main text.
\subsection{$\langle \Phi(X_1) \Phi(X_2)\rangle$}\label{scalar2pt}
Using the NN ansatz \eqref{scalaransatzfirst} for $\Delta>0$ we use Schwinger parameterization \footnote{This is formal because of the divergences arising from $\Theta \cdot X$ being negative, but we are interested in extracting the finite results here.} inspired by \cite{Robinson:2025ybg}
\begin{equation}
A^{-\Delta} = \frac{1}{\Gamma(\Delta)} \int_0^\infty ds \, s^{\Delta-1} e^{-sA}
\end{equation}
 to get 
\begin{align}
    \langle \Phi(X_1) \Phi(X_2) \rangle& = 
 \langle (\Theta \cdot X_1)^{-\Delta} (\Theta \cdot X_2)^{-\Delta} \rangle \notag\\
&=\frac{1}{\Gamma(\Delta)^2} \int_0^\infty ds \int_0^\infty dt \, s^{\Delta-1} t^{\Delta-1} \langle e^{-s(\Theta \cdot X_1)} e^{-t(\Theta \cdot X_2)} \rangle \notag \\
&= \frac{1}{\Gamma(\Delta)^2} \int_0^\infty ds \int_0^\infty dt \, s^{\Delta-1} t^{\Delta-1} \langle e^{-\Theta \cdot (sX_1 + tX_2)} \rangle
\end{align}
Now let $J = -(sX_1 + tX_2)$. Using 
$\langle e^{\Theta \cdot J} \rangle = e^{\frac{1}{2}J^2}$ (See Appendix \ref{FFfromFF}), and 
\begin{equation}
    J^2 = (-(sX_1 + tX_2))^2 = s^2 X_1^2 + t^2 X_2^2 + 2st(X_1 \cdot X_2)=2st(X_1 \cdot X_2)
\end{equation}
using $X_1^2=0=X_2^2$, the integral becomes 
\begin{equation}
\langle \Phi(X_1) \Phi(X_2) \rangle = \frac{1}{\Gamma(\Delta)^2} \int_0^\infty ds \int_0^\infty dt \, s^{\Delta-1} t^{\Delta-1} e^{st(X_1 \cdot X_2)}
\end{equation}
To decouple the integrals, we change variables to $u = st$ and $v = s/t$.
The integral becomes
\begin{equation}
    \langle \Phi(X_1) \Phi(X_2) \rangle = \frac{1}{\Gamma(\Delta)^2} \int_0^\infty du \int_0^\infty \frac{dv}{2v} \, u^{\Delta-1} e^{u(X_1 \cdot X_2)}
\end{equation}

The divergent piece i.e. $v$ integral is the Haar measure of dilatation group $\mathbb{R}$ i.e. $\int_0^\infty \frac{dv}{v}$, representing the redundancy of the Schwinger parameterization of the integral which can be modded out. The divergences arising from $\Theta \cdot X=0$ are washed out in the prescription used above allowing us to extract the finite results outside the window (in scaling dimensions) of convergence.  (Also see Appendix \ref{FFfromFF} and \cite{Capuozzo:2025ozt} for a more elaborate discussion). We are left with the finite $u$ integral:

\begin{equation}
    \langle \Phi(X_1) \Phi(X_2) \rangle \propto \frac{1}{\Gamma(\Delta)^2} \int_0^\infty du \, u^{\Delta-1} e^{-u(-X_1 \cdot X_2)}
\end{equation}

Since $(-X_1 \cdot X_2) =\frac{1}{2}(x_1-x_2)^2> 0$  this reduces to a standard Gamma function integral which evaluates to 
\begin{equation}
\langle \Phi(X_1) \Phi(X_2) \rangle = \frac{c_\Delta}{(X_1 \cdot X_2)^\Delta}
\end{equation}
where $c_\Delta=\frac{(-1)^\Delta}{2 \Gamma(\Delta)}$ is a normalization constant which can be reabsorbed into the field redefinition of the scalar $\Phi(X)$ via a rescaling. For $\Delta \le0$ the Schwinger parameterization does not converge but in that case we can use the usual Wick-contraction technique as reviewed in the beginning of section \ref{sec:main}.

\subsection{ $\langle A(X_1,Z_1) A(X_2,Z_2) \rangle$}\label{2ptappendix}
Using the NN ansatz \eqref{spin1ansatz} we get 
\begin{align}
\left\langle A(X_1,Z_1) A(X_2,Z_2)\right\rangle
&=
\Big\langle
(\tilde{\Theta}\!\cdot\! X_1)^{-\Delta-1}
(\tilde{\Theta}\!\cdot\! X_2)^{-\Delta-1}
\nonumber\\
&\qquad\times
\left((\Theta\!\cdot\! X_1)(\eta\!\cdot\! Z_1)-(\eta\!\cdot\! X_1)(\Theta\!\cdot\! Z_1)\right)
\left((\Theta\!\cdot\! X_2)(\eta\!\cdot\! Z_2)-(\eta\!\cdot\! X_2)(\Theta\!\cdot\! Z_2)\right)
\Big\rangle
\nonumber\\[0.5em]
& \sim
(X_1\!\cdot\! X_2)^{-\Delta-1}
\Big\langle
(\Theta\!\cdot\! X_1)(\eta\!\cdot\! Z_1)(\Theta\!\cdot\! X_2)(\eta\!\cdot\! Z_2)
\nonumber\\
&\qquad
-(\Theta\!\cdot\! X_1)(\eta\!\cdot\! Z_1)(\eta\!\cdot\! X_2)(\Theta\!\cdot\! Z_2)
\nonumber\\
&\qquad
-(\eta\!\cdot\! X_1)(\Theta\!\cdot\! Z_1)(\Theta\!\cdot\! X_2)(\eta\!\cdot\! Z_2)
\nonumber\\
&\qquad
+(\eta\!\cdot\! X_1)(\Theta\!\cdot\! Z_1)(\eta\!\cdot\! X_2)(\Theta\!\cdot\! Z_2)
\Big\rangle
\nonumber\\[0.5em]
&=
(X_1\!\cdot\! X_2)^{-\Delta-1}
\Big\langle
(\Theta^\mu X_{1\mu})(\eta^\nu Z_{1\nu})
(\Theta^\rho X_{2\rho})(\eta^\sigma Z_{2\sigma})
\nonumber\\
&\qquad
-(\Theta^\mu X_{1\mu})(\eta^\nu Z_{1\nu})
(\eta^\rho X_{2\rho})(\Theta^\sigma Z_{2\sigma})
\nonumber\\
&\qquad
-(\eta^\mu X_{1\mu})(\Theta^\nu Z_{1\nu})
(\Theta^\rho X_{2\rho})(\eta^\sigma Z_{2\sigma})
\nonumber\\
&\qquad
+(\eta^\mu X_{1\mu})(\Theta^\nu Z_{1\nu})
(\eta^\rho X_{2\rho})(\Theta^\sigma Z_{2\sigma})
\Big\rangle
\nonumber\\[0.5em]
&=
(X_1\!\cdot\! X_2)^{-\Delta-1}
\Big[
\langle \Theta^\mu \Theta^\rho\rangle
\langle \eta^\nu \eta^\sigma\rangle
X_{1\mu} X_{2\rho} Z_{1\nu} Z_{2\sigma}
\nonumber\\
&\qquad
-\langle \Theta^\mu \Theta^\sigma\rangle
\langle \eta^\nu \eta^\rho\rangle
X_{1\mu} Z_{2\sigma} Z_{1\nu} X_{2\rho}
\nonumber\\
&\qquad
-\langle \Theta^\nu \Theta^\rho\rangle
\langle \eta^\mu \eta^\sigma\rangle
Z_{1\nu} X_{2\rho} X_{1\mu} Z_{2\sigma}
\nonumber\\
&\qquad
+\langle \Theta^\nu \Theta^\sigma\rangle
\langle \eta^\mu \eta^\rho\rangle
Z_{1\nu} Z_{2\sigma} X_{1\mu} X_{2\rho}
\Big]
\nonumber\\[0.5em]
&=
(X_1\!\cdot\! X_2)^{-\Delta-1}
\Big[
\delta^{\mu\rho}\delta^{\nu\sigma}
X_{1\mu} X_{2\rho} Z_{1\nu} Z_{2\sigma}
\nonumber\\
&\qquad
-\delta^{\mu\sigma}\delta^{\nu\rho}
X_{1\mu} Z_{2\sigma} Z_{1\nu} X_{2\rho}
\nonumber\\
&\qquad
-\delta^{\nu\rho}\delta^{\mu\sigma}
Z_{1\nu} X_{2\rho} X_{1\mu} Z_{2\sigma}
\nonumber\\
&\qquad
+\delta^{\nu\sigma}\delta^{\mu\rho}
Z_{1\nu} Z_{2\sigma} X_{1\mu} X_{2\rho}
\Big]
\nonumber\\[0.5em]
&=
(X_1\!\cdot\! X_2)^{-\Delta-1}
\Big[
(X_1\!\cdot\! X_2)(Z_1\!\cdot\! Z_2)
-(X_1\!\cdot\! Z_2)(X_2\!\cdot\! Z_1)
\nonumber\\
&\qquad
-(X_1\!\cdot\! Z_2)(X_2\!\cdot\! Z_1)
+(X_1\!\cdot\! X_2)(Z_1\!\cdot\! Z_2)
\Big]
\nonumber\\[0.5em]
&=
2 (X_1\!\cdot\! X_2)^{-\Delta-1}
\Big[
(X_1\!\cdot\! X_2)(Z_1\!\cdot\! Z_2)
-(X_1\!\cdot\! Z_2)(X_2\!\cdot\! Z_1)
\Big].
\end{align}
which was mentioned in \eqref{eq:AA_final}. In the first step we have used $\langle (\tilde{\Theta} \cdot X_1)^{-\Delta-1} (\tilde{\Theta}\cdot X_2)^{-\Delta-1} \rangle \sim(X_1 \cdot X_2)^{-\Delta-1}$ for generic $\Delta$ which is reviewed in section \ref{sec:main} (with special treatment of $\Delta>0$ in Appendix \ref{scalar2pt}).
\subsection{Spin-2 2-point correlator}\label{TTApendix}
\begin{align}
\langle T(X_1,Z_1) T(X_2,Z_2)&\rangle
=
\Big\langle (\tilde{\Theta}\!\cdot\! X_1)^{-\Delta-2} (\tilde{\Theta}\!\cdot\! X_2)^{-\Delta-2}
\nonumber\\
&\qquad\times
\left[ (\Theta\!\cdot\! X_1)(\eta\!\cdot\! Z_1) - (\eta\!\cdot\! X_1)(\Theta\!\cdot\! Z_1) \right]^2
\left[ (\Theta\!\cdot\! X_2)(\eta\!\cdot\! Z_2) - (\eta\!\cdot\! X_2)(\Theta\!\cdot\! Z_2) \right]^2 \Big\rangle
\nonumber\\[0.5em]
&=
(X_1\!\cdot\! X_2)^{-\Delta-2}
\Big\langle
\left[ (\Theta\!\cdot\! X_1)^2(\eta\!\cdot\! Z_1)^2 - 2(\Theta\!\cdot\! X_1)(\eta\!\cdot\! Z_1)(\eta\!\cdot\! X_1)(\Theta\!\cdot\! Z_1) + (\eta\!\cdot\! X_1)^2(\Theta\!\cdot\! Z_1)^2 \right]
\nonumber\\
&\qquad\times
\left[ (\Theta\!\cdot\! X_2)^2(\eta\!\cdot\! Z_2)^2 - 2(\Theta\!\cdot\! X_2)(\eta\!\cdot\! Z_2)(\eta\!\cdot\! X_2)(\Theta\!\cdot\! Z_2) + (\eta\!\cdot\! X_2)^2(\Theta\!\cdot\! Z_2)^2 \right]
\Big\rangle
\nonumber\\[0.5em]
&=
(X_1\!\cdot\! X_2)^{-\Delta-2}
\Big[
\langle (\Theta\!\cdot\! X_1)^2 (\Theta\!\cdot\! X_2)^2 \rangle \langle (\eta\!\cdot\! Z_1)^2 (\eta\!\cdot\! Z_2)^2 \rangle
\nonumber\\
&\qquad
- 2\langle (\Theta\!\cdot\! X_1)^2 (\Theta\!\cdot\! X_2)(\Theta\!\cdot\! Z_2) \rangle \langle (\eta\!\cdot\! Z_1)^2 (\eta\!\cdot\! X_2)(\eta\!\cdot\! Z_2) \rangle
\nonumber\\
&\qquad
+ \langle (\Theta\!\cdot\! X_1)^2 (\Theta\!\cdot\! Z_2)^2 \rangle \langle (\eta\!\cdot\! Z_1)^2 (\eta\!\cdot\! X_2)^2 \rangle
\nonumber\\
&\qquad
- 2\langle (\Theta\!\cdot\! X_1)(\Theta\!\cdot\! Z_1) (\Theta\!\cdot\! X_2)^2 \rangle \langle (\eta\!\cdot\! X_1)(\eta\!\cdot\! Z_1) (\eta\!\cdot\! Z_2)^2 \rangle
\nonumber\\
&\qquad
+ 4\langle (\Theta\!\cdot\! X_1)(\Theta\!\cdot\! Z_1) (\Theta\!\cdot\! X_2)(\Theta\!\cdot\! Z_2) \rangle \langle (\eta\!\cdot\! X_1)(\eta\!\cdot\! Z_1) (\eta\!\cdot\! X_2)(\eta\!\cdot\! Z_2) \rangle
\nonumber\\
&\qquad
- 2\langle (\Theta\!\cdot\! X_1)(\Theta\!\cdot\! Z_1) (\Theta\!\cdot\! Z_2)^2 \rangle \langle (\eta\!\cdot\! X_1)(\eta\!\cdot\! Z_1) (\eta\!\cdot\! X_2)^2 \rangle
\nonumber\\
&\qquad
+ \langle (\Theta\!\cdot\! Z_1)^2 (\Theta\!\cdot\! X_2)^2 \rangle \langle (\eta\!\cdot\! X_1)^2 (\eta\!\cdot\! Z_2)^2 \rangle
\nonumber\\
&\qquad
- 2\langle (\Theta\!\cdot\! Z_1)^2 (\Theta\!\cdot\! X_2)(\Theta\!\cdot\! Z_2) \rangle \langle (\eta\!\cdot\! X_1)^2 (\eta\!\cdot\! X_2)(\eta\!\cdot\! Z_2) \rangle
\nonumber\\
&\qquad
+ \langle (\Theta\!\cdot\! Z_1)^2 (\Theta\!\cdot\! Z_2)^2 \rangle \langle (\eta\!\cdot\! X_1)^2 (\eta\!\cdot\! X_2)^2 \rangle
\Big]
\nonumber\\[0.5em]
&=
(X_1\!\cdot\! X_2)^{-\Delta-2}
\Big[
\big(\langle \Theta^\mu \Theta^\nu \Theta^\rho \Theta^\sigma \rangle X_{1\mu} X_{1\nu} X_{2\rho} X_{2\sigma} \big) \big(\langle \eta^\alpha \eta^\beta \eta^\gamma \eta^\delta \rangle Z_{1\alpha} Z_{1\beta} Z_{2\gamma} Z_{2\delta} \big)
\nonumber\\
&\qquad
- 2\big(\langle \Theta^\mu \Theta^\nu \Theta^\rho \Theta^\sigma \rangle X_{1\mu} X_{1\nu} X_{2\rho} Z_{2\sigma} \big) \big(\langle \eta^\alpha \eta^\beta \eta^\gamma \eta^\delta \rangle Z_{1\alpha} Z_{1\beta} X_{2\gamma} Z_{2\delta} \big)
\nonumber\\
&\qquad
+ \big(\langle \Theta^\mu \Theta^\nu \Theta^\rho \Theta^\sigma \rangle X_{1\mu} X_{1\nu} Z_{2\rho} Z_{2\sigma} \big) \big(\langle \eta^\alpha \eta^\beta \eta^\gamma \eta^\delta \rangle Z_{1\alpha} Z_{1\beta} X_{2\gamma} X_{2\delta} \big)
\nonumber\\
&\qquad
- 2\big(\langle \Theta^\mu \Theta^\nu \Theta^\rho \Theta^\sigma \rangle X_{1\mu} Z_{1\nu} X_{2\rho} X_{2\sigma} \big) \big(\langle \eta^\alpha \eta^\beta \eta^\gamma \eta^\delta \rangle X_{1\alpha} Z_{1\beta} Z_{2\gamma} Z_{2\delta} \big)
\nonumber\\
&\qquad
+ 4\big(\langle \Theta^\mu \Theta^\nu \Theta^\rho \Theta^\sigma \rangle X_{1\mu} Z_{1\nu} X_{2\rho} Z_{2\sigma} \big) \big(\langle \eta^\alpha \eta^\beta \eta^\gamma \eta^\delta \rangle X_{1\alpha} Z_{1\beta} X_{2\gamma} Z_{2\delta} \big)
\nonumber\\
&\qquad
- 2\big(\langle \Theta^\mu \Theta^\nu \Theta^\rho \Theta^\sigma \rangle X_{1\mu} Z_{1\nu} Z_{2\rho} Z_{2\sigma} \big) \big(\langle \eta^\alpha \eta^\beta \eta^\gamma \eta^\delta \rangle X_{1\alpha} Z_{1\beta} X_{2\gamma} X_{2\delta} \big)
\nonumber\\
&\qquad
+ \big(\langle \Theta^\mu \Theta^\nu \Theta^\rho \Theta^\sigma \rangle Z_{1\mu} Z_{1\nu} X_{2\rho} X_{2\sigma} \big) \big(\langle \eta^\alpha \eta^\beta \eta^\gamma \eta^\delta \rangle X_{1\alpha} X_{1\beta} Z_{2\gamma} Z_{2\delta} \big)
\nonumber\\
&\qquad
- 2\big(\langle \Theta^\mu \Theta^\nu \Theta^\rho \Theta^\sigma \rangle Z_{1\mu} Z_{1\nu} X_{2\rho} Z_{2\sigma} \big) \big(\langle \eta^\alpha \eta^\beta \eta^\gamma \eta^\delta \rangle X_{1\alpha} X_{1\beta} X_{2\gamma} Z_{2\delta} \big)
\nonumber\\
&\qquad
+ \big(\langle \Theta^\mu \Theta^\nu \Theta^\rho \Theta^\sigma \rangle Z_{1\mu} Z_{1\nu} Z_{2\rho} Z_{2\sigma} \big) \big(\langle \eta^\alpha \eta^\beta \eta^\gamma \eta^\delta \rangle X_{1\alpha} X_{1\beta} X_{2\gamma} X_{2\delta} \big)
\Big]
\nonumber\\[0.5em]
&=
(X_1\!\cdot\! X_2)^{-\Delta-2}
\Big[
\big( X_1^2 X_2^2 + 2(X_1\!\cdot\! X_2)^2 \big) \big( Z_1^2 Z_2^2 + 2(Z_1\!\cdot\! Z_2)^2 \big)
\nonumber\\
&\qquad
- 2\big( X_1^2 (X_2\!\cdot\! Z_2) + 2(X_1\!\cdot\! X_2)(X_1\!\cdot\! Z_2) \big) \big( Z_1^2 (X_2\!\cdot\! Z_2) + 2(Z_1\!\cdot\! X_2)(Z_1\!\cdot\! Z_2) \big)
\nonumber\\
&\qquad
+ \big( X_1^2 Z_2^2 + 2(X_1\!\cdot\! Z_2)^2 \big) \big( Z_1^2 X_2^2 + 2(Z_1\!\cdot\! X_2)^2 \big)
\nonumber\\
&\qquad
- 2\big( (X_1\!\cdot\! Z_1) X_2^2 + 2(X_1\!\cdot\! X_2)(Z_1\!\cdot\! X_2) \big) \big( (X_1\!\cdot\! Z_1) Z_2^2 + 2(X_1\!\cdot\! Z_2)(Z_1\!\cdot\! Z_2) \big)
\nonumber\\
&\qquad
+ 4\big( (X_1\!\cdot\! Z_1)(X_2\!\cdot\! Z_2) + (X_1\!\cdot\! X_2)(Z_1\!\cdot\! Z_2) + (X_1\!\cdot\! Z_2)(Z_1\!\cdot\! X_2) \big)
\nonumber\\
&\hspace{1.5cm} \times
\big( (X_1\!\cdot\! Z_1)(X_2\!\cdot\! Z_2) + (X_1\!\cdot\! X_2)(Z_1\!\cdot\! Z_2) + (X_1\!\cdot\! Z_2)(Z_1\!\cdot\! X_2) \big)
\nonumber\\
&\qquad
- 2\big( (X_1\!\cdot\! Z_1) Z_2^2 + 2(X_1\!\cdot\! Z_2)(Z_1\!\cdot\! Z_2) \big) \big( (X_1\!\cdot\! Z_1) X_2^2 + 2(X_1\!\cdot\! X_2)(Z_1\!\cdot\! X_2) \big)
\nonumber\\
&\qquad
+ \big( Z_1^2 X_2^2 + 2(Z_1\!\cdot\! X_2)^2 \big) \big( X_1^2 Z_2^2 + 2(X_1\!\cdot\! Z_2)^2 \big)
\nonumber\\
&\qquad
- 2\big( Z_1^2 (X_2\!\cdot\! Z_2) + 2(Z_1\!\cdot\! X_2)(Z_1\!\cdot\! Z_2) \big) \big( X_1^2 (X_2\!\cdot\! Z_2) + 2(X_1\!\cdot\! X_2)(X_1\!\cdot\! Z_2) \big)
\nonumber\\
&\qquad
+ \big( Z_1^2 Z_2^2 + 2(Z_1\!\cdot\! Z_2)^2 \big) \big( X_1^2 X_2^2 + 2(X_1\!\cdot\! X_2)^2 \big)
\Big]
\nonumber\\[0.5em]
\end{align}

\begin{align}
&\text{Imposing the null cone constraints } X_i^2 = 0, Z_i^2 = 0 \text{ and transversality we get} X_i\!\cdot\! Z_i = 0
\nonumber\\[0.5em]
&
(X_1\!\cdot\! X_2)^{-\Delta-2}
\Big[
\left( 2(X_1\!\cdot\! X_2)^2 \right)\left( 2(Z_1\!\cdot\! Z_2)^2 \right)
- 2\left( 2(X_1\!\cdot\! X_2)(X_1\!\cdot\! Z_2) \right)\left( 2(Z_1\!\cdot\! X_2)(Z_1\!\cdot\! Z_2) \right)
\nonumber\\
&\qquad
+ \left( 2(X_1\!\cdot\! Z_2)^2 \right)\left( 2(Z_1\!\cdot\! X_2)^2 \right)
\nonumber\\
&\qquad
- 2\left( 2(X_1\!\cdot\! X_2)(Z_1\!\cdot\! X_2) \right)\left( 2(X_1\!\cdot\! Z_2)(Z_1\!\cdot\! Z_2) \right)
\nonumber\\
&\qquad
+ 4\left( (X_1\!\cdot\! X_2)(Z_1\!\cdot\! Z_2) + (X_1\!\cdot\! Z_2)(Z_1\!\cdot\! X_2) \right)\left( (X_1\!\cdot\! X_2)(Z_1\!\cdot\! Z_2) + (X_1\!\cdot\! Z_2)(Z_1\!\cdot\! X_2) \right)
\nonumber\\
&\qquad
- 2\left( 2(X_1\!\cdot\! Z_2)(Z_1\!\cdot\! Z_2) \right)\left( 2(X_1\!\cdot\! X_2)(Z_1\!\cdot\! X_2) \right)
\nonumber\\
&\qquad
+ \left( 2(Z_1\!\cdot\! X_2)^2 \right)\left( 2(X_1\!\cdot\! Z_2)^2 \right)
\nonumber\\
&\qquad
- 2\left( 2(Z_1\!\cdot\! X_2)(Z_1\!\cdot\! Z_2) \right)\left( 2(X_1\!\cdot\! X_2)(X_1\!\cdot\! Z_2) \right)
\nonumber\\
&\qquad
+ \left( 2(Z_1\!\cdot\! Z_2)^2 \right)\left( 2(X_1\!\cdot\! X_2)^2 \right)
\Big]
\nonumber\\[0.5em]
&=
(X_1\!\cdot\! X_2)^{-\Delta-2}
\Big[
4(X_1\!\cdot\! X_2)^2(Z_1\!\cdot\! Z_2)^2
- 8(X_1\!\cdot\! X_2)(X_1\!\cdot\! Z_2)(Z_1\!\cdot\! X_2)(Z_1\!\cdot\! Z_2)
+ 4(X_1\!\cdot\! Z_2)^2(Z_1\!\cdot\! X_2)^2
\nonumber\\
&\qquad
- 8(X_1\!\cdot\! X_2)(Z_1\!\cdot\! X_2)(X_1\!\cdot\! Z_2)(Z_1\!\cdot\! Z_2)
\nonumber\\
&\qquad
+ 4\left( (X_1\!\cdot\! X_2)^2(Z_1\!\cdot\! Z_2)^2 + 2(X_1\!\cdot\! X_2)(Z_1\!\cdot\! Z_2)(X_1\!\cdot\! Z_2)(Z_1\!\cdot\! X_2) + (X_1\!\cdot\! Z_2)^2(Z_1\!\cdot\! X_2)^2 \right)
\nonumber\\
&\qquad
- 8(X_1\!\cdot\! Z_2)(Z_1\!\cdot\! Z_2)(X_1\!\cdot\! X_2)(Z_1\!\cdot\! X_2)
+ 4(Z_1\!\cdot\! X_2)^2(X_1\!\cdot\! Z_2)^2
\nonumber\\
&\qquad
- 8(Z_1\!\cdot\! X_2)(Z_1\!\cdot\! Z_2)(X_1\!\cdot\! X_2)(X_1\!\cdot\! Z_2)
+ 4(Z_1\!\cdot\! Z_2)^2(X_1\!\cdot\! X_2)^2
\Big]
\nonumber\\[0.5em]
&=
(X_1\!\cdot\! X_2)^{-\Delta-2}
\Big[
12(X_1\!\cdot\! X_2)^2(Z_1\!\cdot\! Z_2)^2
- 24(X_1\!\cdot\! X_2)(Z_1\!\cdot\! Z_2)(X_1\!\cdot\! Z_2)(Z_1\!\cdot\! X_2)
+ 12(X_1\!\cdot\! Z_2)^2(Z_1\!\cdot\! X_2)^2
\Big]
\nonumber\\[0.5em]
&=
12(X_1\!\cdot\! X_2)^{-\Delta-2}
\Big[
(X_1\!\cdot\! X_2)(Z_1\!\cdot\! Z_2) - (X_1\!\cdot\! Z_2)(X_2\!\cdot\! Z_1)
\Big]^2
\nonumber\\[0.5em]
&=
3(X_1\!\cdot\! X_2)^{-\Delta-2}
\Big[
-2\big( (X_1\!\cdot\! X_2)(Z_1\!\cdot\! Z_2) - (X_1\!\cdot\! Z_2)(X_2\!\cdot\! Z_1) \big)
\Big]^2
\nonumber\\[0.5em]
&=
3(X_1\!\cdot\! X_2)^{-\Delta-2} H_{12}^2.
\end{align}

Here Wick's theorem $\langle \Theta^A \Theta^B \Theta^C \Theta^D \rangle = \delta^{AB}\delta^{CD} + \delta^{AC}\delta^{BD} + \delta^{AD}\delta^{BC}$ and similar relations for the $\eta$s have been used. Also the null cone constraints $X_i^2=0=Z_i^2$ and transversality $X_i \cdot Z_i=0$ were crucial for the above derivation to hold.

\subsection{$\langle \Phi(X_1) \Phi(X_2) A(X_3) \rangle$}\label{PhiPhiAappendix}
For the $3$-point correlator $\langle \Phi(X_1) \Phi(X_2) A(X_3) \rangle$ we use \eqref{PhiPhiAansatz} with $\Delta_i=-1$ for $i=1,2,3$. We find,
\begin{align}
\langle \Phi(X_1) \Phi(X_2) A(X_3) \rangle 
&=
\Big\langle
(\Theta\!\cdot\! X_1)(\eta\!\cdot\! X_2)
\left[ (\Theta\!\cdot\! X_3)(\eta\!\cdot\! Z_3) - (\eta\!\cdot\! X_3)(\Theta\!\cdot\! Z_3) \right]
\Big\rangle
\nonumber\\[0.5em]
&=
\Big\langle
(\Theta\!\cdot\! X_1)(\eta\!\cdot\! X_2)(\Theta\!\cdot\! X_3)(\eta\!\cdot\! Z_3)
\nonumber\\
&\qquad
-(\Theta\!\cdot\! X_1)(\eta\!\cdot\! X_2)(\eta\!\cdot\! X_3)(\Theta\!\cdot\! Z_3)
\Big\rangle
\nonumber\\[0.5em]
&=
\Big\langle
(\Theta^\mu X_{1\mu})(\eta^\nu X_{2\nu})(\Theta^\rho X_{3\rho})(\eta^\sigma Z_{3\sigma})
\nonumber\\
&\qquad
-(\Theta^\mu X_{1\mu})(\eta^\nu X_{2\nu})(\eta^\rho X_{3\rho})(\Theta^\sigma Z_{3\sigma})
\Big\rangle
\nonumber\\[0.5em]
&=
\Big[
\langle \Theta^\mu \Theta^\rho\rangle \langle \eta^\nu \eta^\sigma\rangle
X_{1\mu} X_{2\nu} X_{3\rho} Z_{3\sigma}
\nonumber\\
&\qquad
-\langle \Theta^\mu \Theta^\sigma\rangle \langle \eta^\nu \eta^\rho\rangle
X_{1\mu} X_{2\nu} X_{3\rho} Z_{3\sigma}
\Big]
\nonumber\\[0.5em]
&=
\Big[
\delta^{\mu\rho}\delta^{\nu\sigma}
X_{1\mu} X_{2\nu} X_{3\rho} Z_{3\sigma}
\nonumber\\
&\qquad
-\delta^{\mu\sigma}\delta^{\nu\rho}
X_{1\mu} X_{2\nu} X_{3\rho} Z_{3\sigma}
\Big]
\nonumber\\[0.5em]
&=
\Big[
(X_{1\mu} X_3^\mu)(X_{2\nu} Z_3^\nu)
-(X_{1\mu} Z_3^\mu)(X_{2\nu} X_3^\nu)
\Big]
\nonumber\\[0.5em]
&=
(X_1\!\cdot\! X_3)(X_2\!\cdot\! Z_3) - (X_1\!\cdot\! Z_3)(X_2\!\cdot\! X_3).
\label{eq:PhiPhiA_expanded_app}
\end{align}
which is the result in \eqref{eq:PhiPhiA_expanded}
\subsection{$\langle A_1(X_1,Z_1) A_2(X_2,Z_2) A_3(X_3,Z_3)\rangle$}\label{3ptappendix}
For the $3$-point correlator $\langle A_1(X_1,Z_1) A_2(X_2,Z_2) A_3(X_3,Z_3) \rangle$ we use \eqref{AAAansatz} and get 
\begin{align}
\langle A_1(X_1) A_2(X_2) A_3(X_3) \rangle 
&=
\Big\langle
\left[ (\Theta\!\cdot\! X_1)(\eta\!\cdot\! Z_1) - (\eta\!\cdot\! X_1)(\Theta\!\cdot\! Z_1) \right] \nonumber\\
&\qquad \times \left[ (\tilde{\Theta}\!\cdot\! X_2)(\Theta\!\cdot\! Z_2) - (\Theta\!\cdot\! X_2)(\tilde{\Theta}\!\cdot\! Z_2) \right] \nonumber\\
&\qquad \times \left[ (\eta\!\cdot\! X_3)(\tilde{\Theta}\!\cdot\! Z_3) - (\tilde{\Theta}\!\cdot\! X_3)(\eta\!\cdot\! Z_3) \right]
\Big\rangle
\nonumber\\[0.5em]
&=
\Big\langle
(\Theta\!\cdot\! X_1)(\eta\!\cdot\! Z_1)(\tilde{\Theta}\!\cdot\! X_2)(\Theta\!\cdot\! Z_2)(\eta\!\cdot\! X_3)(\tilde{\Theta}\!\cdot\! Z_3)
\nonumber\\
&\qquad
-(\Theta\!\cdot\! X_1)(\eta\!\cdot\! Z_1)(\tilde{\Theta}\!\cdot\! X_2)(\Theta\!\cdot\! Z_2)(\tilde{\Theta}\!\cdot\! X_3)(\eta\!\cdot\! Z_3)
\nonumber\\
&\qquad
-(\Theta\!\cdot\! X_1)(\eta\!\cdot\! Z_1)(\Theta\!\cdot\! X_2)(\tilde{\Theta}\!\cdot\! Z_2)(\eta\!\cdot\! X_3)(\tilde{\Theta}\!\cdot\! Z_3)
\nonumber\\
&\qquad
+(\Theta\!\cdot\! X_1)(\eta\!\cdot\! Z_1)(\Theta\!\cdot\! X_2)(\tilde{\Theta}\!\cdot\! Z_2)(\tilde{\Theta}\!\cdot\! X_3)(\eta\!\cdot\! Z_3)
\nonumber\\
&\qquad
-(\eta\!\cdot\! X_1)(\Theta\!\cdot\! Z_1)(\tilde{\Theta}\!\cdot\! X_2)(\Theta\!\cdot\! Z_2)(\eta\!\cdot\! X_3)(\tilde{\Theta}\!\cdot\! Z_3)
\nonumber\\
&\qquad
+(\eta\!\cdot\! X_1)(\Theta\!\cdot\! Z_1)(\tilde{\Theta}\!\cdot\! X_2)(\Theta\!\cdot\! Z_2)(\tilde{\Theta}\!\cdot\! X_3)(\eta\!\cdot\! Z_3)
\nonumber\\
&\qquad
+(\eta\!\cdot\! X_1)(\Theta\!\cdot\! Z_1)(\Theta\!\cdot\! X_2)(\tilde{\Theta}\!\cdot\! Z_2)(\eta\!\cdot\! X_3)(\tilde{\Theta}\!\cdot\! Z_3)
\nonumber\\
&\qquad
-(\eta\!\cdot\! X_1)(\Theta\!\cdot\! Z_1)(\Theta\!\cdot\! X_2)(\tilde{\Theta}\!\cdot\! Z_2)(\tilde{\Theta}\!\cdot\! X_3)(\eta\!\cdot\! Z_3)
\Big\rangle
\nonumber\\[0.5em]
&=
\Big\langle
(\Theta^\mu X_{1\mu})(\eta^\nu Z_{1\nu})(\tilde{\Theta}^\rho X_{2\rho})(\Theta^\sigma Z_{2\sigma})(\eta^\alpha X_{3\alpha})(\tilde{\Theta}^\beta Z_{3\beta})
\nonumber\\
&\qquad
-(\Theta^\mu X_{1\mu})(\eta^\nu Z_{1\nu})(\tilde{\Theta}^\rho X_{2\rho})(\Theta^\sigma Z_{2\sigma})(\tilde{\Theta}^\alpha X_{3\alpha})(\eta^\beta Z_{3\beta})
\nonumber\\
&\qquad
-(\Theta^\mu X_{1\mu})(\eta^\nu Z_{1\nu})(\Theta^\rho X_{2\rho})(\tilde{\Theta}^\sigma Z_{2\sigma})(\eta^\alpha X_{3\alpha})(\tilde{\Theta}^\beta Z_{3\beta})
\nonumber\\
&\qquad
+(\Theta^\mu X_{1\mu})(\eta^\nu Z_{1\nu})(\Theta^\rho X_{2\rho})(\tilde{\Theta}^\sigma Z_{2\sigma})(\tilde{\Theta}^\alpha X_{3\alpha})(\eta^\beta Z_{3\beta})
\nonumber\\
&\qquad
-(\eta^\mu X_{1\mu})(\Theta^\nu Z_{1\nu})(\tilde{\Theta}^\rho X_{2\rho})(\Theta^\sigma Z_{2\sigma})(\eta^\alpha X_{3\alpha})(\tilde{\Theta}^\beta Z_{3\beta})
\nonumber\\
&\qquad
+(\eta^\mu X_{1\mu})(\Theta^\nu Z_{1\nu})(\tilde{\Theta}^\rho X_{2\rho})(\Theta^\sigma Z_{2\sigma})(\tilde{\Theta}^\alpha X_{3\alpha})(\eta^\beta Z_{3\beta})
\nonumber\\
&\qquad
+(\eta^\mu X_{1\mu})(\Theta^\nu Z_{1\nu})(\Theta^\rho X_{2\rho})(\tilde{\Theta}^\sigma Z_{2\sigma})(\eta^\alpha X_{3\alpha})(\tilde{\Theta}^\beta Z_{3\beta})
\nonumber\\
&\qquad
-(\eta^\mu X_{1\mu})(\Theta^\nu Z_{1\nu})(\Theta^\rho X_{2\rho})(\tilde{\Theta}^\sigma Z_{2\sigma})(\tilde{\Theta}^\alpha X_{3\alpha})(\eta^\beta Z_{3\beta})
\Big\rangle
\nonumber\\[0.5em]
&=
\Big[
\langle \Theta^\mu \Theta^\sigma \rangle \langle \eta^\nu \eta^\alpha \rangle \langle \tilde{\Theta}^\rho \tilde{\Theta}^\beta \rangle X_{1\mu} Z_{1\nu} X_{2\rho} Z_{2\sigma} X_{3\alpha} Z_{3\beta}
\nonumber\\
&\qquad
-\langle \Theta^\mu \Theta^\sigma \rangle \langle \eta^\nu \eta^\beta \rangle \langle \tilde{\Theta}^\rho \tilde{\Theta}^\alpha \rangle X_{1\mu} Z_{1\nu} X_{2\rho} Z_{2\sigma} X_{3\alpha} Z_{3\beta}
\nonumber\\
&\qquad
-\langle \Theta^\mu \Theta^\rho \rangle \langle \eta^\nu \eta^\alpha \rangle \langle \tilde{\Theta}^\sigma \tilde{\Theta}^\beta \rangle X_{1\mu} Z_{1\nu} X_{2\rho} Z_{2\sigma} X_{3\alpha} Z_{3\beta}
\nonumber\\
&\qquad
+\langle \Theta^\mu \Theta^\rho \rangle \langle \eta^\nu \eta^\beta \rangle \langle \tilde{\Theta}^\sigma \tilde{\Theta}^\alpha \rangle X_{1\mu} Z_{1\nu} X_{2\rho} Z_{2\sigma} X_{3\alpha} Z_{3\beta}
\nonumber\\
&\qquad
-\langle \Theta^\nu \Theta^\sigma \rangle \langle \eta^\mu \eta^\alpha \rangle \langle \tilde{\Theta}^\rho \tilde{\Theta}^\beta \rangle X_{1\mu} Z_{1\nu} X_{2\rho} Z_{2\sigma} X_{3\alpha} Z_{3\beta}
\nonumber\\
&\qquad
+\langle \Theta^\nu \Theta^\sigma \rangle \langle \eta^\mu \eta^\beta \rangle \langle \tilde{\Theta}^\rho \tilde{\Theta}^\alpha \rangle X_{1\mu} Z_{1\nu} X_{2\rho} Z_{2\sigma} X_{3\alpha} Z_{3\beta}
\nonumber\\
&\qquad
+\langle \Theta^\nu \Theta^\rho \rangle \langle \eta^\mu \eta^\alpha \rangle \langle \tilde{\Theta}^\sigma \tilde{\Theta}^\beta \rangle X_{1\mu} Z_{1\nu} X_{2\rho} Z_{2\sigma} X_{3\alpha} Z_{3\beta}
\nonumber\\
&\qquad
-\langle \Theta^\nu \Theta^\rho \rangle \langle \eta^\mu \eta^\beta \rangle \langle \tilde{\Theta}^\sigma \tilde{\Theta}^\alpha \rangle X_{1\mu} Z_{1\nu} X_{2\rho} Z_{2\sigma} X_{3\alpha} Z_{3\beta}
\Big]
\nonumber\\[0.5em]
\end{align}

\begin{align}
&=
\Big[
\delta^{\mu\sigma}\delta^{\nu\alpha}\delta^{\rho\beta} X_{1\mu} Z_{1\nu} X_{2\rho} Z_{2\sigma} X_{3\alpha} Z_{3\beta}
\nonumber\\
&\qquad
-\delta^{\mu\sigma}\delta^{\nu\beta}\delta^{\rho\alpha} X_{1\mu} Z_{1\nu} X_{2\rho} Z_{2\sigma} X_{3\alpha} Z_{3\beta}
\nonumber\\
&\qquad
-\delta^{\mu\rho}\delta^{\nu\alpha}\delta^{\sigma\beta} X_{1\mu} Z_{1\nu} X_{2\rho} Z_{2\sigma} X_{3\alpha} Z_{3\beta}
\nonumber\\
&\qquad
+\delta^{\mu\rho}\delta^{\nu\beta}\delta^{\sigma\alpha} X_{1\mu} Z_{1\nu} X_{2\rho} Z_{2\sigma} X_{3\alpha} Z_{3\beta}
\nonumber\\
&\qquad
-\delta^{\nu\sigma}\delta^{\mu\alpha}\delta^{\rho\beta} X_{1\mu} Z_{1\nu} X_{2\rho} Z_{2\sigma} X_{3\alpha} Z_{3\beta}
\nonumber\\
&\qquad
+\delta^{\nu\sigma}\delta^{\mu\beta}\delta^{\rho\alpha} X_{1\mu} Z_{1\nu} X_{2\rho} Z_{2\sigma} X_{3\alpha} Z_{3\beta}
\nonumber\\
&\qquad
+\delta^{\nu\rho}\delta^{\mu\alpha}\delta^{\sigma\beta} X_{1\mu} Z_{1\nu} X_{2\rho} Z_{2\sigma} X_{3\alpha} Z_{3\beta}
\nonumber\\
&\qquad
-\delta^{\nu\rho}\delta^{\mu\beta}\delta^{\sigma\alpha} X_{1\mu} Z_{1\nu} X_{2\rho} Z_{2\sigma} X_{3\alpha} Z_{3\beta}
\Big]
\nonumber\\[0.5em]
&=
\Big[
(X_{1\mu} Z_2^\mu)(Z_{1\nu} X_3^\nu)(X_{2\rho} Z_3^\rho)
-(X_{1\mu} Z_2^\mu)(Z_{1\nu} Z_3^\nu)(X_{2\rho} X_3^\rho)
\nonumber\\
&\qquad
-(X_{1\mu} X_2^\mu)(Z_{1\nu} X_3^\nu)(Z_{2\sigma} Z_3^\sigma)
+(X_{1\mu} X_2^\mu)(Z_{1\nu} Z_3^\nu)(Z_{2\sigma} X_3^\sigma)
\nonumber\\
&\qquad
-(X_{1\mu} X_3^\mu)(Z_{1\nu} Z_2^\nu)(X_{2\rho} Z_3^\rho)
+(X_{1\mu} Z_3^\mu)(Z_{1\nu} Z_2^\nu)(X_{2\rho} X_3^\rho)
\nonumber\\
&\qquad
+(X_{1\mu} X_3^\mu)(Z_{1\nu} X_2^\nu)(Z_{2\sigma} Z_3^\sigma)
-(X_{1\mu} Z_3^\mu)(Z_{1\nu} X_2^\nu)(Z_{2\sigma} X_3^\sigma)
\Big]
\nonumber\\[0.5em]
&=
(X_1\!\cdot\! Z_2)(Z_1\!\cdot\! X_3)(X_2\!\cdot\! Z_3) 
- (X_1\!\cdot\! Z_2)(Z_1\!\cdot\! Z_3)(X_2\!\cdot\! X_3) 
\nonumber\\
&\qquad
- (X_1\!\cdot\! X_2)(Z_1\!\cdot\! X_3)(Z_2\!\cdot\! Z_3) 
+ (X_1\!\cdot\! X_2)(Z_1\!\cdot\! Z_3)(Z_2\!\cdot\! X_3) 
\nonumber\\
&\qquad
- (X_1\!\cdot\! X_3)(Z_1\!\cdot\! Z_2)(X_2\!\cdot\! Z_3) 
+ (X_1\!\cdot\! Z_3)(Z_1\!\cdot\! Z_2)(X_2\!\cdot\! X_3) 
\nonumber\\
&\qquad
+ (X_1\!\cdot\! X_3)(Z_1\!\cdot\! X_2)(Z_2\!\cdot\! Z_3) 
- (X_1\!\cdot\! Z_3)(Z_1\!\cdot\! X_2)(Z_2\!\cdot\! X_3).
\label{eq:AAA_expanded_app}
\end{align}
which is \eqref{AAAexpanded}.
\subsubsection{Expansion in terms of tensor structures}\label{expansion3pt}
To verify the expansion in \eqref{AAAtensorexpand}, we will substitute the explicit definitions of the tensor structures $H_{ij}$ and $V_{i,jk}$ from \eqref{Hij} and \eqref{Vijk} into the right-hand side of the proposed decomposition:
\begin{equation}\label{AAAtensorexpandapp}
    \text{RHS} = -\frac{1}{2} H_{12} V_{3,12} + \frac{1}{2} H_{13} V_{2,13} - \frac{1}{2} H_{23} V_{1,23} + V_{1,23} V_{2,13} V_{3,12}
\end{equation}
Expanding the first term in \eqref{AAAtensorexpandapp} $-\frac{1}{2} H_{12} V_{3,12}$ yields:
\begin{align}
    -\frac{1}{2} H_{12} V_{3,12} &= \frac{1}{(X_1 \cdot X_2)} \big[ (Z_1 \cdot Z_2)(X_1 \cdot X_2) - (Z_1 \cdot X_2)(Z_2 \cdot X_1) \big] \big[ (Z_3 \cdot X_1)(X_2 \cdot X_3) - (Z_3 \cdot X_2)(X_1 \cdot X_3) \big] \notag \\
    &= (Z_1 \cdot Z_2)(Z_3 \cdot X_1)(X_2 \cdot X_3) - (Z_1 \cdot Z_2)(Z_3 \cdot X_2)(X_1 \cdot X_3) \notag \\
    &\quad - \frac{(X_2 \cdot X_3)}{(X_1 \cdot X_2)} (Z_1 \cdot X_2)(Z_2 \cdot X_1)(Z_3 \cdot X_1) + \frac{(X_1 \cdot X_3)}{(X_1 \cdot X_2)} (Z_1 \cdot X_2)(Z_2 \cdot X_1)(Z_3 \cdot X_2)
    \label{term1}
\end{align}
Similarly, expanding the 2nd term in \eqref{AAAtensorexpandapp} i.e. $\frac{1}{2} H_{13} V_{2,13}$ gives:
\begin{align}
    \frac{1}{2} H_{13} V_{2,13} &= -\frac{1}{(X_1 \cdot X_3)} \big[ (Z_1 \cdot Z_3)(X_1 \cdot X_3) - (Z_1 \cdot X_3)(Z_3 \cdot X_1) \big] \big[ (Z_2 \cdot X_1)(X_2 \cdot X_3) - (Z_2 \cdot X_3)(X_1 \cdot X_2) \big] \notag \\
    &= -(Z_1 \cdot Z_3)(Z_2 \cdot X_1)(X_2 \cdot X_3) + (Z_1 \cdot Z_3)(Z_2 \cdot X_3)(X_1 \cdot X_2) \notag \\
    &\quad + \frac{(X_2 \cdot X_3)}{(X_1 \cdot X_3)} (Z_1 \cdot X_3)(Z_3 \cdot X_1)(Z_2 \cdot X_1) - \frac{(X_1 \cdot X_2)}{(X_1 \cdot X_3)} (Z_1 \cdot X_3)(Z_3 \cdot X_1)(Z_2 \cdot X_3)
    \label{term2}
\end{align}
Expanding $-\frac{1}{2} H_{23} V_{1,23}$ gives:
\begin{align}
    -\frac{1}{2} H_{23} V_{1,23} &= \frac{1}{(X_2 \cdot X_3)} \big[ (Z_2 \cdot Z_3)(X_2 \cdot X_3) - (Z_2 \cdot X_3)(Z_3 \cdot X_2) \big] \big[ (Z_1 \cdot X_2)(X_1 \cdot X_3) - (Z_1 \cdot X_3)(X_1 \cdot X_2) \big] \\
    &= (Z_2 \cdot Z_3)(Z_1 \cdot X_2)(X_1 \cdot X_3) - (Z_2 \cdot Z_3)(Z_1 \cdot X_3)(X_1 \cdot X_2) \notag \\
    &\quad - \frac{(X_1 \cdot X_3)}{(X_2 \cdot X_3)} (Z_2 \cdot X_3)(Z_3 \cdot X_2)(Z_1 \cdot X_2) + \frac{(X_1 \cdot X_2)}{(X_2 \cdot X_3)} (Z_2 \cdot X_3)(Z_3 \cdot X_2)(Z_1 \cdot X_3)
    \label{term3}
\end{align}
Finally, we expand the product of the three $V$ structures. This gives the following eight terms:
\begin{align}
    V_{1,23} V_{2,13} V_{3,12} &= \frac{1}{(X_1 \cdot X_2)(X_1 \cdot X_3)(X_2 \cdot X_3)} \Big[ (Z_1 \cdot X_2)(X_1 \cdot X_3) - (Z_1 \cdot X_3)(X_1 \cdot X_2) \Big] \notag \\
    &\qquad \times \Big[ (Z_2 \cdot X_1)(X_2 \cdot X_3) - (Z_2 \cdot X_3)(X_1 \cdot X_2) \Big] \Big[ (Z_3 \cdot X_1)(X_2 \cdot X_3) - (Z_3 \cdot X_2)(X_1 \cdot X_3) \Big] \notag \\
    &= \frac{(X_2 \cdot X_3)}{(X_1 \cdot X_2)} (Z_1 \cdot X_2)(Z_2 \cdot X_1)(Z_3 \cdot X_1) - \frac{(X_1 \cdot X_3)}{(X_1 \cdot X_2)} (Z_1 \cdot X_2)(Z_2 \cdot X_1)(Z_3 \cdot X_2) \notag \\
    &\quad - (Z_1 \cdot X_2)(Z_2 \cdot X_3)(Z_3 \cdot X_1) + \frac{(X_1 \cdot X_3)}{(X_2 \cdot X_3)} (Z_1 \cdot X_2)(Z_2 \cdot X_3)(Z_3 \cdot X_2) \notag \\
    &\quad - \frac{(X_2 \cdot X_3)}{(X_1 \cdot X_3)} (Z_1 \cdot X_3)(Z_2 \cdot X_1)(Z_3 \cdot X_1) + (Z_1 \cdot X_3)(Z_2 \cdot X_1)(Z_3 \cdot X_2) \notag \\
    &\quad + \frac{(X_1 \cdot X_2)}{(X_1 \cdot X_3)} (Z_1 \cdot X_3)(Z_2 \cdot X_3)(Z_3 \cdot X_1) - \frac{(X_1 \cdot X_2)}{(X_2 \cdot X_3)} (Z_1 \cdot X_3)(Z_2 \cdot X_3)(Z_3 \cdot X_2)
    \label{term4}
\end{align}
Note that the six fractional terms generated by the $V_{1,23} V_{2,13} V_{3,12}$ product in \eqref{term4} perfectly cancel the six fractional terms produced by the $H V$ expansions in \eqref{term1}-\eqref{term3}. This is crucial for consistency because our architecture for $\Delta_i=-1$ does not have any fractional term. Collecting the remaining terms from all four components, we are left with:
\begin{align*}
    \text{RHS} &= (Z_1 \cdot Z_2)(Z_3 \cdot X_1)(X_2 \cdot X_3) - (Z_1 \cdot Z_2)(Z_3 \cdot X_2)(X_1 \cdot X_3) \\
    &\quad - (Z_1 \cdot Z_3)(Z_2 \cdot X_1)(X_2 \cdot X_3) + (Z_1 \cdot Z_3)(Z_2 \cdot X_3)(X_1 \cdot X_2) \\
    &\quad + (Z_2 \cdot Z_3)(Z_1 \cdot X_2)(X_1 \cdot X_3) - (Z_2 \cdot Z_3)(Z_1 \cdot X_3)(X_1 \cdot X_2) \\
    &\quad - (Z_1 \cdot X_2)(Z_2 \cdot X_3)(Z_3 \cdot X_1) + (Z_1 \cdot X_3)(Z_2 \cdot X_1)(Z_3 \cdot X_2)
\end{align*}
This reconstructs the explicit $3$-point function $\langle A_1(X_1) A_2(X_2) A_3(X_3) \rangle$ derived in \eqref{AAAexpanded}.
\subsection{$\langle A(X_1,Z_1) A(X_2,Z_2) \Phi(X_3) \Phi(X_4) \rangle$}\label{AAPhiPhi}

\subsubsection{Factorized correlator}\label{AAPhiPhiGFF}
Using \eqref{GFFansatz} we can compute the $4$-point correlator as 
\begin{align}
\langle A(X_1) A(X_2) \Phi(X_3) \Phi(X_4) \rangle 
&=
\Big\langle
\left[ (\Theta\!\cdot\! X_1)(\eta\!\cdot\! Z_1) - (\eta\!\cdot\! X_1)(\Theta\!\cdot\! Z_1) \right] \nonumber\\
&\qquad \times \left[ (\Theta\!\cdot\! X_2)(\eta\!\cdot\! Z_2) - (\eta\!\cdot\! X_2)(\Theta\!\cdot\! Z_2) \right]
(\tilde{\Theta}\!\cdot\! X_3)(\tilde{\Theta}\!\cdot\! X_4)
\Big\rangle
\nonumber\\[0.5em]
&=
\Big\langle
(\Theta\!\cdot\! X_1)(\eta\!\cdot\! Z_1)(\Theta\!\cdot\! X_2)(\eta\!\cdot\! Z_2)(\tilde{\Theta}\!\cdot\! X_3)(\tilde{\Theta}\!\cdot\! X_4)
\nonumber\\
&\qquad
-(\Theta\!\cdot\! X_1)(\eta\!\cdot\! Z_1)(\eta\!\cdot\! X_2)(\Theta\!\cdot\! Z_2)(\tilde{\Theta}\!\cdot\! X_3)(\tilde{\Theta}\!\cdot\! X_4)
\nonumber\\
&\qquad
-(\eta\!\cdot\! X_1)(\Theta\!\cdot\! Z_1)(\Theta\!\cdot\! X_2)(\eta\!\cdot\! Z_2)(\tilde{\Theta}\!\cdot\! X_3)(\tilde{\Theta}\!\cdot\! X_4)
\nonumber\\
&\qquad
+(\eta\!\cdot\! X_1)(\Theta\!\cdot\! Z_1)(\eta\!\cdot\! X_2)(\Theta\!\cdot\! Z_2)(\tilde{\Theta}\!\cdot\! X_3)(\tilde{\Theta}\!\cdot\! X_4)
\Big\rangle
\nonumber\\[0.5em]
&=
\Big\langle
(\Theta^\mu X_{1\mu})(\eta^\nu Z_{1\nu})(\Theta^\rho X_{2\rho})(\eta^\sigma Z_{2\sigma})(\tilde{\Theta}^\alpha X_{3\alpha})(\tilde{\Theta}^\beta X_{4\beta})
\nonumber\\
&\qquad
-(\Theta^\mu X_{1\mu})(\eta^\nu Z_{1\nu})(\eta^\rho X_{2\rho})(\Theta^\sigma Z_{2\sigma})(\tilde{\Theta}^\alpha X_{3\alpha})(\tilde{\Theta}^\beta X_{4\beta})
\nonumber\\
&\qquad
-(\eta^\mu X_{1\mu})(\Theta^\nu Z_{1\nu})(\Theta^\rho X_{2\rho})(\eta^\sigma Z_{2\sigma})(\tilde{\Theta}^\alpha X_{3\alpha})(\tilde{\Theta}^\beta X_{4\beta})
\nonumber\\
&\qquad
+(\eta^\mu X_{1\mu})(\Theta^\nu Z_{1\nu})(\eta^\rho X_{2\rho})(\Theta^\sigma Z_{2\sigma})(\tilde{\Theta}^\alpha X_{3\alpha})(\tilde{\Theta}^\beta X_{4\beta})
\Big\rangle
\nonumber\\[0.5em]
&=
\Big[
\langle \Theta^\mu \Theta^\rho\rangle \langle \eta^\nu \eta^\sigma\rangle \langle \tilde{\Theta}^\alpha \tilde{\Theta}^\beta\rangle
X_{1\mu} Z_{1\nu} X_{2\rho} Z_{2\sigma} X_{3\alpha} X_{4\beta}
\nonumber\\
&\qquad
-\langle \Theta^\mu \Theta^\sigma\rangle \langle \eta^\nu \eta^\rho\rangle \langle \tilde{\Theta}^\alpha \tilde{\Theta}^\beta\rangle
X_{1\mu} Z_{1\nu} X_{2\rho} Z_{2\sigma} X_{3\alpha} X_{4\beta}
\nonumber\\
&\qquad
-\langle \eta^\mu \eta^\sigma\rangle \langle \Theta^\nu \Theta^\rho\rangle \langle \tilde{\Theta}^\alpha \tilde{\Theta}^\beta\rangle
X_{1\mu} Z_{1\nu} X_{2\rho} Z_{2\sigma} X_{3\alpha} X_{4\beta}
\nonumber\\
&\qquad
+\langle \eta^\mu \eta^\rho\rangle \langle \Theta^\nu \Theta^\sigma\rangle \langle \tilde{\Theta}^\alpha \tilde{\Theta}^\beta\rangle
X_{1\mu} Z_{1\nu} X_{2\rho} Z_{2\sigma} X_{3\alpha} X_{4\beta}
\Big]
\nonumber\\[0.5em]
&=
\Big[
\delta^{\mu\rho}\delta^{\nu\sigma}\delta^{\alpha\beta}
X_{1\mu} Z_{1\nu} X_{2\rho} Z_{2\sigma} X_{3\alpha} X_{4\beta}
\nonumber\\
&\qquad
-\delta^{\mu\sigma}\delta^{\nu\rho}\delta^{\alpha\beta}
X_{1\mu} Z_{1\nu} X_{2\rho} Z_{2\sigma} X_{3\alpha} X_{4\beta}
\nonumber\\
&\qquad
-\delta^{\mu\sigma}\delta^{\nu\rho}\delta^{\alpha\beta}
X_{1\mu} Z_{1\nu} X_{2\rho} Z_{2\sigma} X_{3\alpha} X_{4\beta}
\nonumber\\
&\qquad
+\delta^{\mu\rho}\delta^{\nu\sigma}\delta^{\alpha\beta}
X_{1\mu} Z_{1\nu} X_{2\rho} Z_{2\sigma} X_{3\alpha} X_{4\beta}
\Big]
\nonumber\\[0.5em]
&=
\Big[
(X_{1\mu} X_2^\mu)(Z_{1\nu} Z_2^\nu)(X_{3\alpha} X_4^\alpha)
-(X_{1\mu} Z_2^\mu)(Z_{1\nu} X_2^\nu)(X_{3\alpha} X_4^\alpha)
\nonumber\\
&\qquad
-(X_{1\mu} Z_2^\mu)(Z_{1\nu} X_2^\nu)(X_{3\alpha} X_4^\alpha)
+(X_{1\mu} X_2^\mu)(Z_{1\nu} Z_2^\nu)(X_{3\alpha} X_4^\alpha)
\Big]
\nonumber\\[0.5em]
&=
2(X_1\!\cdot\! X_2)(Z_1\!\cdot\! Z_2)(X_3\!\cdot\! X_4) - 2(X_1\!\cdot\! Z_2)(Z_1\!\cdot\! X_2)(X_3\!\cdot\! X_4).
\label{eq:AAPhiPhi_expanded_app}
\end{align}
which is \eqref{eq:AAPhiPhi_expanded}
\subsubsection{Connected Correlator}\label{AAPhiPhiInteracting}
If instead we use \eqref{AAPhiPhiansatz2} we get 
$$
\begin{aligned}
G_4 &= \big[ (\eta_1 \cdot Z_1)(\Theta_1 \cdot X_1) - (\eta_1 \cdot X_1)(\Theta_1 \cdot Z_1) \big] \big[ (\Theta_1 \cdot Z_2)(\Theta_2 \cdot X_2) - (\Theta_1 \cdot X_2)(\Theta_2 \cdot Z_2) \big] (\Theta_2 \cdot X_3) (\eta_1 \cdot X_4) \\
&= (\eta_1 \cdot Z_1)(\Theta_1 \cdot X_1)(\Theta_1 \cdot Z_2)(\Theta_2 \cdot X_2)(\Theta_2 \cdot X_3)(\eta_1 \cdot X_4) \\
&\quad - (\eta_1 \cdot Z_1)(\Theta_1 \cdot X_1)(\Theta_1 \cdot X_2)(\Theta_2 \cdot Z_2)(\Theta_2 \cdot X_3)(\eta_1 \cdot X_4) \\
&\quad - (\eta_1 \cdot X_1)(\Theta_1 \cdot Z_1)(\Theta_1 \cdot Z_2)(\Theta_2 \cdot X_2)(\Theta_2 \cdot X_3)(\eta_1 \cdot X_4) \\
&\quad + (\eta_1 \cdot X_1)(\Theta_1 \cdot Z_1)(\Theta_1 \cdot X_2)(\Theta_2 \cdot Z_2)(\Theta_2 \cdot X_3)(\eta_1 \cdot X_4)\\
&= (\eta_1^{\mu_1} Z_{1,\mu_1} \eta_1^{\mu_2} X_{4,\mu_2}) (\Theta_1^{\nu_1} X_{1,\nu_1} \Theta_1^{\nu_2} Z_{2,\nu_2}) (\Theta_2^{\rho_1} X_{2,\rho_1} \Theta_2^{\rho_2} X_{3,\rho_2}) \\
&\quad - (\eta_1^{\mu_1} Z_{1,\mu_1} \eta_1^{\mu_2} X_{4,\mu_2}) (\Theta_1^{\nu_1} X_{1,\nu_1} \Theta_1^{\nu_2} X_{2,\nu_2}) (\Theta_2^{\rho_1} Z_{2,\rho_1} \Theta_2^{\rho_2} X_{3,\rho_2}) \\
&\quad - (\eta_1^{\mu_1} X_{1,\mu_1} \eta_1^{\mu_2} X_{4,\mu_2}) (\Theta_1^{\nu_1} Z_{1,\nu_1} \Theta_1^{\nu_2} Z_{2,\nu_2}) (\Theta_2^{\rho_1} X_{2,\rho_1} \Theta_2^{\rho_2} X_{3,\rho_2}) \\
&\quad + (\eta_1^{\mu_1} X_{1,\mu_1} \eta_1^{\mu_2} X_{4,\mu_2}) (\Theta_1^{\nu_1} Z_{1,\nu_1} \Theta_1^{\nu_2} X_{2,\nu_2}) (\Theta_2^{\rho_1} Z_{2,\rho_1} \Theta_2^{\rho_2} X_{3,\rho_2}) \\
&= (\delta^{\mu_1 \mu_2} Z_{1,\mu_1} X_{4,\mu_2}) (\delta^{\nu_1 \nu_2} X_{1,\nu_1} Z_{2,\nu_2}) (\delta^{\rho_1 \rho_2} X_{2,\rho_1} X_{3,\rho_2}) \\
&\quad - (\delta^{\mu_1 \mu_2} Z_{1,\mu_1} X_{4,\mu_2}) (\delta^{\nu_1 \nu_2} X_{1,\nu_1} X_{2,\nu_2}) (\delta^{\rho_1 \rho_2} Z_{2,\rho_1} X_{3,\rho_2}) \\
&\quad - (\delta^{\mu_1 \mu_2} X_{1,\mu_1} X_{4,\mu_2}) (\delta^{\nu_1 \nu_2} Z_{1,\nu_1} Z_{2,\nu_2}) (\delta^{\rho_1 \rho_2} X_{2,\rho_1} X_{3,\rho_2}) \\
&\quad + (\delta^{\mu_1 \mu_2} X_{1,\mu_1} X_{4,\mu_2}) (\delta^{\nu_1 \nu_2} Z_{1,\nu_1} X_{2,\nu_2}) (\delta^{\rho_1 \rho_2} Z_{2,\rho_1} X_{3,\rho_2})\\
&= (Z_1 \cdot X_4)(X_1 \cdot Z_2)(X_2 \cdot X_3) - (Z_1 \cdot X_4)(X_1 \cdot X_2)(Z_2 \cdot X_3) \\
&\quad - (X_1 \cdot X_4)(Z_1 \cdot Z_2)(X_2 \cdot X_3) + (X_1 \cdot X_4)(Z_1 \cdot X_2)(Z_2 \cdot X_3)
\end{aligned}
$$
which is \eqref{obtained}.
\subsubsection{Expansion in terms of tensor structures}\label{expansion4pt}

 Let us evaluate the prefactor $P$ in \eqref{target} explicitly. Put $\Delta_i=-1$ for $i=1,2,3,4$ and $l=(1,1,0,0)$ we have $\tau_1 = \tau_2 = 0, \tau_3 = \tau_4 = -1$ using which we find from \eqref{alphas}
$$
\begin{aligned}
\Sigma:=\Sigma_{k=1}^4 \tau_k &= \tau_1 + \tau_2 + \tau_3 + \tau_4 = 0 + 0 - 1 - 1 = -2 \\
\alpha_{12} &= \frac{0 + 0}{2} - \frac{-2}{6} = \frac{1}{3} \\
\alpha_{34} &= \frac{-1 - 1}{2} - \frac{-2}{6} = -1 + \frac{1}{3} = -\frac{2}{3} \\
\alpha_{13} &= \frac{0 - 1}{2} - \frac{-2}{6} = -\frac{1}{2} + \frac{1}{3} = -\frac{1}{6} \\
\alpha_{14} &= \frac{0 - 1}{2} - \frac{-2}{6} = -\frac{1}{2} + \frac{1}{3} = -\frac{1}{6} \\
\alpha_{23} &= \frac{0 - 1}{2} - \frac{-2}{6} = -\frac{1}{2} + \frac{1}{3} = -\frac{1}{6} \\
\alpha_{24} &= \frac{0 - 1}{2} - \frac{-2}{6} = -\frac{1}{2} + \frac{1}{3} = -\frac{1}{6} 
\end{aligned}
$$
Using this we can find the expression for $P$
\begin{align}
P &= (-2 X_1 \cdot X_2)^{-1/3} (-2 X_3 \cdot X_4)^{2/3} (-2 X_1 \cdot X_3)^{1/6} (-2 X_1 \cdot X_4)^{1/6} (-2 X_2 \cdot X_3)^{1/6} (-2 X_2 \cdot X_4)^{1/6} \notag \\
&= (-2)^{-1/3 + 2/3 + 1/6 + 1/6 + 1/6 + 1/6} (X_1 \cdot X_2)^{-1/3} (X_3 \cdot X_4)^{2/3} (X_1 \cdot X_3)^{1/6} \notag \\
&\qquad \times (X_1 \cdot X_4)^{1/6} (X_2 \cdot X_3)^{1/6} (X_2 \cdot X_4)^{1/6} \notag \\
&= -2 (X_1 \cdot X_2)^{-1/3} (X_3 \cdot X_4)^{2/3} (X_1 \cdot X_3)^{1/6} (X_1 \cdot X_4)^{1/6} \times (X_2 \cdot X_3)^{1/6} (X_2 \cdot X_4)^{1/6} \label{prefactor4}
\end{align}
The coefficient of the first term on RHS of \eqref{target} is $-\frac{1}{4} P u^{-2/3} v^{5/6}$. Using $P=\prod_{i<j}^4 X_{ij}^{-\alpha_{ij}}$ with the exponents just obtained, together with the definitions of the cross ratios from \eqref{crossratios},
\begin{align}
-\frac{1}{4} P u^{-2/3} v^{5/6} &= -\frac{1}{4} \Big[ X_{12}^{-1/3} X_{34}^{2/3} X_{13}^{1/6} X_{14}^{1/6} X_{23}^{1/6} X_{24}^{1/6} \Big] \left( \frac{X_{12}X_{34}}{X_{13}X_{24}} \right)^{-2/3} \left( \frac{X_{14}X_{23}}{X_{13}X_{24}} \right)^{5/6} \notag \\
&= -\frac{1}{4} X_{12}^{-1/3 - 2/3} X_{34}^{2/3 - 2/3} X_{13}^{1/6 - (-2/3) - 5/6} X_{24}^{1/6 - (-2/3) - 5/6} X_{14}^{1/6 + 5/6} X_{23}^{1/6 + 5/6} \notag \\
&= -\frac{1}{4} X_{12}^{-1} X_{34}^{0} X_{13}^{0} X_{24}^{0} X_{14}^{1} X_{23}^{1} \notag \\
&= -\frac{X_{14} X_{23}}{4 X_{12}}
\end{align}
Similarly, the coefficient of the second term in $\eqref{target}$ is $\frac{1}{2} P u^{-2/3} v^{-1/6}$, which is
\begin{align}
\frac{1}{2} P u^{-2/3} v^{-1/6} &= \frac{1}{2} \Big[ X_{12}^{-1/3} X_{34}^{2/3} X_{13}^{1/6} X_{14}^{1/6} X_{23}^{1/6} X_{24}^{1/6} \Big] \left( \frac{X_{12}X_{34}}{X_{13}X_{24}} \right)^{-2/3} \left( \frac{X_{14}X_{23}}{X_{13}X_{24}} \right)^{-1/6} \notag \\
&= \frac{1}{2} X_{12}^{-1/3 - 2/3} X_{34}^{2/3 - 2/3} X_{13}^{1/6 - (-2/3) - (-1/6)} X_{24}^{1/6 - (-2/3) - (-1/6)} X_{14}^{1/6 - 1/6} X_{23}^{1/6 - 1/6} \notag \\
&= \frac{1}{2} X_{12}^{-1} X_{34}^{0} X_{13}^{1} X_{24}^{1} X_{14}^{0} X_{23}^{0} \notag \\
&= \frac{1}{2} X_{12}^{-1} X_{13} X_{24} \notag \\
&= \frac{X_{13} X_{24}}{2 X_{12}}
\end{align}
Therefore the RHS of \eqref{target} simplifies to 
$$G_4 = -\left( \frac{X_{14}X_{23}}{4 X_{12}} \right) H_{12} + \left( \frac{X_{13}X_{24}}{2 X_{12}} \right) V_{1,24}V_{2,13}$$
Now recall,
$$H_{12} = -2 \big[ (Z_1 \cdot Z_2)(X_1 \cdot X_2) - (Z_1 \cdot X_2)(Z_2 \cdot X_1) \big]
        = X_{12} (Z_1 \cdot Z_2) + 2 (Z_1 \cdot X_2)(Z_2 \cdot X_1)$$
$$V_{1,24} = \frac{(Z_1 \cdot X_2)X_{14} - (Z_1 \cdot X_4)X_{12}}{X_{24}}$$
$$V_{2,13} = \frac{(Z_2 \cdot X_1)X_{23} - (Z_2 \cdot X_3)X_{12}}{X_{13}}$$
Therefore $G_4$ simplifies to 
$$
\begin{aligned}
G_4 &= -\left( \frac{X_{14}X_{23}}{4 X_{12}} \right) \big[ X_{12} (Z_1 \cdot Z_2) + 2 (Z_1 \cdot X_2)(Z_2 \cdot X_1) \big] \\
&\quad + \left( \frac{X_{13}X_{24}}{2 X_{12}} \right) \left( \frac{(Z_1 \cdot X_2)X_{14} - (Z_1 \cdot X_4)X_{12}}{X_{24}} \right) \left( \frac{(Z_2 \cdot X_1)X_{23} - (Z_2 \cdot X_3)X_{12}}{X_{13}} \right) \\
&= -\frac{X_{14}X_{23}}{4} (Z_1 \cdot Z_2) - \frac{X_{14}X_{23}}{2X_{12}} (Z_1 \cdot X_2)(Z_2 \cdot X_1) \\
&\quad + \frac{1}{2X_{12}} \big[ (Z_1 \cdot X_2)X_{14} - (Z_1 \cdot X_4)X_{12} \big] \big[ (Z_2 \cdot X_1)X_{23} - (Z_2 \cdot X_3)X_{12} \big] \\
&= -\frac{X_{14}X_{23}}{4} (Z_1 \cdot Z_2) - \frac{X_{14}X_{23}}{2X_{12}} (Z_1 \cdot X_2)(Z_2 \cdot X_1) + \frac{X_{14}X_{23}}{2X_{12}} (Z_1 \cdot X_2)(Z_2 \cdot X_1) \\
&\quad - \frac{X_{14}}{2} (Z_1 \cdot X_2)(Z_2 \cdot X_3) - \frac{X_{23}}{2} (Z_1 \cdot X_4)(Z_2 \cdot X_1) + \frac{X_{12}}{2} (Z_1 \cdot X_4)(Z_2 \cdot X_3) \\
&= -\frac{X_{14}X_{23}}{4} (Z_1 \cdot Z_2) - \frac{X_{14}}{2}(Z_1 \cdot X_2)(Z_2 \cdot X_3) \\
&\quad - \frac{X_{23}}{2}(Z_1 \cdot X_4)(Z_2 \cdot X_1) + \frac{X_{12}}{2}(Z_1 \cdot X_4)(Z_2 \cdot X_3) \\
&= (Z_1 \cdot X_4)(X_1 \cdot Z_2)(X_2 \cdot X_3) - (Z_1 \cdot X_4)(X_1 \cdot X_2)(Z_2 \cdot X_3) \\
&\quad - (X_1 \cdot X_4)(Z_1 \cdot Z_2)(X_2 \cdot X_3) + (X_1 \cdot X_4)(Z_1 \cdot X_2)(Z_2 \cdot X_3)
\end{aligned}
$$
where $X_{ij}=-2X_i \cdot X_j$ has been used in the last step.
This is what we obtained in \eqref{obtained}.
\subsubsection{Expansions of other substitutions}\label{AAPhiPhiAll}

If we start with the architecture \eqref{ansatz} with all the parameters in the field taken to be distinct i.e. 
\begin{align}
    A_1(X_1,Z_1)&=(\Theta_1 \cdot X_1) (\eta_1 \cdot Z_1)-(\eta_1 \cdot X_1)(\Theta_1 \cdot Z_1)\\ 
    A_2(X_2,Z_2)&=(\Theta_2 \cdot X_2) (\eta_2 \cdot Z_2)-(\eta_2 \cdot X_2)(\Theta_2 \cdot Z_2)\\
    \Phi_3(X_3)&=\Theta_3 \cdot X_3\\
    \Phi_4(X_4)&=\Theta_4 \cdot X_4
\end{align}
and make the following replacements $\Theta_1 \to \eta_2, \eta_1 \to \Theta_2,$ and $\Theta_3 \to \Theta_4$, the architecture reduces to \eqref{GFFansatz} up to an overall sign, since the replacement interchanges the two terms of $A_1$; we therefore get the negative of \eqref{eq:AAPhiPhi_expanded}, which we call $G_4^{(1)}$ here. This can be expressed in terms of the tensor structures as \begin{equation}G_4^{(1)} = P \left( -\frac{1}{2} u^{1/3} v^{-1/6} H_{12} \right) \label{eq:G4_1}\end{equation} where $P$ is the prefactor $\prod_{i<j}^4 X_{ij}^{-\alpha_{ij}}$ evaluated in \eqref{prefactor4}. Similarly, the substitutions $\{\eta_1 \to \eta_2, \Theta_1 \to \Theta_2, \Theta_3 \to \Theta_4\}$ gives the exact negative of \eqref{eq:G4_1}, which is just $-G_4^{(1)}$.

The replacements $\{\Theta_1 \to \eta_2, \Theta_2 \to \Theta_3, \eta_1 \to \Theta_4\}$ or $\{\eta_1 \to \Theta_2, \eta_2 \to \Theta_3, \Theta_1 \to \Theta_4\}$ yield the structure we call $G_4^{(2)}$ which was found in \eqref{target}
This can be expressed as 
\begin{equation} \label{eq:G4_2}
G_4^{(2)} = P \left(-\frac{1}{4} u^{-2/3} v^{5/6} H_{12} + \frac{1}{2} u^{-2/3} v^{-1/6} V_{1,24} V_{2,13}\right)
\end{equation} The associated replacements $\{\eta_1 \to \eta_2, \Theta_2 \to \Theta_3, \Theta_1 \to \Theta_4\}$ or $\{\Theta_1 \to \Theta_2, \eta_2 \to \Theta_3, \eta_1 \to \Theta_4\}$ yield $-G_4^{(2)}$. These were the cases that were discussed earlier.

Additionally, the replacements $\{\Theta_1 \to \eta_2, \eta_1 \to \Theta_3, \Theta_2 \to \Theta_4\}$ or $\{\eta_1 \to \Theta_2, \Theta_1 \to \Theta_3, \eta_2 \to \Theta_4\}$ yield $G_4^{(3)}$:
\begin{equation} \label{eq:G4_3}
\begin{split}
    G_4^{(3)} =& (X_1 \cdot Z_2)(X_2 \cdot X_4)(X_3 \cdot Z_1) + (X_1 \cdot X_3)(X_2 \cdot Z_1)(X_4 \cdot Z_2) \\
    &- (X_1 \cdot X_2)(X_3 \cdot Z_1)(X_4 \cdot Z_2) - (X_1 \cdot X_3)(X_2 \cdot X_4)(Z_1 \cdot Z_2)
\end{split}
\end{equation}
which can be expressed as 
\begin{equation}
    G_4^{(3)} = P \left(-\frac{1}{4} u^{-2/3} v^{-1/6} H_{12} + \frac{1}{2} u^{-2/3} v^{5/6} V_{1,23} V_{2,14}\right)
\end{equation}
and the replacements $\{\eta_1 \to \eta_2, \Theta_1 \to \Theta_3, \Theta_2 \to \Theta_4\}$ or $\{\Theta_1 \to \Theta_2, \eta_1 \to \Theta_3, \eta_2 \to \Theta_4\}$ yield $-G_4^{(3)}$. 

Finally, the replacements $\{\Theta_1 \to \eta_2, \eta_1 \to \Theta_4, \Theta_2 \to \Theta_4, \Theta_3 \to \Theta_4\}$ or $\{\eta_1 \to \Theta_2, \Theta_1 \to \Theta_4, \eta_2 \to \Theta_4, \Theta_3 \to \Theta_4\}$ yield $G_4^{(4)}$:
\begin{equation} \label{eq:G4_final}
\begin{split}
    G_4^{(4)} =& (X_1 \cdot Z_2)[(X_2 \cdot Z_1)(X_3 \cdot X_4) + (X_2 \cdot X_4)(X_3 \cdot Z_1) + (X_2 \cdot X_3)(X_4 \cdot Z_1)] \\
    &+ (X_2 \cdot Z_1)[(X_1 \cdot Z_2)(X_3 \cdot X_4) + (X_1 \cdot X_4)(X_3 \cdot Z_2) + (X_1 \cdot X_3)(X_4 \cdot Z_2)] \\
    &- [(X_1 \cdot X_4)(X_2 \cdot X_3) + (X_1 \cdot X_3)(X_2 \cdot X_4) + (X_1 \cdot X_2)(X_3 \cdot X_4)](Z_1 \cdot Z_2) \\
    &- (X_1 \cdot X_2)[(X_3 \cdot Z_2)(X_4 \cdot Z_1) + (X_3 \cdot Z_1)(X_4 \cdot Z_2) + (X_3 \cdot X_4)(Z_1 \cdot Z_2)]
\end{split}
\end{equation}
which can be expressed as $G_4^{(4)} = P \Big( \big(-\frac{1}{2} u^{1/3} v^{-1/6} - \frac{1}{4} u^{-2/3} v^{5/6} - \frac{1}{4} u^{-2/3} v^{-1/6}\big) H_{12} + \frac{1}{2} u^{-2/3} v^{-1/6} V_{1,24} V_{2,13} + \frac{1}{2} u^{-2/3} v^{5/6} V_{1,23} V_{2,14} \Big)$, and the replacements $\{\eta_1 \to \eta_2, \Theta_1 \to \Theta_4, \Theta_2 \to \Theta_4, \Theta_3 \to \Theta_4\}$ yield the negative of \eqref{eq:G4_final}. 
These are the only relabellings of the architecture which yield non-zero results. The last result \eqref{eq:G4_final} is actually a direct algebraic sum of the other three, i.e.,
\begin{equation}
    G_4 ^{(4)}= G_4^{(1)} + G_4^{(2)} + G_4^{(3)}
\end{equation}
It is to be noted that the tensor structures $V_{1,23}V_{2,13}$ and $V_{1,24}V_{2,14}$ never appear in this NN architecture no matter what parameter relabelling is done.  However, these structures can be easily obtained by using a different NN architecture.
\subsection{$\langle A(X_1,Z_1) A(X_2,Z_2) A(X_3,Z_3) A(X_4,Z_4) \rangle$}\label{AAAAresults}

For the $4$-point correlator $\langle A(X_1,Z_1) A(X_2,Z_2) A(X_3,Z_3) A(X_4,Z_4) \rangle$ we expect 43 tensor structures of the schematic form $HH$, $HVV$ and $VVVV$. Out of these $28$ contribute to the correlator computed from our proposed architecture \eqref{spin1ansatz}. The neural network computation yields
\begin{align}
\langle A(X_1) A(X_2) A(X_3) A(X_4)\rangle
&= 2 \bigl( X_2\!\cdot\!Z_4\, X_3\!\cdot\!Z_1 + X_2\!\cdot\!Z_1\, X_3\!\cdot\!Z_4 + X_2\!\cdot\!X_3\, Z_1\!\cdot\!Z_4 \bigr) \nonumber \\
&\quad \times \bigl( X_1\!\cdot\!Z_3\, X_4\!\cdot\!Z_2 + X_1\!\cdot\!Z_2\, X_4\!\cdot\!Z_3 + X_1\!\cdot\!X_4\, Z_2\!\cdot\!Z_3 \bigr) \nonumber \\
&\quad + \bigl( X_1\!\cdot\!Z_4\, X_2\!\cdot\!X_3 + X_1\!\cdot\!X_3\, X_2\!\cdot\!Z_4 + X_1\!\cdot\!X_2\, X_3\!\cdot\!Z_4 \bigr) \nonumber \\
&\quad \times \bigl( - X_4\!\cdot\!Z_3\, Z_1\!\cdot\!Z_2 - X_4\!\cdot\!Z_2\, Z_1\!\cdot\!Z_3 - X_4\!\cdot\!Z_1\, Z_2\!\cdot\!Z_3 \bigr) \nonumber \\
&\quad + \bigl( - X_1\!\cdot\!Z_4\, X_2\!\cdot\!X_3 - X_1\!\cdot\!X_3\, X_2\!\cdot\!Z_4 - X_1\!\cdot\!X_2\, X_3\!\cdot\!Z_4 \bigr) \nonumber \\
&\quad \times \bigl( X_4\!\cdot\!Z_3\, Z_1\!\cdot\!Z_2 + X_4\!\cdot\!Z_2\, Z_1\!\cdot\!Z_3 + X_4\!\cdot\!Z_1\, Z_2\!\cdot\!Z_3 \bigr) \nonumber \\
&\quad + 2 \bigl( X_2\!\cdot\!Z_3\, X_4\!\cdot\!Z_1 + X_2\!\cdot\!Z_1\, X_4\!\cdot\!Z_3 + X_2\!\cdot\!X_4\, Z_1\!\cdot\!Z_3 \bigr) \nonumber \\
&\quad \times \bigl( X_1\!\cdot\!Z_4\, X_3\!\cdot\!Z_2 + X_1\!\cdot\!Z_2\, X_3\!\cdot\!Z_4 + X_1\!\cdot\!X_3\, Z_2\!\cdot\!Z_4 \bigr) \nonumber \\
&\quad + \bigl( X_1\!\cdot\!Z_3\, X_2\!\cdot\!X_4 + X_1\!\cdot\!X_4\, X_2\!\cdot\!Z_3 + X_1\!\cdot\!X_2\, X_4\!\cdot\!Z_3 \bigr) \nonumber \\
&\quad \times \bigl( - X_3\!\cdot\!Z_4\, Z_1\!\cdot\!Z_2 - X_3\!\cdot\!Z_2\, Z_1\!\cdot\!Z_4 - X_3\!\cdot\!Z_1\, Z_2\!\cdot\!Z_4 \bigr) \nonumber \\
&\quad + \bigl( - X_1\!\cdot\!Z_3\, X_2\!\cdot\!X_4 - X_1\!\cdot\!X_4\, X_2\!\cdot\!Z_3 - X_1\!\cdot\!X_2\, X_4\!\cdot\!Z_3 \bigr) \nonumber \\
&\quad \times \bigl( X_3\!\cdot\!Z_4\, Z_1\!\cdot\!Z_2 + X_3\!\cdot\!Z_2\, Z_1\!\cdot\!Z_4 + X_3\!\cdot\!Z_1\, Z_2\!\cdot\!Z_4 \bigr) \nonumber \\
&\quad + 2 \bigl( X_3\!\cdot\!Z_2\, X_4\!\cdot\!Z_1 + X_3\!\cdot\!Z_1\, X_4\!\cdot\!Z_2 + X_3\!\cdot\!X_4\, Z_1\!\cdot\!Z_2 \bigr) \nonumber \\
&\quad \times \bigl( X_1\!\cdot\!Z_4\, X_2\!\cdot\!Z_3 + X_1\!\cdot\!Z_3\, X_2\!\cdot\!Z_4 + X_1\!\cdot\!X_2\, Z_3\!\cdot\!Z_4 \bigr) \nonumber \\
&\quad + \bigl( X_2\!\cdot\!Z_1\, X_3\!\cdot\!X_4 + X_2\!\cdot\!X_4\, X_3\!\cdot\!Z_1 + X_2\!\cdot\!X_3\, X_4\!\cdot\!Z_1 \bigr) \nonumber \\
&\quad \times \bigl( - X_1\!\cdot\!Z_4\, Z_2\!\cdot\!Z_3 - X_1\!\cdot\!Z_3\, Z_2\!\cdot\!Z_4 - X_1\!\cdot\!Z_2\, Z_3\!\cdot\!Z_4 \bigr) \nonumber \\
&\quad + \bigl( - X_2\!\cdot\!Z_1\, X_3\!\cdot\!X_4 - X_2\!\cdot\!X_4\, X_3\!\cdot\!Z_1 - X_2\!\cdot\!X_3\, X_4\!\cdot\!Z_1 \bigr) \nonumber \\
&\quad \times \bigl( X_1\!\cdot\!Z_4\, Z_2\!\cdot\!Z_3 + X_1\!\cdot\!Z_3\, Z_2\!\cdot\!Z_4 + X_1\!\cdot\!Z_2\, Z_3\!\cdot\!Z_4 \bigr) \nonumber \\
&\quad + \bigl( X_1\!\cdot\!Z_2\, X_3\!\cdot\!X_4 + X_1\!\cdot\!X_4\, X_3\!\cdot\!Z_2 + X_1\!\cdot\!X_3\, X_4\!\cdot\!Z_2 \bigr) \nonumber \\
&\quad \times \bigl( - X_2\!\cdot\!Z_4\, Z_1\!\cdot\!Z_3 - X_2\!\cdot\!Z_3\, Z_1\!\cdot\!Z_4 - X_2\!\cdot\!Z_1\, Z_3\!\cdot\!Z_4 \bigr) \nonumber \\
&\quad + \bigl( - X_1\!\cdot\!Z_2\, X_3\!\cdot\!X_4 - X_1\!\cdot\!X_4\, X_3\!\cdot\!Z_2 - X_1\!\cdot\!X_3\, X_4\!\cdot\!Z_2 \bigr) \nonumber \\
&\quad \times \bigl( X_2\!\cdot\!Z_4\, Z_1\!\cdot\!Z_3 + X_2\!\cdot\!Z_3\, Z_1\!\cdot\!Z_4 + X_2\!\cdot\!Z_1\, Z_3\!\cdot\!Z_4 \bigr) \nonumber \\
&\quad + 2 \bigl( X_1\!\cdot\!X_4\, X_2\!\cdot\!X_3 + X_1\!\cdot\!X_3\, X_2\!\cdot\!X_4 + X_1\!\cdot\!X_2\, X_3\!\cdot\!X_4 \bigr) \nonumber \\
&\quad \times \bigl( Z_1\!\cdot\!Z_4\, Z_2\!\cdot\!Z_3 + Z_1\!\cdot\!Z_3\, Z_2\!\cdot\!Z_4 + Z_1\!\cdot\!Z_2\, Z_3\!\cdot\!Z_4 \bigr).
\end{align}
which can be expressed as a linear combination of $28$ tensor structures with the coefficients being functions of cross ratios $u$ and $v$ as follows:
$$
\begin{aligned}
\langle A(X_1) A(X_2) A(X_3) A(X_4)\rangle
&= \left(\frac{1+u+2v}{2v}\right) H_{14}H_{23}
+ \left(\frac{u+v+2}{2}\right) H_{13}H_{24}
+ \left(\frac{1+2u+v}{2u}\right) H_{12}H_{34} \\[6pt]
&\quad - \left(\frac{1+v}{u}\right) H_{12}V_{3,14}V_{4,13}
+ H_{12}V_{3,14}V_{4,12}
+ H_{12}V_{3,12}V_{4,13}
+ H_{13}V_{2,14}V_{4,13} \\
&\quad - (u+v) H_{13}V_{2,14}V_{4,12}
+ H_{14}V_{2,14}V_{3,12}
+ H_{13}V_{2,13}V_{4,12}
+ H_{14}V_{2,13}V_{3,14} \\
&\quad - \left(\frac{1+u}{v}\right) H_{14}V_{2,13}V_{3,12}
+ \left(\frac{1}{v}\right) H_{23}V_{1,24}V_{4,13}
+ \left(\frac{1}{v}\right) H_{23}V_{1,24}V_{4,12}
\\
&\quad - H_{24}V_{1,24}V_{3,12} - \left(\frac{1}{u}\right) H_{34}V_{1,24}V_{2,13}
- H_{23}V_{1,23}V_{4,12}
+ v\, H_{24}V_{1,23}V_{3,14} \\
&\quad + v\, H_{24}V_{1,23}V_{3,12}- \left(\frac{v}{u}\right) H_{34}V_{1,23}V_{2,14} \\[6pt]
&\quad + 2\, V_{1,24}V_{2,14}V_{3,12}V_{4,12}
+ \left(\frac{2}{u}\right) V_{1,24}V_{2,13}V_{3,14}V_{4,13}
- \left(\frac{2}{v}\right) V_{1,24}V_{2,13}V_{3,12}V_{4,13} \\
&\quad - \left(\frac{2}{v}\right) V_{1,24}V_{2,13}V_{3,12}V_{4,12}
+ \left(\frac{2v}{u}\right) V_{1,23}V_{2,14}V_{3,14}V_{4,13}\\
&\quad - 2v\, V_{1,23}V_{2,14}V_{3,12}V_{4,12}
+ 2\, V_{1,23}V_{2,13}V_{3,12}V_{4,12} - 2v\, V_{1,23}V_{2,14}V_{3,14}V_{4,12}
\end{aligned}
$$
If we instead compute the $4$-point correlator $\langle A_1(X_1,Z_1) A_1(X_2,Z_2) A_2(X_3,Z_3) A_2(X_4,Z_4)$ where $A_1$ and $A_2$ are defined in \eqref{AAPhiPhiansatz2} we get

\begin{align}
\langle A_1(X_1,Z_1) A_1(X_2,Z_2)  A_2(X_3,Z_3) &A_2(X_4,Z_4)\rangle \notag \\
&= - \bigl( X_1\!\cdot\!Z_4\, X_2\!\cdot\!X_3 + X_1\!\cdot\!X_3\, X_2\!\cdot\!Z_4 + X_1\!\cdot\!X_2\, X_3\!\cdot\!Z_4 \bigr)\, X_4\!\cdot\!Z_3\, Z_1\!\cdot\!Z_2 \nonumber \\
&\quad - X_3\!\cdot\!Z_4 \bigl( X_1\!\cdot\!Z_3\, X_2\!\cdot\!X_4 + X_1\!\cdot\!X_4\, X_2\!\cdot\!Z_3 + X_1\!\cdot\!X_2\, X_4\!\cdot\!Z_3 \bigr)\, Z_1\!\cdot\!Z_2 \nonumber \\
&\quad + X_1\!\cdot\!Z_2\, X_3\!\cdot\!Z_4 \bigl( X_2\!\cdot\!Z_3\, X_4\!\cdot\!Z_1 + X_2\!\cdot\!Z_1\, X_4\!\cdot\!Z_3 + X_2\!\cdot\!X_4\, Z_1\!\cdot\!Z_3 \bigr) \nonumber \\
&\quad + X_1\!\cdot\!Z_2\, X_4\!\cdot\!Z_3 \bigl( X_2\!\cdot\!Z_4\, X_3\!\cdot\!Z_1 + X_2\!\cdot\!Z_1\, X_3\!\cdot\!Z_4 + X_2\!\cdot\!X_3\, Z_1\!\cdot\!Z_4 \bigr) \nonumber \\
&\quad + X_2\!\cdot\!Z_1\, X_3\!\cdot\!Z_4 \bigl( X_1\!\cdot\!Z_3\, X_4\!\cdot\!Z_2 + X_1\!\cdot\!Z_2\, X_4\!\cdot\!Z_3 + X_1\!\cdot\!X_4\, Z_2\!\cdot\!Z_3 \bigr) \nonumber \\
&\quad - X_1\!\cdot\!X_2\, X_3\!\cdot\!Z_4 \bigl( X_4\!\cdot\!Z_3\, Z_1\!\cdot\!Z_2 + X_4\!\cdot\!Z_2\, Z_1\!\cdot\!Z_3 + X_4\!\cdot\!Z_1\, Z_2\!\cdot\!Z_3 \bigr) \nonumber \\
&\quad + X_2\!\cdot\!Z_1\, X_4\!\cdot\!Z_3 \bigl( X_1\!\cdot\!Z_4\, X_3\!\cdot\!Z_2 + X_1\!\cdot\!Z_2\, X_3\!\cdot\!Z_4 + X_1\!\cdot\!X_3\, Z_2\!\cdot\!Z_4 \bigr) \nonumber \\
&\quad - X_1\!\cdot\!X_2\, X_4\!\cdot\!Z_3 \bigl( X_3\!\cdot\!Z_4\, Z_1\!\cdot\!Z_2 + X_3\!\cdot\!Z_2\, Z_1\!\cdot\!Z_4 + X_3\!\cdot\!Z_1\, Z_2\!\cdot\!Z_4 \bigr) \nonumber \\
&\quad - X_1\!\cdot\!Z_2 \bigl( X_2\!\cdot\!Z_1\, X_3\!\cdot\!X_4 + X_2\!\cdot\!X_4\, X_3\!\cdot\!Z_1 + X_2\!\cdot\!X_3\, X_4\!\cdot\!Z_1 \bigr)\, Z_3\!\cdot\!Z_4 \nonumber \\
&\quad - X_2\!\cdot\!Z_1 \bigl( X_1\!\cdot\!Z_2\, X_3\!\cdot\!X_4 + X_1\!\cdot\!X_4\, X_3\!\cdot\!Z_2 + X_1\!\cdot\!X_3\, X_4\!\cdot\!Z_2 \bigr)\, Z_3\!\cdot\!Z_4 \nonumber \\
&\quad + \bigl( X_1\!\cdot\!X_4\, X_2\!\cdot\!X_3 + X_1\!\cdot\!X_3\, X_2\!\cdot\!X_4 + X_1\!\cdot\!X_2\, X_3\!\cdot\!X_4 \bigr)\, Z_1\!\cdot\!Z_2\, Z_3\!\cdot\!Z_4 \nonumber \\
&\quad + X_1\!\cdot\!X_2 \bigl( X_3\!\cdot\!Z_2\, X_4\!\cdot\!Z_1 + X_3\!\cdot\!Z_1\, X_4\!\cdot\!Z_2 + X_3\!\cdot\!X_4\, Z_1\!\cdot\!Z_2 \bigr)\, Z_3\!\cdot\!Z_4 \nonumber \\
&\quad + X_3\!\cdot\!X_4\, Z_1\!\cdot\!Z_2 \bigl( X_1\!\cdot\!Z_4\, X_2\!\cdot\!Z_3 + X_1\!\cdot\!Z_3\, X_2\!\cdot\!Z_4 + X_1\!\cdot\!X_2\, Z_3\!\cdot\!Z_4 \bigr) \nonumber \\
&\quad - X_2\!\cdot\!Z_1\, X_3\!\cdot\!X_4 \bigl( X_1\!\cdot\!Z_4\, Z_2\!\cdot\!Z_3 + X_1\!\cdot\!Z_3\, Z_2\!\cdot\!Z_4 + X_1\!\cdot\!Z_2\, Z_3\!\cdot\!Z_4 \bigr) \nonumber \\
&\quad - X_1\!\cdot\!Z_2\, X_3\!\cdot\!X_4 \bigl( X_2\!\cdot\!Z_4\, Z_1\!\cdot\!Z_3 + X_2\!\cdot\!Z_3\, Z_1\!\cdot\!Z_4 + X_2\!\cdot\!Z_1\, Z_3\!\cdot\!Z_4 \bigr) \nonumber \\
&\quad + X_1\!\cdot\!X_2\, X_3\!\cdot\!X_4 \bigl( Z_1\!\cdot\!Z_4\, Z_2\!\cdot\!Z_3 + Z_1\!\cdot\!Z_3\, Z_2\!\cdot\!Z_4 + Z_1\!\cdot\!Z_2\, Z_3\!\cdot\!Z_4 \bigr).
\end{align}
whose correctness is implied by the existence of the following expansion:
$$\begin{aligned}
\langle A_1(X_1,Z_1) A_1(X_2,Z_2) A_2(X_3,Z_3) A_2(X_4,Z_4) \rangle &= \left(\frac{u}{4v}\right) H_{14}H_{23} + \left(\frac{u}{4}\right) H_{13}H_{24} + \left(\frac{4u+v+1}{4u}\right) H_{12}H_{34} \\
&- \left(\frac{v+1}{2u}\right) H_{12}V_{3,14}V_{4,13} + \frac{1}{2} H_{12}V_{3,14}V_{4,12} + \frac{1}{2} H_{12}V_{3,12}V_{4,13} \\
&\quad + \frac{1}{2} H_{13}V_{2,14}V_{4,13} - \frac{u}{2} H_{13}V_{2,14}V_{4,12} \\
&\quad + \frac{1}{2} H_{14}V_{2,13}V_{3,14} - \left(\frac{u}{2v}\right) H_{14}V_{2,13}V_{3,12} \\
&\quad + \left(\frac{1}{2v}\right) H_{23}V_{1,24}V_{4,13} - \left(\frac{1}{2u}\right) H_{34}V_{1,24}V_{2,13} \\
&\quad + \left(\frac{1}{u}\right) V_{1,24}V_{2,13}V_{3,14}V_{4,13} - \left(\frac{1}{v}\right) V_{1,24}V_{2,13}V_{3,12}V_{4,13} \\
&\quad + \frac{v}{2} H_{24}V_{1,23}V_{3,14} - \left(\frac{v}{2u}\right) H_{34}V_{1,23}V_{2,14} \\
&\quad + \left(\frac{v}{u}\right) V_{1,23}V_{2,14}V_{3,14}V_{4,13} - v V_{1,23}V_{2,14}V_{3,14}V_{4,12}
\end{aligned}$$
These $4$-point functions are crossing invariant by construction but can also be verified explicitly. Also for example, under exchange of $1 \leftrightarrow 2$ the tensor structure in the first term above becomes $H_{24}H_{13}$ which is the tensor structure in the 2nd term. Accordingly using the fact that $f(u,v) \to f(\frac{u}{v},\frac{1}{v})$ under $1 \leftrightarrow 2$, the coefficient of the first term i.e. $\frac{u}{4v}$ becomes $\frac{u/v}{4\times 1/v}=\frac{u}{4}$ which is indeed the coefficient of the 2nd term.
\subsection{Generalized Free Fields}
\label{sec:exact_gff}

As demonstrated in section \ref{sec:main}, our main architecture \eqref{ansatz} for $\Delta=-1$ is bilinear in independent Gaussian parameters naturally yield non-Gaussian cross-contractions, obstructing strict factorization into a Generalized Free Field (GFF) in general. To isolate a true GFF sector, the neural network field architecture is to be made linear in the parameter space. We construct a spin-1 conformal primary $A(X,Z)$ of dimension $\Delta = -1$ on the embedding space $\mathbb{R}^{d+1,1}$ using $d+2$ independent Gaussian vector parameters $\Theta^{(a)} \in \mathbb{R}^{d+2}$, where $a \in \{1, \dots, d+2\}$. These parameters are initialized with zero mean and unit covariance, ensuring mutual independence:
\begin{equation}
    \langle \Theta_M^{(a)} \rangle = 0, \quad \langle \Theta_M^{(a)} \Theta_N^{(b)} \rangle = \delta^{ab} \delta_{MN}.
\end{equation}
where $M,N$ labels the embedding space components.
The spin-1 field is then defined as a linear combination of the Gaussian parameters contracted with this form factor:
\begin{equation}
    A(X_i, Z_i) := A_i = \sum_{a, M} \Theta_M^{(a)} f_i^{a, M}.
\end{equation}
where
\begin{equation}
    f_i^{a, M} := X_{i,a} Z_i^M - Z_{i,a} X_i^M.
\end{equation}
with $i$ labelling different embedding space points.

By construction, this field is homogeneous in $X$ with degree $-1$, linear in $Z$, and satisfies the transversality condition $A(X,X) = 0$. Because the field is linear in the Gaussian parameters, it is a Gaussian random field itself, allowing Wick's theorem to apply directly to the field operators.

\subsubsection{Two-Point Function}
We compute the two-point function $\langle A(X_1) A(X_2) \rangle$ as
\begin{equation}
\begin{aligned}
\langle A(X_1) A(X_2) \rangle &= \left\langle \left( \sum_{a, M} \Theta_M^{(a)} f_1^{a, M} \right) \left( \sum_{b, N} \Theta_N^{(b)} f_2^{b, N} \right) \right\rangle \\
&= \sum_{a, b} \sum_{M, N} \langle \Theta_M^{(a)} \Theta_N^{(b)} \rangle f_1^{a, M} f_2^{b, N} \\
&= \sum_{a, b} \sum_{M, N} (\delta^{ab} \delta_{MN}) f_1^{a, M} f_2^{b, N} \\
&= \sum_{a, M} f_1^{a, M} f_2^{a, M} \\
&= \sum_{a, M} (X_{1,a} Z_1^M - Z_{1,a} X_1^M)(X_{2,a} Z_2^M - Z_{2,a} X_2^M) \\
&= \sum_{a, M} \Big( X_{1,a} X_{2,a} Z_1^M Z_2^M - X_{1,a} Z_{2,a} Z_1^M X_2^M - Z_{1,a} X_{2,a} X_1^M Z_2^M + Z_{1,a} Z_{2,a} X_1^M X_2^M \Big) \\
&= (X_1 \cdot X_2)(Z_1 \cdot Z_2) - (X_1 \cdot Z_2)(Z_1 \cdot X_2) - (Z_1 \cdot X_2)(X_1 \cdot Z_2) + (Z_1 \cdot Z_2)(X_1 \cdot X_2) \\
&= 2 \Big[ (X_1 \cdot X_2)(Z_1 \cdot Z_2) - (X_1 \cdot Z_2)(X_2 \cdot Z_1) \Big] \\
&= -H_{12}.
\end{aligned}
\end{equation}

\subsubsection{Three-Point Function}
For any parity-preserving GFF, the three-point function of the field with itself must vanish. In our construction, this arises directly from the vanishing of the third moment of the zero-mean Gaussian distribution:
\begin{equation}
\begin{aligned}
\langle A_1 A_2 A_3 \rangle &= \left\langle \left( \sum_{a, M} \Theta_M^{(a)} f_1^{a, M} \right) \left( \sum_{b, N} \Theta_N^{(b)} f_2^{b, N} \right) \left( \sum_{c, P} \Theta_P^{(c)} f_3^{c, P} \right) \right\rangle \\
&= \sum_{a, b, c} \sum_{M, N, P} \langle \Theta_M^{(a)} \Theta_N^{(b)} \Theta_P^{(c)} \rangle f_1^{a, M} f_2^{b, N} f_3^{c, P} \\
&= 0.
\end{aligned}
\end{equation}

\subsubsection{Four-Point Function}
Similarly, applying Wick's theorem we get:
\begin{equation}
\begin{aligned}
\langle A_1 A_2 A_3 A_4 \rangle &= \sum_{\substack{a_1, a_2, a_3, a_4 \\ M_1, M_2, M_3, M_4}} \langle \Theta_{M_1}^{(a_1)} \Theta_{M_2}^{(a_2)} \Theta_{M_3}^{(a_3)} \Theta_{M_4}^{(a_4)} \rangle f_1^{a_1, M_1} f_2^{a_2, M_2} f_3^{a_3, M_3} f_4^{a_4, M_4} \\
&= \left( \sum_{a_1, a_2, M_1, M_2} \langle \Theta_{M_1}^{(a_1)} \Theta_{M_2}^{(a_2)} \rangle f_1^{a_1, M_1} f_2^{a_2, M_2} \right) \left( \sum_{a_3, a_4, M_3, M_4} \langle \Theta_{M_3}^{(a_3)} \Theta_{M_4}^{(a_4)} \rangle f_3^{a_3, M_3} f_4^{a_4, M_4} \right) \\
&\quad + \left( \sum_{a_1, a_3, M_1, M_3} \langle \Theta_{M_1}^{(a_1)} \Theta_{M_3}^{(a_3)} \rangle f_1^{a_1, M_1} f_3^{a_3, M_3} \right) \left( \sum_{a_2, a_4, M_2, M_4} \langle \Theta_{M_2}^{(a_2)} \Theta_{M_4}^{(a_4)} \rangle f_2^{a_2, M_2} f_4^{a_4, M_4} \right) \\
&\quad + \left( \sum_{a_1, a_4, M_1, M_4} \langle \Theta_{M_1}^{(a_1)} \Theta_{M_4}^{(a_4)} \rangle f_1^{a_1, M_1} f_4^{a_4, M_4} \right) \left( \sum_{a_2, a_3, M_2, M_3} \langle \Theta_{M_2}^{(a_2)} \Theta_{M_3}^{(a_3)} \rangle f_2^{a_2, M_2} f_3^{a_3, M_3} \right) \\
&= \langle A_1 A_2 \rangle \langle A_3 A_4 \rangle + \langle A_1 A_3 \rangle \langle A_2 A_4 \rangle + \langle A_1 A_4 \rangle \langle A_2 A_3 \rangle \\
&= (-H_{12})(-H_{34}) + (-H_{13})(-H_{24}) + (-H_{14})(-H_{23}) \\
&= H_{12}H_{34} + H_{13}H_{24} + H_{14}H_{23}.
\end{aligned}
\end{equation}
This demonstrates that replacing the bilinear ansatz with an architecture linear in a single collection of Gaussian parameters produces a Generalized Free Field at $\Delta = -1$. However, for generic $\Delta$ the most straightforward way to get a GFF would be to use the large-$N$ techniques discussed in \cite{Halverson:2024axc}. A particular application to the case of Maxwell theory is discussed in Appendix \ref{sec:largeN}
\subsection{$d=4$ Maxwell CFT}
\subsubsection{Review of some results from CFT}\label{cftresults}
In this section we prove two statements for generic CFTs and apply them to our scenario to describe the strategy behind obtaining the two point function of Maxwell CFT in $d=4$ from a CFT with vector primaries.\\\\
1. Level-$1$ two-form descendant of a vector primary has vanishing two point correlators for $\Delta=1$\\
Proof: Let $A_\mu$ be a vector primary and $F_{\mu \nu}=\partial_\mu A_nu-\partial_\nu A_\mu$ be its level-1$1$ antisymmetric descendant. By conformal invariance, the two point function of $A_\mu$ is fixed to be 
$$\langle A_\mu(x) A_\nu(0) \rangle = \frac{I_{\mu\nu}(x)}{x^{2\Delta}}$$ where $$I_{\mu\nu}(x) = \delta_{\mu\nu} - 2\frac{x_\mu x_\nu}{x^2}.$$ Then $$\langle F_{\mu\nu}(x) F_{\rho\sigma}(0) \rangle = - \left( \partial_\mu \delta_\nu^\alpha - \partial_\nu \delta_\mu^\alpha \right) \left( \partial_\rho \delta_\sigma^\beta - \partial_\sigma \delta_\rho^\beta \right) \langle A_\alpha(x) A_\beta(0) \rangle$$ which can be simplified as 
$$
\begin{aligned}
\langle F_{\mu\nu}(x) F_{\rho\sigma}(0) \rangle &= -\partial_\mu \partial_\rho \left( \frac{I_{\nu\sigma}(x)}{x^{2\Delta}} \right) + \partial_\mu \partial_\sigma \left( \frac{I_{\nu\rho}(x)}{x^{2\Delta}} \right) + \partial_\nu \partial_\rho \left( \frac{I_{\mu\sigma}(x)}{x^{2\Delta}} \right) - \partial_\nu \partial_\sigma \left( \frac{I_{\mu\rho}(x)}{x^{2\Delta}} \right) \\
&= -\Bigg[ \frac{-2\Delta \delta_{\mu\rho} \delta_{\nu\sigma} - 2 \delta_{\nu\rho} \delta_{\mu\sigma} - 2 \delta_{\mu\nu} \delta_{\sigma\rho}}{x^{2\Delta+2}} \\
&\quad + \frac{4(\Delta+1)}{x^{2\Delta+4}} \Big( \Delta \delta_{\nu\sigma} x_\mu x_\rho + \delta_{\nu\rho} x_\sigma x_\mu + \delta_{\sigma\rho} x_\nu x_\mu + \delta_{\mu\nu} x_\sigma x_\rho + \delta_{\mu\sigma} x_\nu x_\rho + \delta_{\mu\rho} x_\nu x_\sigma \Big) \\
&\quad - \frac{8(\Delta+1)(\Delta+2)}{x^{2\Delta+6}} x_\mu x_\nu x_\rho x_\sigma \Bigg] \\
&\quad + \Bigg[ \frac{-2\Delta \delta_{\mu\sigma} \delta_{\nu\rho} - 2 \delta_{\nu\sigma} \delta_{\mu\rho} - 2 \delta_{\mu\nu} \delta_{\rho\sigma}}{x^{2\Delta+2}} \\
&\quad + \frac{4(\Delta+1)}{x^{2\Delta+4}} \Big( \Delta \delta_{\nu\rho} x_\mu x_\sigma + \delta_{\nu\sigma} x_\rho x_\mu + \delta_{\rho\sigma} x_\nu x_\mu + \delta_{\mu\nu} x_\rho x_\sigma + \delta_{\mu\rho} x_\nu x_\sigma + \delta_{\mu\sigma} x_\nu x_\rho \Big) \\
&\quad - \frac{8(\Delta+1)(\Delta+2)}{x^{2\Delta+6}} x_\mu x_\nu x_\rho x_\sigma \Bigg] \\
&\quad + \Bigg[ \frac{-2\Delta \delta_{\nu\rho} \delta_{\mu\sigma} - 2 \delta_{\mu\rho} \delta_{\nu\sigma} - 2 \delta_{\nu\mu} \delta_{\sigma\rho}}{x^{2\Delta+2}} \\
&\quad + \frac{4(\Delta+1)}{x^{2\Delta+4}} \Big( \Delta \delta_{\mu\sigma} x_\nu x_\rho + \delta_{\mu\rho} x_\sigma x_\nu + \delta_{\sigma\rho} x_\mu x_\nu + \delta_{\nu\mu} x_\sigma x_\rho + \delta_{\nu\sigma} x_\mu x_\rho + \delta_{\nu\rho} x_\mu x_\sigma \Big) \\
&\quad - \frac{8(\Delta+1)(\Delta+2)}{x^{2\Delta+6}} x_\mu x_\nu x_\rho x_\sigma \Bigg] \\
&\quad - \Bigg[ \frac{-2\Delta \delta_{\nu\sigma} \delta_{\mu\rho} - 2 \delta_{\mu\sigma} \delta_{\nu\rho} - 2 \delta_{\nu\mu} \delta_{\rho\sigma}}{x^{2\Delta+2}} \\
&\quad + \frac{4(\Delta+1)}{x^{2\Delta+4}} \Big( \Delta \delta_{\mu\rho} x_\nu x_\sigma + \delta_{\mu\sigma} x_\rho x_\nu + \delta_{\rho\sigma} x_\mu x_\nu + \delta_{\nu\mu} x_\rho x_\sigma + \delta_{\nu\rho} x_\mu x_\sigma + \delta_{\nu\sigma} x_\mu x_\rho \Big) \\
&\quad - \frac{8(\Delta+1)(\Delta+2)}{x^{2\Delta+6}} x_\mu x_\nu x_\rho x_\sigma \Bigg] \\
&= \frac{1}{x^{2\Delta+2}} \Big( 2\Delta \delta_{\mu\rho} \delta_{\nu\sigma} + 2 \delta_{\nu\rho} \delta_{\mu\sigma} + 2 \delta_{\mu\nu} \delta_{\sigma\rho} - 2\Delta \delta_{\mu\sigma} \delta_{\nu\rho} - 2 \delta_{\nu\sigma} \delta_{\mu\rho} - 2 \delta_{\mu\nu} \delta_{\rho\sigma} \\
&\quad - 2\Delta \delta_{\nu\rho} \delta_{\mu\sigma} - 2 \delta_{\mu\rho} \delta_{\nu\sigma} - 2 \delta_{\nu\mu} \delta_{\sigma\rho} + 2\Delta \delta_{\nu\sigma} \delta_{\mu\rho} + 2 \delta_{\mu\sigma} \delta_{\nu\rho} + 2 \delta_{\nu\mu} \delta_{\rho\sigma} \Big) \\
&\quad + \frac{4(\Delta+1)}{x^{2\Delta+4}} \Big( -\Delta \delta_{\nu\sigma} x_\mu x_\rho - \delta_{\nu\rho} x_\sigma x_\mu - \delta_{\sigma\rho} x_\nu x_\mu - \delta_{\mu\nu} x_\sigma x_\rho - \delta_{\mu\sigma} x_\nu x_\rho - \delta_{\mu\rho} x_\nu x_\sigma \\
&\quad + \Delta \delta_{\nu\rho} x_\mu x_\sigma + \delta_{\nu\sigma} x_\rho x_\mu + \delta_{\rho\sigma} x_\nu x_\mu + \delta_{\mu\nu} x_\rho x_\sigma + \delta_{\mu\rho} x_\nu x_\sigma + \delta_{\mu\sigma} x_\nu x_\rho \\
&\quad + \Delta \delta_{\mu\sigma} x_\nu x_\rho + \delta_{\mu\rho} x_\sigma x_\nu + \delta_{\sigma\rho} x_\mu x_\nu + \delta_{\nu\mu} x_\sigma x_\rho + \delta_{\nu\sigma} x_\mu x_\rho + \delta_{\nu\rho} x_\mu x_\sigma \\
&\quad - \Delta \delta_{\mu\rho} x_\nu x_\sigma - \delta_{\mu\sigma} x_\rho x_\nu - \delta_{\rho\sigma} x_\mu x_\nu - \delta_{\nu\mu} x_\rho x_\sigma - \delta_{\nu\rho} x_\mu x_\sigma - \delta_{\nu\sigma} x_\mu x_\rho \Big) \\
&= \frac{4(\Delta - 1)}{x^{2\Delta+2}} (\delta_{\mu\rho} \delta_{\nu\sigma} - \delta_{\mu\sigma} \delta_{\nu\rho}) - \frac{4(\Delta^2 - 1)}{x^{2\Delta+4}} (\delta_{\mu\rho} x_\nu x_\sigma + \delta_{\nu\sigma} x_\mu x_\rho - \delta_{\mu\sigma} x_\nu x_\rho - \delta_{\nu\rho} x_\mu x_\sigma)
\end{aligned}
$$
This is the physical space projection of \eqref{eq:full_correlator_generic} and clearly it vanishes at $\Delta=1$. This vanishing is related to the fact that the unitarity bound obtained from the level-1 2-form descendant of a vector primary is exactly $\Delta \ge 1$.\\\\
2. If a level-1 two-form descendant of a vector primary is itself also a primary, then the vector has scaling dimension $\Delta=1$\\
Proof: To determine the condition for the level-1 2-form descendant $F_{\mu\nu}$ to be a conformal primary, we need to demand that it is annihilated by the Special Conformal Transformation (SCT) generator $K_\lambda$ at the origin. That is, we must solve for the condition where $K_\lambda F_{\mu\nu}(0) = 0$
Using the conformal algebra
\begin{align}
    [K_\lambda, P_\mu] &= 2(\delta_{\lambda\mu} D - M_{\lambda\mu})\\
    [M_{\mu\nu}, P_\rho] &= \delta_{\nu\rho} P_\mu - \delta_{\mu\rho} P_\nu
\end{align}  we get,

$$\begin{aligned}
K_\lambda F_{\mu\nu}(0) &= K_\lambda \big( \partial_\mu A_\nu(0) - \partial_\nu A_\mu(0) \big) \\
&= [K_\lambda, \partial_\mu] A_\nu(0) - [K_\lambda, \partial_\nu] A_\mu(0) \\
&= 2(\delta_{\lambda\mu} D - M_{\lambda\mu}) A_\nu(0) - 2(\delta_{\lambda\nu} D - M_{\lambda\nu}) A_\mu(0) \\
&= \Big( 2\Delta \delta_{\lambda\mu} A_\nu - 2(\delta_{\mu\nu} A_\lambda - \delta_{\lambda\nu} A_\mu) \Big) - \Big( 2\Delta \delta_{\lambda\nu} A_\mu - 2(\delta_{\nu\mu} A_\lambda - \delta_{\lambda\mu} A_\nu) \Big) \\
&= 2\Delta \delta_{\lambda\mu} A_\nu - 2\delta_{\mu\nu} A_\lambda + 2\delta_{\lambda\nu} A_\mu - 2\Delta \delta_{\lambda\nu} A_\mu + 2\delta_{\nu\mu} A_\lambda - 2\delta_{\lambda\mu} A_\nu \\
&= 2(\Delta - 1) \delta_{\lambda\mu} A_\nu - 2(\Delta - 1) \delta_{\lambda\nu} A_\mu \\
&= 2(\Delta - 1) (\delta_{\lambda\mu} A_\nu - \delta_{\lambda\nu} A_\mu)
\end{aligned}$$
Therefore, 
\begin{equation}
    K_\lambda F_{\mu\nu}(0) = 0 \quad \Rightarrow\quad \Delta = 1
\end{equation}

The point of these two statements in our context is that $F$ being a descendant forces the two-point function to have an overall factor of $\Delta-1$ whereas $F$ being a primary forces $\Delta=1$ and together these two imply that the two point correlator vanishes. But here too, to decouple the primary $A_\mu$ one can rescale the $F_{\mu \nu}$ correlators by $1-\Delta$ and obtain

$$\frac{1}{\Delta-1} \langle F_{\mu\nu}(x) F_{\rho\sigma}(0) \rangle = \frac{2}{x^{2\Delta+2}} \Big[ (1-\Delta) (\delta_{\mu\rho} \delta_{\nu\sigma} - \delta_{\mu\sigma} \delta_{\nu\rho}) + (\Delta+1) (I_{\mu\rho} I_{\nu\sigma} - I_{\mu\sigma} I_{\nu\rho}) \Big]$$
which collapses to Maxwell upto a numerical factor after taking the limit $\Delta \to 1$. This is the mechanism that we use to get Maxwell out of a primary $A$ even though the Maxwell connection is not a conformal primary \cite{El-Showk:2011xbs}. Naively it might be a bit surprising to see Maxwell theory emerge out of some ordinary vector CFT, but conformal invariance fixes the form of the two point correlators of a $2$-form primary: so one can anticipate getting at least the two point correlator of Maxwell theory. Also note that the mechanism described above is similar to how $\langle \Box \phi \Box \phi\rangle$ correlator for some \emph{free} scalar $\phi$ vanishes, but if we wish to think of the $\Box \phi$ as not being derived from the free scalar but a GFF which doesn't obey an EOM, so that we might want to recover the structure of the two point function, then we can rescale by exactly that expression that appears in unitarity bounds (in this case $\Delta-(d-2/2)$) to get $\langle \Box \phi \Box \phi \rangle \sim 1/|x|^{2\Delta+4}$.

\subsubsection{Uplift of field strength correlators to embedding space}\label{DirectMaxwell}\label{FFproject}

$d=4$ pure Maxwell theory is a CFT where $F_{\mu \nu}(x)$ is a primary operator whose 2-point correlator is  determined by conformal invariance to be 
\begin{align}\label{MaxwellFF4D}
\langle F_{\mu\nu}(x_1) F_{\lambda\sigma}(x_2) \rangle = \frac{C}{(x_{12}^2)^2} \big( I_{\mu\lambda}I_{\nu\sigma} - I_{\mu\sigma}I_{\nu\lambda} \big),
\end{align}
where $x_{12} = x_1 - x_2$, $C$ is a normalization constant which is equal to $\frac{1}{\pi^2}$ in Dolan-Osborn conventions \cite{Dolan:2000ut} but we will instead temporarily set to $1$ for simplicity in this section. The inversion tensor is defined as 
\begin{equation}
    I_{\mu\nu} := \delta_{\mu\nu} - 2\frac{x_{12\mu} x_{12\nu}}{x_{12}^2}.
\end{equation}
For an index-free representation, we contract the free indices with auxiliary polarization vectors. Since $F_{\mu\nu}$ is an anti-symmetric rank-2 tensor, we introduce two independent polarization vectors at each point: $z_1$ and $z_2$ at $x_1$, and $z_3$ and $z_4$ at $x_2$. Contracting these vectors with the correlator yields the physical index-free polynomial:
\begin{align}
z_1^\mu z_2^\nu z_3^\lambda z_4^\sigma (I_{\mu\lambda}I_{\nu\sigma} - I_{\mu\sigma}I_{\nu\lambda}) = (z_1 \cdot I \cdot z_3)(z_2 \cdot I \cdot z_4) - (z_1 \cdot I \cdot z_4)(z_2 \cdot I \cdot z_3). \label{intm}
\end{align}
Let us explicitly evaluate the building block $(z_A \cdot I \cdot z_B)$ by expanding the inversion tensor:
\begin{align}
z_A \cdot I \cdot z_B &= z_A^\mu \left( \delta_{\mu\nu} - 2\frac{x_{12\mu} x_{12\nu}}{x_{12}^2} \right) z_B^\nu \notag \\
&= (z_A \cdot z_B) - 2 \frac{(z_A \cdot x_{12})(z_B \cdot x_{12})}{x_{12}^2}. \label{eq:physical_I_contraction}
\end{align}
We now lift this structure to the embedding space $\mathbb{R}^{d+2}$. In the standard embedding formalism as described in \cite{Costa:2011mg}, primary operators are described using a single polarization vector $Z_i$ per point $X_i$. However, to accommodate the anti-symmetric structure of $F_{MN}$, we generalize the standard tensor structure $H_{ij}$ to allow for two arbitrary polarizations $Z_A$ and $Z_B$ originating from different points $X_1$ and $X_2$:
\begin{align}
H(Z_A, Z_B ; X_1, X_2) := -2 \big[ (Z_A \cdot Z_B)(X_1 \cdot X_2) - (Z_A \cdot X_2)(Z_B \cdot X_1) \big].
\end{align}

For $Z_A=Z_1,Z_B=Z_2$ this is the same as $H_{12}$ in \eqref{Hij}. It is important to note that previously $X_i\cdot Z_i=0$ was required by transversality, but since there are two polarization vectors per point this does not hold. Keeping this in mind we can project down the above tensor structure with an overall factor of $-2 X_1 \cdot X_2$ to get \begin{align}\label{eq:intermediatereally}
\frac{H(Z_A, Z_B ; X_1, X_2)}{-2(X_1 \cdot X_2)} &= (Z_A \cdot Z_B) - \frac{(Z_A \cdot X_2)(Z_B \cdot X_1)}{X_1 \cdot X_2} \notag \\
&\longrightarrow (z_A \cdot z_B) - \frac{(-z_A \cdot x_{12})(z_B \cdot x_{12})}{-\frac{1}{2}x_{12}^2} \notag \\
&= (z_A \cdot z_B) - 2 \frac{(z_A \cdot x_{12})(z_B \cdot x_{12})}{x_{12}^2}.
\end{align}
In the above the projection rules from \eqref{projections} have been used. Alternatively one can go the other way round: to project the embedding tensor to the physical space one can use the pullback

\begin{align}
    H_{M_1 M_2} \vert_{PS} := h_{\mu\nu}(x_1, x_2) &= \frac{\partial X_1^M}{\partial x_1^\mu} \frac{\partial X_2^N}{\partial x_2^\nu} H_{MN}(X_1, X_2) \notag \\
    &= \frac{\partial X_1^M}{\partial x_1^\mu} \frac{\partial X_2^N}{\partial x_2^\nu} \Big( -2(X_1 \cdot X_2)\delta_{MN} + 2 X_{2M} X_{1N} \Big) \notag \\
    &= -2\left(-\frac{1}{2} x_{12}^2\right) \left( \frac{\partial X_1^M}{\partial x_1^\mu} \frac{\partial X_2^N}{\partial x_2^\nu} \delta_{MN} \right) + 2 \left( \frac{\partial X_1^M}{\partial x_1^\mu} X_{2M} \right) \left( \frac{\partial X_2^N}{\partial x_2^\nu} X_{1N} \right) \notag \\
    &= x_{12}^2 (\partial_\mu X_1 \cdot \partial_\nu X_2) + 2 (\partial_\mu X_1 \cdot X_2) (\partial_\nu X_2 \cdot X_1) \notag \\
    &= x_{12}^2 (\delta_{\mu\nu}) + 2 (-x_{12\mu}) (x_{12\nu}) \notag \\
    &= x_{12}^2 \delta_{\mu\nu} - 2 x_{12\mu} x_{12\nu} \notag \\
    &= x_{12}^2 \left( \delta_{\mu\nu} - 2 \frac{x_{12\mu} x_{12\nu}}{x_{12}^2} \right) \notag \\
    &=x_{12}^2 I_{\mu \nu}
\end{align}
Either way this reproduces the physical inversion tensor contraction $z_A \cdot I \cdot z_B$ from Eq. \eqref{eq:physical_I_contraction}. Using \eqref{eq:intermediatereally} in \eqref{intm} and \eqref{MaxwellFF4D} and $1/(x_{12}^2)^2 \to 1/(-2X_1 \cdot X_2)^2$, we obtain the final embedding space expression (omitting numerical factors) \footnote{For a careful consideration of these numerical factors see Appendix \ref{sec:phiphifromtruncnatedansatz}-\ref{sec:largeN}}:
\begin{align}
\langle F(X_1, Z_1, Z_2) F(X_2, Z_3, Z_4) \rangle &\sim \frac{1}{(X_1 \cdot X_2)^2} \left[ \frac{H(Z_1, Z_3 ; X_1, X_2)}{X_1 \cdot X_2} \frac{H(Z_2, Z_4 ; X_1, X_2)}{X_1 \cdot X_2} - (3 \leftrightarrow 4) \right] \notag \\
&= \frac{H(Z_1, Z_3 ; X_1, X_2) H(Z_2, Z_4 ; X_1, X_2) - H(Z_1, Z_4 ; X_1, X_2) H(Z_2, Z_3 ; X_1, X_2)}{(X_1 \cdot X_2)^4}\label{Maxwellexpectedans}
\end{align}
This is just $d=4$ pure Maxwell CFT field strength $2$-point correlator written using embedding space polynomials. We will now compute the $\langle FF\rangle$ correlator using NN methods and compare with this result.

\subsubsection{Maxwell from non-primary $A$}\label{MaxwellfromnonprimaryA}
In this section, we outline a relatively straightforward way to obtain the Maxwell theory correlators with details about this particular construction spread across the other appendices. The idea is to start with a non-primary $A_M(X)$ in the first place (like in actual Maxwell theory), such that we get the correct $\langle A(X_1,Z_1) A(X_2,Z_2)\rangle$ correlators at least in some gauge. Then the field strength correlators can be computed either by taking appropriate derivatives of this result or a direct computation using NN expressions of $F(X,Z)$ obtained from this new architecture. A non-primary can be obtained simply by breaking the transversality condition $X^M A_M=0$. To see what happens if we break transversality, we assume $X^M A_M \ne 0$ and check how the projected field $A_\mu= \frac{\partial X^M}{\partial X^\mu}A_M$ transforms. A conformal transformation $x \to x'$ on the physical spacetime corresponds to a linear Lorentz transformation $X \to \Lambda X$ in the embedding space. However, $\Lambda X(x)$ will generally not lie on our chosen Poincaré section (where $X^+ = 1$). To project it back, we must rescale it by a conformal scale factor $\Omega(x)$ which multiplies the metric under a conformal transformation \cite{simmonsduffin2018ph229,Rychkov:2016cft} i.e. $\delta_{\mu \nu}\frac{\partial x'^\mu}{\partial x^\alpha}\frac{\partial x'^\nu}{\partial x^\beta}=\Omega(x)^2\delta_{\alpha \beta}$. So we can write 
\begin{equation}
   X^M(x') = \Omega(x)^{-1} \Lambda^M_{\;\;N} X^N(x) 
\end{equation}
Because $A_M(X)$ is an embedding space tensor, it transforms under the Lorentz group as $A'_M(X') = \Lambda_M^{\;\;N} A_N(X)$. Using the homogeneity property, we can evaluate the transformed tensor on the new section
$$A'_M(X(x')) = A'_M(\Omega(x)^{-1} \Lambda X(x)) = \Omega(x)^{\Delta} \Lambda_M^{\;\;N} A_N(X(x))$$
where $\Delta$ is the scaling dimension of $A_\mu(x)$. To find the transformed physical field $A'_\mu(x')$, we  pullback to the physical space and find 
\begin{align}
A'_\mu(x') &= \frac{\partial X^M(x')}{\partial x'^\mu} A'_M(X(x')) \nonumber \\
&= \frac{\partial x^\nu}{\partial x'^\mu} \frac{\partial}{\partial x^\nu} \left[ \Omega(x)^{-1} \Lambda^M_{\;\;K} X^K(x) \right] A'_M(X(x')) \nonumber \\
&= \frac{\partial x^\nu}{\partial x'^\mu} \left[ -\Omega(x)^{-2} (\partial_\nu \Omega(x)) \Lambda^M_{\;\;K} X^K(x) + \Omega(x)^{-1} \Lambda^M_{\;\;K} \partial_\nu X^K(x) \right] \Omega(x)^{\Delta} \Lambda_M^{\;\;N} A_N(X) \nonumber \\
&= \Omega(x)^{\Delta-1} \frac{\partial x^\nu}{\partial x'^\mu} \left[ \partial_\nu X^N A_N(X) - \Omega(x)^{-1} (\partial_\nu \Omega(x)) X^N A_N(X) \right] \nonumber \\
&= \Omega(x)^{\Delta-1} \frac{\partial x^\nu}{\partial x'^\mu} \left[ A_\nu(x) - \Omega(x)^{-1}\partial_\nu \Omega(x) \left( X^N A_N(X) \right) \right] \label{anomtransf}
\end{align}
That is $A_\mu(x)$ transforms as a connection with respect to $\Omega(x)$ with the non-transversality factor $X^N A_N$ acting as the ``charge''. Note that since $\Omega(x)$ is $x$ dependent only for special conformal transformations \cite{simmonsduffin2018ph229}, transversality $X^N A_N=0$ is a requirement to ensure that the physical field $A_\mu(x)$ transforms tensorially even under special conformal transformations. For our purposes here, since we know the Maxwell field is not a conformal primary, we deliberately construct an architecture that breaks transversality in order to get the correct $\langle A(X_1,Z_1) A(X_2,Z_2)\rangle$ in some gauge. Namely, we consider the \emph {truncated ansatz} obtained by removing the 2nd term within brackets of \eqref{spin1ansatzMaxwell}, imposing $\Theta=\tilde{\Theta}$ \footnote{Omitting this constraint introduces extraneous algebraic complexity while leaving the underlying physical content unchanged.}, and $\Delta=1$:
\begin{equation}\label{truncatedansatz}
    A(X,Z)=(\Theta \cdot X)^{-1} (\eta \cdot Z)
\end{equation}
This can also be viewed as an implementation of the removal of pure gauge terms in the Maxwell propagator:
\begin{align}
    \langle A(X_1,Z_A) A(X_2,Z_B) \rangle=(X_1 \cdot X_2)^{-1} (Z_A \cdot Z_B)
\end{align}
where we used results from Appendix \ref{scalar2pt} i.e. we recover the Feynman gauge Maxwell propagator. The $1$-point function vanishes due to the vanishing $1$-point function of $\eta$.
The architecture \eqref{truncatedansatz} manifestly breaks transversality, but note that under correlator brackets it restores the same. Namely,
\begin{equation}
    \langle X^M A_M\rangle=\langle A(X,X) \rangle=0
\end{equation} where the last step again follows due to the vanishing $1$-point function of $\eta$. This becomes crucial when we compute the field strength because it is a genuine primary in $d=4$ Maxwell theory. The field strength is 
\begin{align}
F(X, Z_1, Z_2) &= \left(Z_1 \cdot \frac{\partial}{\partial X}\right) A(X,Z_2) - \left(Z_2 \cdot \frac{\partial}{\partial X}\right) A(X,Z_1) \notag \\
&= \left(Z_1 \cdot \frac{\partial}{\partial X}\right) \big[ (\Theta \cdot X)^{-1} (\eta \cdot Z_2) \big] - \left(Z_2 \cdot \frac{\partial}{\partial X}\right) \big[ (\Theta \cdot X)^{-1} (\eta \cdot Z_1) \big] \notag\\
&= \big[ -(\Theta \cdot X)^{-2} (\Theta \cdot Z_1) (\eta \cdot Z_2) \big] - \big[ -(\Theta \cdot X)^{-2} (\Theta \cdot Z_2) (\eta \cdot Z_1) \big] \notag \\ 
&= -(\Theta \cdot X)^{-2} \big[ (\Theta \cdot Z_1)(\eta \cdot Z_2) - (\Theta \cdot Z_2)(\eta \cdot Z_1) \big]\label{truncatedF}
\end{align}
Clearly it does not satisfy transversality:
\begin{equation}
    F(X,X,Z_2)= -(\Theta \cdot X)^{-2} \big[ (\Theta \cdot X)(\eta \cdot Z_2) - (\Theta \cdot Z_2)(\eta \cdot X) \big] \ne 0
\end{equation}
and similarly $F(X,Z_1,X) \ne 0$. But under the correlator brackets, transversality is recovered again because $ \langle \eta \rangle=0$.
This should also be obviously true for the $2$-point function $\langle F(X_1,Z_1,Z_2) F(X_2,Z_3,Z_4)\rangle$ for consistency. These are calculated in Appendix \ref{FFfromAA} and Appendix \ref{FFfromFF}
using the two different methods mentioned previously:
\begin{align}
    \langle F(X_1,Z_1,& Z_2) F(X_2,Z_3,Z_4) \rangle \nonumber\\&=2(X_1 \cdot X_2)^{-2} \big[ (Z_1 \cdot Z_3)(Z_2 \cdot Z_4) - (Z_1 \cdot Z_4)(Z_2 \cdot Z_3) \big] \notag \\
    &\quad - 2(X_1 \cdot X_2)^{-3} \big[ (Z_2 \cdot Z_4)(X_2 \cdot Z_1)(X_1 \cdot Z_3) - (Z_2 \cdot Z_3)(X_2 \cdot Z_1)(X_1 \cdot Z_4) \notag \\
    &\qquad \qquad \qquad \quad - (Z_1 \cdot Z_4)(X_2 \cdot Z_2)(X_1 \cdot Z_3) + (Z_1 \cdot Z_3)(X_2 \cdot Z_2)(X_1 \cdot Z_4) \big]
\end{align}
which implies 
$$\begin{aligned}
\langle F(X_1,X_1,Z_2)& F(X_2,Z_3,Z_4) \rangle \\&= 2(X_1 \cdot X_2)^{-2} \big[ (X_1 \cdot Z_3)(Z_2 \cdot Z_4) - (X_1 \cdot Z_4)(Z_2 \cdot Z_3) \big] \\
&\quad - 2(X_1 \cdot X_2)^{-3} \big[ (Z_2 \cdot Z_4)(X_2 \cdot X_1)(X_1 \cdot Z_3) - (Z_2 \cdot Z_3)(X_2 \cdot X_1)(X_1 \cdot Z_4) \\
&\qquad\qquad\qquad\qquad\quad - (X_1 \cdot Z_4)(X_2 \cdot Z_2)(X_1 \cdot Z_3) + (X_1 \cdot Z_3)(X_2 \cdot Z_2)(X_1 \cdot Z_4) \big] \\
&= 2(X_1 \cdot X_2)^{-2} \big[ (X_1 \cdot Z_3)(Z_2 \cdot Z_4) - (X_1 \cdot Z_4)(Z_2 \cdot Z_3) \big] \\
&\quad - 2(X_1 \cdot X_2)^{-3} \big[ (Z_2 \cdot Z_4)(X_1 \cdot X_2)(X_1 \cdot Z_3) - (Z_2 \cdot Z_3)(X_1 \cdot X_2)(X_1 \cdot Z_4) \big] \\
&= 2(X_1 \cdot X_2)^{-2} \big[ (X_1 \cdot Z_3)(Z_2 \cdot Z_4) - (X_1 \cdot Z_4)(Z_2 \cdot Z_3) \big] \\
&\quad - 2(X_1 \cdot X_2)^{-2} \big[ (X_1 \cdot Z_3)(Z_2 \cdot Z_4) - (X_1 \cdot Z_4)(Z_2 \cdot Z_3) \big] \\
&= 0
\end{aligned}$$
That transversality is obeyed for $1$-point and $2$-point functions of the field strength is enough because the higher point functions will be in terms of these by Gaussianity which will be implemented by a large-$N$ technique discussed in details in \ref{sec:largeN}. The anomalous term in the transformation \eqref{anomtransf} (and similar term for $F(X,Z)$) vanishes precisely due to the choice of the architecture and probability density of NN parameters, making the generating functional invariant under conformal transformations even if the fields are not. So, in particular, for the purposes of calculating correlators, $F(X,Z)$ essentially behaves like a primary as in pure Maxwell.
One might still insist on some expression for $F$ which is primary even before taking the correlator brackets. In particular, one can compute the correlators of the scalar primary $\Phi:=\frac{1}{4}F_{MN}^2$ and find that they don't match those of Maxwell theory.  Perhaps surprisingly, the two-point function $\langle \Phi(X_1) \Phi(X_2)\rangle$ vanishes for the truncated ansatz (See \ref{sec:phiphifromtruncnatedansatz}). One might think that these deviations are due to the fact that $F(X,Z)$ does not satisfy transversality in these theories unlike the genuine Maxwell theory. Then a field strength architecture satisfying transversality can be constructed by appropriately ``subtracting'' the non-zero $X^MF_{MN}$ as
\begin{equation}
    F^\perp_{MN} = F_{MN} + \bar{X}_M V_N - \bar{X}_N V_M
\end{equation}
where $V_N := X^M F_{MN}$ and $\bar{X}$ is any PNC vector satisfying $X \cdot \bar{X}=-1$. This $F^\perp_{MN}$ has identical correlators as $F_{MN}$ and is further discussed in \ref{sec:phiphifromprimaryF}.  This gives us a non-vanishing scalar primary 2-point correlator which matches Maxwell theory $\langle \Phi(X_1) \Phi(X_2) \rangle$ upto a constant numerical factor \footnote{A simple rescaling doesn't fix this simply because to get the correct $\langle FF\rangle$ correlator normalization, we already had to do a rescaling: there's simply no freedom to rescale to get the correct numerical coefficient for $\langle \Phi \Phi \rangle$ correlator in this approach.}. Thus there is no guarantee that this ``primary'' $F$ would yield consistent higher point correlations, and for the purpose of construction of free Maxwell theory, it is natural to consider a large-$N$ ensemble of the single channel fields as discussed in Appendix \ref{sec:largeN}.

As a short aside, for $\Delta=1$ using the expression of $F_{MN}(X)$ computed from the above ansatz \eqref{truncatedansatz} we get,
\begin{equation}\label{eq:EOMnonidentical}
    \partial^M F_{MN}=  2(\Theta \cdot X)^{-3} [\Theta^2 \eta_N - (\Theta \cdot \eta) \Theta_N] 
\end{equation}
which vanishes after taking the expectation value w.r.t. $\eta$. Projecting to physical spacetime, we obtain the $1$-point Schwinger-Dyson equation in Maxwell theory.
\begin{equation}\label{eq:SDone}
    \langle \partial^\mu F_{\mu \nu} \rangle=0 
\end{equation}
There are no contact terms on the RHS because there is no time ordering in the definition of $\langle \cdot \rangle$ as expected in an Euclidean theory.

\subsubsection{$\langle FF\rangle$ from $\langle AA \rangle$ correlator}\label{FFfromAA}
In this section, we find the $\langle FF\rangle$ correlator using $\langle AA\rangle$. Recall that $\langle F(X_1, Z_1, Z_2) F(X_2, Z_3, Z_4) \rangle$ can be written in terms of $\langle AA \rangle$ correlators as \eqref{FFtoAA} 
which we reproduce here for clarity:

\begin{align}
\langle F(X_1, Z_1, Z_2) F(X_2, Z_3, Z_4) \rangle 
&= \left(Z_1 \cdot \frac{\partial}{\partial X_1}\right) \left(Z_3 \cdot \frac{\partial}{\partial X_2}\right) \langle A(X_1, Z_2) A(X_2, Z_4) \rangle \notag \\
&\quad - \left(Z_1 \cdot \frac{\partial}{\partial X_1}\right) \left(Z_4 \cdot \frac{\partial}{\partial X_2}\right) \langle A(X_1, Z_2) A(X_2, Z_3) \rangle \notag \\
&\quad - \left(Z_2 \cdot \frac{\partial}{\partial X_1}\right) \left(Z_3 \cdot \frac{\partial}{\partial X_2}\right) \langle A(X_1, Z_1) A(X_2, Z_4) \rangle \notag \\
&\quad + \left(Z_2 \cdot \frac{\partial}{\partial X_1}\right) \left(Z_4 \cdot \frac{\partial}{\partial X_2}\right) \langle A(X_1, Z_1) A(X_2, Z_3) \rangle \label{FFintermsofAA}
\end{align}
The two-point function $\langle A(X_1, Z_A) A(X_2,Z_B) \rangle$ is
\begin{align}
G(Z_A, Z_B ; X_1, X_2) = 2 (X_1 \cdot X_2)^{-\Delta-1} \big[ (X_1 \cdot X_2)(Z_A \cdot Z_B) - (X_1 \cdot Z_B)(X_2 \cdot Z_A) \big]
\end{align}
The tensor structure would be $H_{12}$ for $A=1,B=2$, but in general it can be different as seen from \eqref{FFintermsofAA} so that it is actually $H(Z_A,Z_B;X_1,X_2)$ from \eqref{newtensorstructure}.
Hence, we need to evaluate terms of the form 
\begin{equation}
D_{CD} G_{AB} := \left(Z_C \cdot \frac{\partial}{\partial X_1}\right) \left(Z_D \cdot \frac{\partial}{\partial X_2}\right) G(Z_A, Z_B ; X_1, X_2)
\end{equation}
since the equation \eqref{FFintermsofAA} can be written as
\begin{align}
\langle F(X_1, Z_1, Z_2) F(X_2, Z_3, Z_4) \rangle &= D_{13} G(Z_2, Z_4 ; X_1, X_2) - D_{14} G(Z_2, Z_3 ; X_1, X_2) \notag \\
&\quad - D_{23} G(Z_1, Z_4 ; X_1, X_2) + D_{24} G(Z_1, Z_3 ; X_1, X_2)
\end{align}
Noting that $\frac{\partial}{\partial X_1} (X_1 \cdot X_2) = X_2$ and $\frac{\partial}{\partial X_2} (X_1 \cdot X_2) = X_1$, the generic double-derivative evaluated as follows:
$$
\begin{aligned}
\left(Z_D \cdot \frac{\partial}{\partial X_2}\right) G &= -2\Delta (X_1 \cdot X_2)^{-\Delta-1} (X_1 \cdot Z_D) (Z_A \cdot Z_B) \notag \\
&\quad - 2 (X_1 \cdot Z_B) \Big[ -(\Delta+1) (X_1 \cdot X_2)^{-\Delta-2} (X_1 \cdot Z_D) (X_2 \cdot Z_A) + (X_1 \cdot X_2)^{-\Delta-1} (Z_A \cdot Z_D) \Big] \notag \\
&= -2\Delta (X_1 \cdot X_2)^{-\Delta-1} (Z_A \cdot Z_B)(X_1 \cdot Z_D) - 2(X_1 \cdot X_2)^{-\Delta-1} (X_1 \cdot Z_B)(Z_A \cdot Z_D) \notag \\
&\quad + 2(\Delta+1) (X_1 \cdot X_2)^{-\Delta-2} (X_2 \cdot Z_A)(X_1 \cdot Z_B)(X_1 \cdot Z_D) \notag \\
D_{CD} G_{AB} &= \left(Z_C \cdot \frac{\partial}{\partial X_1}\right) \left[ \left(Z_D \cdot \frac{\partial}{\partial X_2}\right) G \right] \notag \\
&= -2\Delta (Z_A \cdot Z_B) \Big[ -(\Delta+1) (X_1 \cdot X_2)^{-\Delta-2} (X_2 \cdot Z_C) (X_1 \cdot Z_D) + (X_1 \cdot X_2)^{-\Delta-1} (Z_C \cdot Z_D) \Big] \notag \\
&\quad - 2(Z_A \cdot Z_D) \Big[ -(\Delta+1) (X_1 \cdot X_2)^{-\Delta-2} (X_2 \cdot Z_C) (X_1 \cdot Z_B) + (X_1 \cdot X_2)^{-\Delta-1} (Z_B \cdot Z_C) \Big] \notag \\
&\quad + 2(\Delta+1)(X_2 \cdot Z_A) \Big[ -(\Delta+2) (X_1 \cdot X_2)^{-\Delta-3} (X_2 \cdot Z_C) (X_1 \cdot Z_B)(X_1 \cdot Z_D) \\
&\qquad \qquad \qquad \qquad + (X_1 \cdot X_2)^{-\Delta-2} (Z_B \cdot Z_C) (X_1 \cdot Z_D) + (X_1 \cdot X_2)^{-\Delta-2} (X_1 \cdot Z_B) (Z_C \cdot Z_D) \Big] \\
&= (X_1 \cdot X_2)^{-\Delta-1} \big[ -2\Delta (Z_A \cdot Z_B)(Z_C \cdot Z_D) - 2 (Z_A \cdot Z_D)(Z_B \cdot Z_C) \big] \\
&\quad + 2(\Delta+1)(X_1 \cdot X_2)^{-\Delta-2} \big[ \Delta (Z_A \cdot Z_B)(X_2 \cdot Z_C)(X_1 \cdot Z_D) + (Z_C \cdot Z_D)(X_1 \cdot Z_B)(X_2 \cdot Z_A) \\
&\qquad \qquad \qquad \qquad \quad + (Z_B \cdot Z_C)(X_1 \cdot Z_D)(X_2 \cdot Z_A) + (Z_A \cdot Z_D)(X_1 \cdot Z_B)(X_2 \cdot Z_C) \big] \\
&\quad - 2(\Delta+1)(\Delta+2) (X_1 \cdot X_2)^{-\Delta-3} \big[ (X_1 \cdot Z_B)(X_1 \cdot Z_D)(X_2 \cdot Z_A)(X_2 \cdot Z_C) \big]
\end{aligned}
$$
Let us explicitly sum these four permutations, grouping them by powers of $(X_1 \cdot X_2)$. \\
\textbf{The $(X_1 \cdot X_2)^{-\Delta-3}$ terms:}
\begin{align}
D_{13}G_{24} &\to -2(\Delta+1)(\Delta+2) \big[ (X_1 \cdot Z_4)(X_1 \cdot Z_3)(X_2 \cdot Z_2)(X_2 \cdot Z_1) \big] \notag \\
-D_{14}G_{23} &\to +2(\Delta+1)(\Delta+2) \big[ (X_1 \cdot Z_3)(X_1 \cdot Z_4)(X_2 \cdot Z_2)(X_2 \cdot Z_1) \big] \notag \\
-D_{23}G_{14} &\to +2(\Delta+1)(\Delta+2) \big[ (X_1 \cdot Z_4)(X_1 \cdot Z_3)(X_2 \cdot Z_1)(X_2 \cdot Z_2) \big] \notag \\
+D_{24}G_{13} &\to -2(\Delta+1)(\Delta+2) \big[ (X_1 \cdot Z_3)(X_1 \cdot Z_4)(X_2 \cdot Z_1)(X_2 \cdot Z_2) \big]
\end{align}
Because scalar products commute, $(X_1 \cdot Z_4)(X_1 \cdot Z_3) = (X_1 \cdot Z_3)(X_1 \cdot Z_4)$ and $(X_2 \cdot Z_2)(X_2 \cdot Z_1) = (X_2 \cdot Z_1)(X_2 \cdot Z_2)$. All four terms are mathematically identical. Factoring out the common term, the sum is  proportional to $(-1 + 1 + 1 - 1) = 0$. That there's no term proportional to $(X_1 \cdot X_2)^{-\Delta-3}$ can be anticipated from the fact that the antisymmetric field strength operator annihilates the manifestly symmetric tensor generated by double derivatives acting purely on the factor of $1/(X_1 \cdot X_2)^{-\Delta}$ in $G_{AB}$.\\\\
\textbf{The $(X_1 \cdot X_2)^{-\Delta-1}$ terms:}
\begin{align}
D_{13}G_{24} &\to -2\Delta(Z_2 \cdot Z_4)(Z_1 \cdot Z_3) - 2(Z_2 \cdot Z_3)(Z_4 \cdot Z_1) \notag \\
-D_{14}G_{23} &\to 2\Delta(Z_2 \cdot Z_3)(Z_1 \cdot Z_4) + 2(Z_2 \cdot Z_4)(Z_3 \cdot Z_1) \notag \\
-D_{23}G_{14} &\to 2\Delta(Z_1 \cdot Z_4)(Z_2 \cdot Z_3) + 2(Z_1 \cdot Z_3)(Z_4 \cdot Z_2) \notag \\
+D_{24}G_{13} &\to -2\Delta(Z_1 \cdot Z_3)(Z_2 \cdot Z_4) - 2(Z_1 \cdot Z_4)(Z_3 \cdot Z_2)
\end{align}
Summing the coefficients of $(Z_1 \cdot Z_3)(Z_2 \cdot Z_4)$, we get $(-2\Delta + 2 + 2 - 2\Delta) = -4(\Delta - 1)$. 
Summing the coefficients of $(Z_1 \cdot Z_4)(Z_2 \cdot Z_3)$, we get $(-2 + 2\Delta + 2\Delta - 2) = +4(\Delta - 1)$.
The total for this group is:
\begin{align}
-4(\Delta-1) (X_1 \cdot X_2)^{-\Delta-1} \big[ (Z_1 \cdot Z_3)(Z_2 \cdot Z_4) - (Z_1 \cdot Z_4)(Z_2 \cdot Z_3) \big]
\end{align}
\textbf{The $(X_1 \cdot X_2)^{-\Delta-2}$ terms:}\\
Let us define $K_{\text{diff}}$ as in \eqref{KDiff}:
\begin{align}
K_{\text{diff}} &:= (Z_1 \cdot Z_3)(X_1 \cdot Z_4)(X_2 \cdot Z_2) + (Z_2 \cdot Z_4)(X_1 \cdot Z_3)(X_2 \cdot Z_1) \notag \\
&\quad - (Z_1 \cdot Z_4)(X_1 \cdot Z_3)(X_2 \cdot Z_2) - (Z_2 \cdot Z_3)(X_1 \cdot Z_4)(X_2 \cdot Z_1) 
\end{align}
Note again that terms like $Z_2 \cdot X_2$ do not vanish identically now because $Z_2$ is the polarization vector associated with $X_1$ and not $X_2$ as in the case of traceless symmetric tensors discussed in rest of the work.
We can group the terms in the correlator which are proportional to $(X_1 \cdot X_2)^{-\Delta-2}$ by the four terms in $K_{\text{diff}}$. For instance, the coefficient of the first term $(Z_1 \cdot Z_3)(X_1 \cdot Z_4)(X_2 \cdot Z_2)$ receives contributions from:
\begin{align}
    D_{13}G_{24}: +2(\Delta+1) \times (1) \notag\\
    -D_{14}G_{23}: -2(\Delta+1) \times (1) \notag\\
    -D_{23}G_{14}: -2(\Delta+1) \times (1)\notag \\
    +D_{24}G_{13}: +2(\Delta+1) \times (\Delta)
\end{align}
Summing these gives $2(\Delta+1)(1 - 1 - 1 + \Delta) = 2(\Delta+1)(\Delta-1)$. This identical coefficient emerges for all four components, allowing us to factor the entire sum compactly:
\begin{align}
+ 2(\Delta-1)(\Delta+1) (X_1 \cdot X_2)^{-\Delta-2} K_{\text{diff}}
\end{align}
Combining all the non-zero contributions we get 
\begin{align}
\langle F(X_1, Z_1, Z_2) F(X_2, Z_3, Z_4) \rangle &= -4(\Delta-1) (X_1 \cdot X_2)^{-\Delta-1} \big[ (Z_1 \cdot Z_3)(Z_2 \cdot Z_4) - (Z_1 \cdot Z_4)(Z_2 \cdot Z_3) \big] \notag \\
&\quad + 2(\Delta-1)(\Delta+1) (X_1 \cdot X_2)^{-\Delta-2} K_{\text{diff}}
\end{align}
which is \eqref{eq:full_correlator_generic}.\\\\
\textbf{$\langle FF\rangle$ from truncated ansatz}\label{FFfromAAtruncated}\\
In this section, we do the above computation  but with the ``truncated ansatz'' \eqref{truncatedansatz} rewritten here for clarity
\begin{equation}
    A(X,Z)=(\Theta \cdot X)^{-1} (\eta \cdot Z)
\end{equation}
which yields the two-point correlator
\begin{align}
    \langle A(X_1,Z_A) A(X_2,Z_B) \rangle=(X_1 \cdot X_2)^{-1} (Z_A \cdot Z_B)
\end{align}
Applying the double derivative $D_{CD} := \left(Z_C \cdot \frac{\partial}{\partial X_1}\right) \left(Z_D \cdot \frac{\partial}{\partial X_2}\right)$ yields:
\begin{align}
\left(Z_D \cdot \frac{\partial}{\partial X_2}\right) \big[ -(X_1 \cdot X_2)^{-1} (Z_A \cdot Z_B) \big] &= (X_1 \cdot X_2)^{-2} (Z_D \cdot X_1)(Z_A \cdot Z_B) \notag \\
\left(Z_C \cdot \frac{\partial}{\partial X_1}\right) \big[ (X_1 \cdot X_2)^{-2} (Z_D \cdot X_1)(Z_A \cdot Z_B) \big] &= -2 (X_1 \cdot X_2)^{-3} (Z_C \cdot X_2)(Z_D \cdot X_1)(Z_A \cdot Z_B) \notag \\
&\quad + (X_1 \cdot X_2)^{-2} (Z_C \cdot Z_D)(Z_A \cdot Z_B)
\end{align}
The $\langle F(X_1, Z_1, Z_2) F(X_2, Z_3, Z_4) \rangle$ correlator becomes
\begin{align}
&\langle F(X_1, Z_1, Z_2) F(X_2, Z_3, Z_4) \rangle \notag \\
&= D_{13}\langle A(X_1,Z_2) A(X_2,Z_4) \rangle  - D_{14}\langle A(X_1,Z_2) A(X_2,Z_3) \rangle - D_{23}\langle A(X_1,Z_1) A(X_2,Z_4) \rangle+ D_{24}\langle A(X_1,Z_1) A(X_2,Z_4) \rangle\notag \\
&= \big[ (X_1 \cdot X_2)^{-2}(Z_1 \cdot Z_3)(Z_2 \cdot Z_4) - 2(X_1 \cdot X_2)^{-3}(X_2 \cdot Z_1)(X_1 \cdot Z_3)(Z_2 \cdot Z_4) \big] \notag \\
&\quad - \big[ (X_1 \cdot X_2)^{-2}(Z_1 \cdot Z_4)(Z_2 \cdot Z_3) - 2(X_1 \cdot X_2)^{-3}(X_2 \cdot Z_1)(X_1 \cdot Z_4)(Z_2 \cdot Z_3) \big] \notag \\
&\quad - \big[ (X_1 \cdot X_2)^{-2}(Z_2 \cdot Z_3)(Z_1 \cdot Z_4) - 2(X_1 \cdot X_2)^{-3}(X_2 \cdot Z_2)(X_1 \cdot Z_3)(Z_1 \cdot Z_4) \big] \notag \\
&\quad + \big[ (X_1 \cdot X_2)^{-2}(Z_2 \cdot Z_4)(Z_1 \cdot Z_3) - 2(X_1 \cdot X_2)^{-3}(X_2 \cdot Z_2)(X_1 \cdot Z_4)(Z_1 \cdot Z_3) \big] \notag \\
&= 2(X_1 \cdot X_2)^{-2} \big[ (Z_1 \cdot Z_3)(Z_2 \cdot Z_4) - (Z_1 \cdot Z_4)(Z_2 \cdot Z_3) \big] \notag \\
&\quad - 2(X_1 \cdot X_2)^{-3} \big[ (Z_2 \cdot Z_4)(X_2 \cdot Z_1)(X_1 \cdot Z_3) - (Z_2 \cdot Z_3)(X_2 \cdot Z_1)(X_1 \cdot Z_4) \notag \\
&\qquad \qquad \qquad \quad - (Z_1 \cdot Z_4)(X_2 \cdot Z_2)(X_1 \cdot Z_3) + (Z_1 \cdot Z_3)(X_2 \cdot Z_2)(X_1 \cdot Z_4) \big]
\label{FFMaxwellintermediate}
\end{align}
\subsubsection{$\langle FF \rangle $ from NN expression for $F$}\label{FFfromFF}
\textbf{Truncated Ansatz}\\
In this subsection, we use the truncated NN ansatz \eqref{truncatedansatz}
\begin{align}
A(X,Z) &= (\tilde{\Theta} \cdot X)^{-1} (\eta \cdot Z)
\end{align}
which gives the field strength \eqref{truncatedF}
\begin{align}
F(X, Z_1, Z_2) 
&= -(\tilde{\Theta} \cdot X)^{-2} \big[ (\tilde{\Theta} \cdot Z_1)(\eta \cdot Z_2) - (\tilde{\Theta} \cdot Z_2)(\eta \cdot Z_1) \big]
\end{align}
to compute the field strength correlator $\langle F(X_1, Z_1, Z_2) F(X_2, Z_3, Z_4) \rangle$. Explicitly,

\begin{align}
     F(X_1, Z_1, Z_2) &= -(\tilde{\Theta} \cdot X_1)^{-2} \big[ (\tilde{\Theta} \cdot Z_1)(\eta \cdot Z_2) - (\tilde{\Theta} \cdot Z_2)(\eta \cdot Z_1) \big] \\
F(X_2, Z_3, Z_4) &= -(\tilde{\Theta} \cdot X_2)^{-2} \big[ (\tilde{\Theta} \cdot Z_3)(\eta \cdot Z_4) - (\tilde{\Theta} \cdot Z_4)(\eta \cdot Z_3) \big] 
\end{align}
Expanding the product of the two field strengths and evaluating the expectation value over $\eta$ using Wick contractions,

\begin{align}
\langle F(X_1, Z_1, Z_2)  F(X_2, Z_3, Z_4) \rangle_\eta &= (\tilde{\Theta} \cdot X_1)^{-2} (\tilde{\Theta} \cdot X_2)^{-2} \Big\{ (\tilde{\Theta} \cdot Z_1)(\tilde{\Theta} \cdot Z_3) \langle (\eta \cdot Z_2)(\eta \cdot Z_4) \rangle_\eta \notag \\
&\qquad \qquad \qquad \qquad \quad - (\tilde{\Theta} \cdot Z_1)(\tilde{\Theta} \cdot Z_4) \langle (\eta \cdot Z_2)(\eta \cdot Z_3) \rangle_\eta \notag \\
&\qquad \qquad \qquad \qquad \quad - (\tilde{\Theta} \cdot Z_2)(\tilde{\Theta} \cdot Z_3) \langle (\eta \cdot Z_1)(\eta \cdot Z_4) \rangle_\eta \notag \\
&\qquad \qquad \qquad \qquad \quad + (\tilde{\Theta} \cdot Z_2)(\tilde{\Theta} \cdot Z_4) \langle (\eta \cdot Z_1)(\eta \cdot Z_3) \rangle_\eta \Big\} \notag \\
&= (\tilde{\Theta} \cdot X_1)^{-2} (\tilde{\Theta} \cdot X_2)^{-2} \Big\{ (\tilde{\Theta} \cdot Z_1)(\tilde{\Theta} \cdot Z_3)(Z_2 \cdot Z_4) \notag \\
&\qquad \qquad \qquad \qquad \quad - (\tilde{\Theta} \cdot Z_1)(\tilde{\Theta} \cdot Z_4)(Z_2 \cdot Z_3) \notag \\
&\qquad \qquad \qquad \qquad \quad - (\tilde{\Theta} \cdot Z_2)(\tilde{\Theta} \cdot Z_3)(Z_1 \cdot Z_4)  \\
&\qquad \qquad \qquad \qquad \quad + (\tilde{\Theta} \cdot Z_2)(\tilde{\Theta} \cdot Z_4)(Z_1 \cdot Z_3) \Big\}
\end{align}

To evaluate the remaining expectation value over the NN parameter $\tilde{\Theta}$, we define the integral $I_{ik}$ for the factors multiplying the $(Z_i \cdot Z_k)$ polarization structures. 

\begin{align}
I_{ik} := \left\langle \frac{(\tilde{\Theta} \cdot Z_i)(\tilde{\Theta} \cdot Z_k)}{(\tilde{\Theta} \cdot X_1)^2 (\tilde{\Theta} \cdot X_2)^2} \right\rangle 
\end{align}
where crucially $i \in \{1,2 \}$ and $k \in \{ 3,4\}$.
Using this we can write the $\langle F_1 F_2\rangle$ correlator as
\begin{align}
\langle F_1 F_2 \rangle &= I_{13}(Z_2 \cdot Z_4) - I_{14}(Z_2 \cdot Z_3) - I_{23}(Z_1 \cdot Z_4) + I_{24}(Z_1 \cdot Z_3)
\end{align}

Introducing Schwinger parameters $s$ and $t$ and using the identity $A^{-2} = \int_0^\infty ds \, s \, e^{-sA}$ and similar for $t$ we get
\begin{align}
I_{ik} := \left\langle \frac{(\tilde{\Theta} \cdot Z_i)(\tilde{\Theta} \cdot Z_k)}{(\tilde{\Theta} \cdot X_1)^2 (\tilde{\Theta} \cdot X_2)^2} \right\rangle &= \int_0^\infty ds \int_0^\infty dt \, s \, t \, \langle (\tilde{\Theta} \cdot Z_i)(\tilde{\Theta} \cdot Z_k) e^{-\tilde{\Theta} \cdot (sX_1 + tX_2)} \rangle \label{Schwinger}
\end{align}
We evaluate the Gaussian expectation value by taking derivatives with respect to a source $J = -(sX_1 + tX_2)$. For this, recall that for a generic $\mathcal{O}(\Theta)$ the Gaussian integral w.r.t. $\Theta$ is defined as

\begin{equation}
    \langle \mathcal{O}(\Theta) \rangle := \frac{\int [d\Theta] \, \mathcal{O}(\Theta) \, e^{-\frac{1}{2} \Theta^2}}{\int [d\Theta] \, e^{-\frac{1}{2} \Theta^2}} \\
\end{equation}
Then the following is true:
\begin{align}
\langle e^{\Theta \cdot J} \rangle &= \frac{\int [d\Theta] \, e^{-\frac{1}{2} \Theta^2 + \Theta \cdot J}}{\int [d\Theta] \, e^{-\frac{1}{2} \Theta^2}} \notag \\
&= \frac{\int [d\Theta] \, e^{-\frac{1}{2} (\Theta - J)^2 + \frac{1}{2} J^2}}{\int [d\Theta] \, e^{-\frac{1}{2} \Theta^2}} \notag \\
&= e^{\frac{1}{2} J^2} \frac{\int [d(\Theta - J)] \, e^{-\frac{1}{2} (\Theta - J)^2}}{\int [d\Theta] \, e^{-\frac{1}{2} \Theta^2}} \notag \\
&= e^{\frac{1}{2} J^2} 
\end{align}
$$
\begin{aligned}
\langle (\Theta \cdot X_1)(\Theta \cdot X_2) e^{\Theta \cdot J} \rangle &= \frac{\int [d\Theta] \, (\Theta \cdot X_1)(\Theta \cdot X_2) \, e^{-\frac{1}{2} \Theta^2 + \Theta \cdot J}}{\int [d\Theta] \, e^{-\frac{1}{2} \Theta^2}} \\
&= \frac{\int [d\Theta] \, \left( X_1 \cdot \frac{\partial}{\partial J} \right) \left( X_2 \cdot \frac{\partial}{\partial J} \right) e^{-\frac{1}{2} \Theta^2 + \Theta \cdot J}}{\int [d\Theta] \, e^{-\frac{1}{2} \Theta^2}} \\
&= \left( X_1 \cdot \frac{\partial}{\partial J} \right) \left( X_2 \cdot \frac{\partial}{\partial J} \right) \left[ \frac{\int [d\Theta] \, e^{-\frac{1}{2} \Theta^2 + \Theta \cdot J}}{\int [d\Theta] \, e^{-\frac{1}{2} \Theta^2}} \right] \\
&= \left( X_1 \cdot \frac{\partial}{\partial J} \right) \left( X_2 \cdot \frac{\partial}{\partial J} \right) \langle e^{\Theta \cdot J} \rangle \\
&= \left( X_1 \cdot \frac{\partial}{\partial J} \right) \left( X_2 \cdot \frac{\partial}{\partial J} \right) e^{\frac{1}{2} J^2} \\
\\
&= \left( X_{1N} \frac{\partial}{\partial J_N} \right) \left[ X_{2M} J^M e^{\frac{1}{2} J^2} \right] \\
&= X_{1N} \left[ \left( \frac{\partial}{\partial J_N} (X_{2M} J^M) \right) e^{\frac{1}{2} J^2} + (X_{2M} J^M) \left( \frac{\partial}{\partial J_N} e^{\frac{1}{2} J^2} \right) \right] \\
&= X_{1N} \left[ \left( X_{2M} \delta^M_N \right) e^{\frac{1}{2} J^2} + (X_2 \cdot J) \left( J_N e^{\frac{1}{2} J^2} \right) \right] \\
&= \left[ (X_{1N} X_{2N}) + (X_{1N} J_N)(X_2 \cdot J) \right] e^{\frac{1}{2} J^2} \\
&= \big[ (X_1 \cdot X_2) + (X_1 \cdot J)(X_2 \cdot J) \big] e^{\frac{1}{2} J^2}
\end{aligned}
$$
Applying this to our case \eqref{Schwinger}
\begin{align}
\langle (\tilde{\Theta} \cdot Z_i)(\tilde{\Theta} \cdot Z_k) e^{\tilde{\Theta} \cdot J} \rangle &= \big[ (Z_i \cdot Z_k) + (J \cdot Z_i)(J \cdot Z_k) \big] e^{\frac{1}{2} J^2} \notag \\
&= \Big[ (Z_i \cdot Z_k) + \big( -(sX_1 + tX_2) \cdot Z_i \big) \big( -(sX_1 + tX_2) \cdot Z_k \big) \Big] e^{\frac{1}{2} (-(sX_1 + tX_2))^2} \notag \\
&= \Big[ (Z_i \cdot Z_k) + \big( -s(X_1 \cdot Z_i) - t(X_2 \cdot Z_i) \big) \big( -s(X_1 \cdot Z_k) - t(X_2 \cdot Z_k) \big) \Big] \notag \\
&\qquad \times e^{\frac{1}{2} (s^2 X_1^2 + t^2 X_2^2 + 2st(X_1 \cdot X_2))} \notag \\
&= \big[ (Z_i \cdot Z_k) + st(X_2 \cdot Z_i)(X_1 \cdot Z_k) \big] e^{st(X_1 \cdot X_2)}
\end{align}
where in the last step we used the PNC condition $X_1^2=X_2^2=0$ and kinematic transversality (which is different from the field transversality that we broke in our ansatz). Note that we can use the PNC because we Wick rotate to Lorentzian signature after taking the NN parameter average.

Next, we use the variables $u = st$ and $v = s/t$, which gives a divergent integral over $v$ i.e. $\int \frac{dv}{v}$. This can be viewed as a redundancy in the Schwinger parameterization space: the integrand depends only on the product of the Schwinger parameters, and not their ratio $-$ this happened after imposing the PNC constraint (the analog of this in usual 1-loop perturbative QFT calculations using Schwinger parameterization is absent because of the loop \emph{integrals}, so we never get these extra divergences in usual QFT computations). This divergence is an artifact of this method of calculation and can be safely ignored since we are interested in extracting the finite piece. Alternatively, interpreting the Schwinger parameters $s$ and $t$  as scale factors ($\in[0,\infty)$) multiplying $X_1$ and $X_2$ respectively, $v$ is the relative scaling and $\int dv/v$ is precisely the Haar group volume \cite{Folland2015} of this relative scaling $v \in (0,\infty)$: since $v$ doesn't appear in the integrand one should formally mod out by the gauge volume which gets rid of the $v$ integral \footnote{The ambiguity in whether to loose $\int dv/v$ or a result merely proportional to that can be fixed by fixing the field normalization in general.}. Therefore, throughout the rest of this section, we will be replacing $\int_0^\infty \frac{dv}{v} \to 1$ while retaining the Jacobian factor $\tfrac{1}{2}$ from $ds\,dt = \frac{du\,dv}{2v}$.
The remaining $u$ integral can be simplified as follows:
\begin{align}
I_{ik} &= \int_0^\infty du \, u \big[ (Z_i \cdot Z_k) + u(X_2 \cdot Z_i)(X_1 \cdot Z_k) \big] e^{u(X_1 \cdot X_2)} \notag \\
&= (Z_i \cdot Z_k) \int_0^\infty u e^{u(X_1 \cdot X_2)} du + (X_2 \cdot Z_i)(X_1 \cdot Z_k) \int_0^\infty u^2 e^{u(X_1 \cdot X_2)} du \notag \\
&= \frac{(Z_i \cdot Z_k)}{(X_1 \cdot X_2)^2} - \frac{2(X_2 \cdot Z_i)(X_1 \cdot Z_k)}{(X_1 \cdot X_2)^3}
\end{align}

The last equation holds if the integrals in the penultimate step converges which does since $X_1 \cdot X_2<0$ which one can easily see recalling $X_1 \cdot X_2=-\frac{1}{2}(x_1-x_2)^2<0$ because $x_k \in \mathbb{R}^d$.
The correlator becomes
\begin{align}
    \langle F(X_1,Z_1,&Z_2) F(X_2,Z_3,Z_4) \rangle \\&=\frac{2}{(X_1 \cdot X_2)^2} \big[ (Z_1 \cdot Z_3)(Z_2 \cdot Z_4) - (Z_1 \cdot Z_4)(Z_2 \cdot Z_3) \big] \notag \\
    &\quad - \frac{2}{(X_1 \cdot X_2)^3} \big[ (X_2 \cdot Z_1)(X_1 \cdot Z_3)(Z_2 \cdot Z_4) - (X_2 \cdot Z_1)(X_1 \cdot Z_4)(Z_2 \cdot Z_3) \notag \\
    &\qquad \qquad \quad - (X_2 \cdot Z_2)(X_1 \cdot Z_3)(Z_1 \cdot Z_4) + (X_2 \cdot Z_2)(X_1 \cdot Z_4)(Z_1 \cdot Z_3) \big]
\end{align}
which is \eqref{FFMaxwellintermediate}. Therefore, both the methods give us the same result for $\langle F(X_1,Z_1,Z_2) F(X_2,Z_3,Z_4) \rangle$.\\\\
\textbf{Full ansatz calculation}\\
Here we do the same computation but now for the ansatz \eqref{spin1ansatzMaxwell}.
The untruncated network field is:
\begin{equation}
A(X,Z) = (\tilde{\Theta}\cdot X)^{-\Delta-1} \big( (\Theta\cdot X)(\eta\cdot Z) - (\eta\cdot X)(\Theta\cdot Z) \big)
\end{equation}
The field strength $F(X, Z_1, Z_2) = Z_1^M \partial_M A(X, Z_2) - Z_2^M \partial_M A(X, Z_1)$ evaluates directly to \eqref{FansatzgenericDelta} which is
\begin{align}
F(X, Z_1, Z_2) &= -(\Delta+1)(\tilde{\Theta}\cdot X)^{-\Delta-2} \Big[ (\tilde{\Theta}\cdot Z_1)(\Theta\cdot X)(\eta\cdot Z_2) - (\tilde{\Theta}\cdot Z_1)(\eta\cdot X)(\Theta\cdot Z_2) \nonumber\\
&\qquad\qquad\qquad\qquad\qquad - (\tilde{\Theta}\cdot Z_2)(\Theta\cdot X)(\eta\cdot Z_1) + (\tilde{\Theta}\cdot Z_2)(\eta\cdot X)(\Theta\cdot Z_1) \Big] \nonumber\\[0.5em]
&\quad + 2(\tilde{\Theta}\cdot X)^{-\Delta-1} \Big[ (\Theta\cdot Z_1)(\eta\cdot Z_2) - (\eta\cdot Z_1)(\Theta\cdot Z_2) \Big]
\end{align}
For the two point correlator, we write down the explicit field strength evaluated at the two points. 
At point $X_1$ with polarizations $Z_1, Z_2$:
\begin{align}
F(X_1, Z_1, Z_2) &= -(\Delta+1)(\tilde{\Theta}\cdot X_1)^{-\Delta-2} \Big[ (\tilde{\Theta}\cdot Z_1)(\Theta\cdot X_1)(\eta\cdot Z_2) - (\tilde{\Theta}\cdot Z_1)(\eta\cdot X_1)(\Theta\cdot Z_2) \nonumber\\
&\qquad\qquad\qquad\qquad\qquad\quad - (\tilde{\Theta}\cdot Z_2)(\Theta\cdot X_1)(\eta\cdot Z_1) + (\tilde{\Theta}\cdot Z_2)(\eta\cdot X_1)(\Theta\cdot Z_1) \Big] \nonumber\\[0.5em]
&\quad + 2(\tilde{\Theta}\cdot X_1)^{-\Delta-1} \Big[ (\Theta\cdot Z_1)(\eta\cdot Z_2) - (\eta\cdot Z_1)(\Theta\cdot Z_2) \Big]
\end{align}
At point $X_2$ with polarizations $Z_3, Z_4$:
\begin{align}
F(X_2, Z_3, Z_4) &= -(\Delta+1)(\tilde{\Theta}\cdot X_2)^{-\Delta-2} \Big[ (\tilde{\Theta}\cdot Z_3)(\Theta\cdot X_2)(\eta\cdot Z_4) - (\tilde{\Theta}\cdot Z_3)(\eta\cdot X_2)(\Theta\cdot Z_4) \nonumber\\
&\qquad\qquad\qquad\qquad\qquad\quad - (\tilde{\Theta}\cdot Z_4)(\Theta\cdot X_2)(\eta\cdot Z_3) + (\tilde{\Theta}\cdot Z_4)(\eta\cdot X_2)(\Theta\cdot Z_3) \Big] \nonumber\\[0.5em]
&\quad + 2(\tilde{\Theta}\cdot X_2)^{-\Delta-1} \Big[ (\Theta\cdot Z_3)(\eta\cdot Z_4) - (\eta\cdot Z_3)(\Theta\cdot Z_4) \Big]
\end{align}
When we multiply these together, we get four terms $T_1,T_2,T_3$ and $T_4$ which are collected according to powers of the prefactors $(\tilde{\Theta}\cdot X_1)$ and $(\tilde{\Theta}\cdot X_2)$ as
\begin{align}
T_1 &= \Big( 2(\tilde{\Theta}\cdot X_1)^{-\Delta-1} \big[ (\Theta\cdot Z_1)(\eta\cdot Z_2) - (\eta\cdot Z_1)(\Theta\cdot Z_2) \big] \Big) \nonumber\\
&\quad \times \Big( 2(\tilde{\Theta}\cdot X_2)^{-\Delta-1} \big[ (\Theta\cdot Z_3)(\eta\cdot Z_4) - (\eta\cdot Z_3)(\Theta\cdot Z_4) \big] \Big) \nonumber\\[0.5em]
&= 4 (\tilde{\Theta}\cdot X_1)^{-\Delta-1} (\tilde{\Theta}\cdot X_2)^{-\Delta-1} \Big[ \nonumber\\
&\qquad\quad (\Theta\cdot Z_1)(\eta\cdot Z_2)(\Theta\cdot Z_3)(\eta\cdot Z_4) \nonumber\\
&\qquad - (\Theta\cdot Z_1)(\eta\cdot Z_2)(\eta\cdot Z_3)(\Theta\cdot Z_4) \nonumber\\
&\qquad - (\eta\cdot Z_1)(\Theta\cdot Z_2)(\Theta\cdot Z_3)(\eta\cdot Z_4) \nonumber\\
&\qquad + (\eta\cdot Z_1)(\Theta\cdot Z_2)(\eta\cdot Z_3)(\Theta\cdot Z_4) \Big]
\end{align}

\begin{align}
T_2 &= \Big( -(\Delta+1)(\tilde{\Theta}\cdot X_1)^{-\Delta-2} \big[ (\tilde{\Theta}\cdot Z_1)(\Theta\cdot X_1)(\eta\cdot Z_2) - (\tilde{\Theta}\cdot Z_1)(\eta\cdot X_1)(\Theta\cdot Z_2) \nonumber\\
&\qquad\qquad\qquad\qquad\qquad\qquad - (\tilde{\Theta}\cdot Z_2)(\Theta\cdot X_1)(\eta\cdot Z_1) + (\tilde{\Theta}\cdot Z_2)(\eta\cdot X_1)(\Theta\cdot Z_1) \big] \Big) \nonumber\\
&\quad \times \Big( 2(\tilde{\Theta}\cdot X_2)^{-\Delta-1} \big[ (\Theta\cdot Z_3)(\eta\cdot Z_4) - (\eta\cdot Z_3)(\Theta\cdot Z_4) \big] \Big) \nonumber\\[0.5em]
&= -2(\Delta+1) (\tilde{\Theta}\cdot X_1)^{-\Delta-2} (\tilde{\Theta}\cdot X_2)^{-\Delta-1} \Big[ \nonumber\\
&\qquad\quad (\tilde{\Theta}\cdot Z_1)(\Theta\cdot X_1)(\eta\cdot Z_2)(\Theta\cdot Z_3)(\eta\cdot Z_4) - (\tilde{\Theta}\cdot Z_1)(\Theta\cdot X_1)(\eta\cdot Z_2)(\eta\cdot Z_3)(\Theta\cdot Z_4) \nonumber\\
&\qquad - (\tilde{\Theta}\cdot Z_1)(\eta\cdot X_1)(\Theta\cdot Z_2)(\Theta\cdot Z_3)(\eta\cdot Z_4) + (\tilde{\Theta}\cdot Z_1)(\eta\cdot X_1)(\Theta\cdot Z_2)(\eta\cdot Z_3)(\Theta\cdot Z_4) \nonumber\\
&\qquad - (\tilde{\Theta}\cdot Z_2)(\Theta\cdot X_1)(\eta\cdot Z_1)(\Theta\cdot Z_3)(\eta\cdot Z_4) + (\tilde{\Theta}\cdot Z_2)(\Theta\cdot X_1)(\eta\cdot Z_1)(\eta\cdot Z_3)(\Theta\cdot Z_4) \nonumber\\
&\qquad + (\tilde{\Theta}\cdot Z_2)(\eta\cdot X_1)(\Theta\cdot Z_1)(\Theta\cdot Z_3)(\eta\cdot Z_4) - (\tilde{\Theta}\cdot Z_2)(\eta\cdot X_1)(\Theta\cdot Z_1)(\eta\cdot Z_3)(\Theta\cdot Z_4) \Big]
\end{align}

\begin{align}
T_3 &= \Big( 2(\tilde{\Theta}\cdot X_1)^{-\Delta-1} \big[ (\Theta\cdot Z_1)(\eta\cdot Z_2) - (\eta\cdot Z_1)(\Theta\cdot Z_2) \big] \Big) \nonumber\\
&\quad \times \Big( -(\Delta+1)(\tilde{\Theta}\cdot X_2)^{-\Delta-2} \big[ (\tilde{\Theta}\cdot Z_3)(\Theta\cdot X_2)(\eta\cdot Z_4) - (\tilde{\Theta}\cdot Z_3)(\eta\cdot X_2)(\Theta\cdot Z_4) \nonumber\\
&\qquad\qquad\qquad\qquad\qquad\qquad - (\tilde{\Theta}\cdot Z_4)(\Theta\cdot X_2)(\eta\cdot Z_3) + (\tilde{\Theta}\cdot Z_4)(\eta\cdot X_2)(\Theta\cdot Z_3) \big] \Big) \nonumber\\[0.5em]
&= -2(\Delta+1) (\tilde{\Theta}\cdot X_1)^{-\Delta-1} (\tilde{\Theta}\cdot X_2)^{-\Delta-2} \Big[ \nonumber\\
&\qquad\quad (\Theta\cdot Z_1)(\eta\cdot Z_2)(\tilde{\Theta}\cdot Z_3)(\Theta\cdot X_2)(\eta\cdot Z_4) - (\Theta\cdot Z_1)(\eta\cdot Z_2)(\tilde{\Theta}\cdot Z_3)(\eta\cdot X_2)(\Theta\cdot Z_4) \nonumber\\
&\qquad - (\Theta\cdot Z_1)(\eta\cdot Z_2)(\tilde{\Theta}\cdot Z_4)(\Theta\cdot X_2)(\eta\cdot Z_3) + (\Theta\cdot Z_1)(\eta\cdot Z_2)(\tilde{\Theta}\cdot Z_4)(\eta\cdot X_2)(\Theta\cdot Z_3) \nonumber\\
&\qquad - (\eta\cdot Z_1)(\Theta\cdot Z_2)(\tilde{\Theta}\cdot Z_3)(\Theta\cdot X_2)(\eta\cdot Z_4) + (\eta\cdot Z_1)(\Theta\cdot Z_2)(\tilde{\Theta}\cdot Z_3)(\eta\cdot X_2)(\Theta\cdot Z_4) \nonumber\\
&\qquad + (\eta\cdot Z_1)(\Theta\cdot Z_2)(\tilde{\Theta}\cdot Z_4)(\Theta\cdot X_2)(\eta\cdot Z_3) - (\eta\cdot Z_1)(\Theta\cdot Z_2)(\tilde{\Theta}\cdot Z_4)(\eta\cdot X_2)(\Theta\cdot Z_3) \Big]
\end{align}

and 

\begin{align}
T_4 &= \Big( -(\Delta+1)(\tilde{\Theta}\cdot X_1)^{-\Delta-2} \big[ (\tilde{\Theta}\cdot Z_1)(\Theta\cdot X_1)(\eta\cdot Z_2) - (\tilde{\Theta}\cdot Z_1)(\eta\cdot X_1)(\Theta\cdot Z_2) \nonumber\\
&\qquad\qquad\qquad\qquad\qquad\qquad - (\tilde{\Theta}\cdot Z_2)(\Theta\cdot X_1)(\eta\cdot Z_1) + (\tilde{\Theta}\cdot Z_2)(\eta\cdot X_1)(\Theta\cdot Z_1) \big] \Big) \nonumber\\
&\quad \times \Big( -(\Delta+1)(\tilde{\Theta}\cdot X_2)^{-\Delta-2} \big[ (\tilde{\Theta}\cdot Z_3)(\Theta\cdot X_2)(\eta\cdot Z_4) - (\tilde{\Theta}\cdot Z_3)(\eta\cdot X_2)(\Theta\cdot Z_4) \nonumber\\
&\qquad\qquad\qquad\qquad\qquad\qquad - (\tilde{\Theta}\cdot Z_4)(\Theta\cdot X_2)(\eta\cdot Z_3) + (\tilde{\Theta}\cdot Z_4)(\eta\cdot X_2)(\Theta\cdot Z_3) \big] \Big) \nonumber\\[0.5em]
&= (\Delta+1)^2 (\tilde{\Theta}\cdot X_1)^{-\Delta-2} (\tilde{\Theta}\cdot X_2)^{-\Delta-2} \Big[ \nonumber\\
&\qquad\quad (\tilde{\Theta}\cdot Z_1)(\Theta\cdot X_1)(\eta\cdot Z_2)(\tilde{\Theta}\cdot Z_3)(\Theta\cdot X_2)(\eta\cdot Z_4) \nonumber\\
&\qquad - (\tilde{\Theta}\cdot Z_1)(\Theta\cdot X_1)(\eta\cdot Z_2)(\tilde{\Theta}\cdot Z_3)(\eta\cdot X_2)(\Theta\cdot Z_4) \nonumber\\
&\qquad - (\tilde{\Theta}\cdot Z_1)(\Theta\cdot X_1)(\eta\cdot Z_2)(\tilde{\Theta}\cdot Z_4)(\Theta\cdot X_2)(\eta\cdot Z_3) \nonumber\\
&\qquad + (\tilde{\Theta}\cdot Z_1)(\Theta\cdot X_1)(\eta\cdot Z_2)(\tilde{\Theta}\cdot Z_4)(\eta\cdot X_2)(\Theta\cdot Z_3) \nonumber\\[0.5em]
&\qquad - (\tilde{\Theta}\cdot Z_1)(\eta\cdot X_1)(\Theta\cdot Z_2)(\tilde{\Theta}\cdot Z_3)(\Theta\cdot X_2)(\eta\cdot Z_4) \nonumber\\
&\qquad + (\tilde{\Theta}\cdot Z_1)(\eta\cdot X_1)(\Theta\cdot Z_2)(\tilde{\Theta}\cdot Z_3)(\eta\cdot X_2)(\Theta\cdot Z_4) \nonumber\\
&\qquad + (\tilde{\Theta}\cdot Z_1)(\eta\cdot X_1)(\Theta\cdot Z_2)(\tilde{\Theta}\cdot Z_4)(\Theta\cdot X_2)(\eta\cdot Z_3) \nonumber\\
&\qquad - (\tilde{\Theta}\cdot Z_1)(\eta\cdot X_1)(\Theta\cdot Z_2)(\tilde{\Theta}\cdot Z_4)(\eta\cdot X_2)(\Theta\cdot Z_3) \nonumber\\[0.5em]
&\qquad - (\tilde{\Theta}\cdot Z_2)(\Theta\cdot X_1)(\eta\cdot Z_1)(\tilde{\Theta}\cdot Z_3)(\Theta\cdot X_2)(\eta\cdot Z_4) \nonumber\\
&\qquad + (\tilde{\Theta}\cdot Z_2)(\Theta\cdot X_1)(\eta\cdot Z_1)(\tilde{\Theta}\cdot Z_3)(\eta\cdot X_2)(\Theta\cdot Z_4) \nonumber\\
&\qquad + (\tilde{\Theta}\cdot Z_2)(\Theta\cdot X_1)(\eta\cdot Z_1)(\tilde{\Theta}\cdot Z_4)(\Theta\cdot X_2)(\eta\cdot Z_3) \nonumber\\
&\qquad - (\tilde{\Theta}\cdot Z_2)(\Theta\cdot X_1)(\eta\cdot Z_1)(\tilde{\Theta}\cdot Z_4)(\eta\cdot X_2)(\Theta\cdot Z_3) \nonumber\\[0.5em]
&\qquad + (\tilde{\Theta}\cdot Z_2)(\eta\cdot X_1)(\Theta\cdot Z_1)(\tilde{\Theta}\cdot Z_3)(\Theta\cdot X_2)(\eta\cdot Z_4) \nonumber\\
&\qquad - (\tilde{\Theta}\cdot Z_2)(\eta\cdot X_1)(\Theta\cdot Z_1)(\tilde{\Theta}\cdot Z_3)(\eta\cdot X_2)(\Theta\cdot Z_4) \nonumber\\
&\qquad - (\tilde{\Theta}\cdot Z_2)(\eta\cdot X_1)(\Theta\cdot Z_1)(\tilde{\Theta}\cdot Z_4)(\Theta\cdot X_2)(\eta\cdot Z_3) \nonumber\\
&\qquad + (\tilde{\Theta}\cdot Z_2)(\eta\cdot X_1)(\Theta\cdot Z_1)(\tilde{\Theta}\cdot Z_4)(\eta\cdot X_2)(\Theta\cdot Z_3) \Big]
\end{align}

where obviously 
\begin{equation}
    F(X_1,Z_1,Z_2)F(X_2,Z_3,Z_4)=T_1+T_2+T_3+T_4
\end{equation}
The averages over $\Theta,\eta$ can be easily performed via Wick contractions, while that over $\tilde{\Theta}$ will require Schwinger parameterization techniques of the kind used in the previous subsections. Performing the Wick contractions we get,
\begin{align}
\langle T_1 \rangle_{\Theta, \eta} &= 4 (\tilde{\Theta}\cdot X_1)^{-\Delta-1} (\tilde{\Theta}\cdot X_2)^{-\Delta-1} \Big[ \nonumber\\
&\qquad\quad \langle(\Theta\cdot Z_1)(\Theta\cdot Z_3)\rangle \langle(\eta\cdot Z_2)(\eta\cdot Z_4)\rangle \nonumber\\
&\qquad - \langle(\Theta\cdot Z_1)(\Theta\cdot Z_4)\rangle \langle(\eta\cdot Z_2)(\eta\cdot Z_3)\rangle \nonumber\\
&\qquad - \langle(\Theta\cdot Z_2)(\Theta\cdot Z_3)\rangle \langle(\eta\cdot Z_1)(\eta\cdot Z_4)\rangle \nonumber\\
&\qquad + \langle(\Theta\cdot Z_2)(\Theta\cdot Z_4)\rangle \langle(\eta\cdot Z_1)(\eta\cdot Z_3)\rangle \Big] \nonumber\\[0.5em]
&= 4 (\tilde{\Theta}\cdot X_1)^{-\Delta-1} (\tilde{\Theta}\cdot X_2)^{-\Delta-1} \Big[ \nonumber\\
&\qquad\quad (Z_1\cdot Z_3)(Z_2\cdot Z_4) - (Z_1\cdot Z_4)(Z_2\cdot Z_3) \nonumber\\
&\qquad - (Z_2\cdot Z_3)(Z_1\cdot Z_4) + (Z_2\cdot Z_4)(Z_1\cdot Z_3) \Big] \nonumber\\[0.5em]
&= 8 (\tilde{\Theta}\cdot X_1)^{-\Delta-1} (\tilde{\Theta}\cdot X_2)^{-\Delta-1} \Big[ (Z_1\cdot Z_3)(Z_2\cdot Z_4) - (Z_1\cdot Z_4)(Z_2\cdot Z_3) \Big]
\end{align}

\begin{align}
\langle T_2 \rangle_{\Theta, \eta} &= -2(\Delta+1) (\tilde{\Theta}\cdot X_1)^{-\Delta-2} (\tilde{\Theta}\cdot X_2)^{-\Delta-1} \Big[ \nonumber\\
&\qquad\quad (\tilde{\Theta}\cdot Z_1)(X_1\cdot Z_3)(Z_2\cdot Z_4) - (\tilde{\Theta}\cdot Z_1)(X_1\cdot Z_4)(Z_2\cdot Z_3) \nonumber\\
&\qquad - (\tilde{\Theta}\cdot Z_1)(X_1\cdot Z_4)(Z_2\cdot Z_3) + (\tilde{\Theta}\cdot Z_1)(X_1\cdot Z_3)(Z_2\cdot Z_4) \nonumber\\
&\qquad - (\tilde{\Theta}\cdot Z_2)(X_1\cdot Z_3)(Z_1\cdot Z_4) + (\tilde{\Theta}\cdot Z_2)(X_1\cdot Z_4)(Z_1\cdot Z_3) \nonumber\\
&\qquad + (\tilde{\Theta}\cdot Z_2)(X_1\cdot Z_4)(Z_1\cdot Z_3) - (\tilde{\Theta}\cdot Z_2)(X_1\cdot Z_3)(Z_1\cdot Z_4) \Big] \nonumber\\[0.5em]
&= -4(\Delta+1) (\tilde{\Theta}\cdot X_1)^{-\Delta-2} (\tilde{\Theta}\cdot X_2)^{-\Delta-1} \Big[ \nonumber\\
&\qquad\quad (\tilde{\Theta}\cdot Z_1)(X_1\cdot Z_3)(Z_2\cdot Z_4) - (\tilde{\Theta}\cdot Z_1)(X_1\cdot Z_4)(Z_2\cdot Z_3) \nonumber\\
&\qquad - (\tilde{\Theta}\cdot Z_2)(X_1\cdot Z_3)(Z_1\cdot Z_4) + (\tilde{\Theta}\cdot Z_2)(X_1\cdot Z_4)(Z_1\cdot Z_3) \Big]
\end{align}

\begin{align}
\langle T_3 \rangle_{\Theta, \eta} &= -2(\Delta+1) (\tilde{\Theta}\cdot X_1)^{-\Delta-1} (\tilde{\Theta}\cdot X_2)^{-\Delta-2} \Big[ \nonumber\\
&\qquad\quad (\tilde{\Theta}\cdot Z_3)(X_2\cdot Z_1)(Z_4\cdot Z_2) - (\tilde{\Theta}\cdot Z_3)(X_2\cdot Z_2)(Z_4\cdot Z_1) \nonumber\\
&\qquad - (\tilde{\Theta}\cdot Z_4)(X_2\cdot Z_1)(Z_3\cdot Z_2) + (\tilde{\Theta}\cdot Z_4)(X_2\cdot Z_2)(Z_3\cdot Z_1) \nonumber\\
&\qquad - (\tilde{\Theta}\cdot Z_3)(X_2\cdot Z_2)(Z_4\cdot Z_1) + (\tilde{\Theta}\cdot Z_3)(X_2\cdot Z_1)(Z_4\cdot Z_2) \nonumber\\
&\qquad + (\tilde{\Theta}\cdot Z_4)(X_2\cdot Z_2)(Z_3\cdot Z_1) - (\tilde{\Theta}\cdot Z_4)(X_2\cdot Z_1)(Z_3\cdot Z_2) \Big] \nonumber\\[0.5em]
&= -4(\Delta+1) (\tilde{\Theta}\cdot X_1)^{-\Delta-1} (\tilde{\Theta}\cdot X_2)^{-\Delta-2} \Big[ \nonumber\\
&\qquad\quad (\tilde{\Theta}\cdot Z_3)(X_2\cdot Z_1)(Z_4\cdot Z_2) - (\tilde{\Theta}\cdot Z_3)(X_2\cdot Z_2)(Z_4\cdot Z_1) \nonumber\\
&\qquad - (\tilde{\Theta}\cdot Z_4)(X_2\cdot Z_1)(Z_3\cdot Z_2) + (\tilde{\Theta}\cdot Z_4)(X_2\cdot Z_2)(Z_3\cdot Z_1) \Big]
\end{align}

\begin{align}
\langle T_4 \rangle_{\Theta, \eta} &= (\Delta+1)^2 (\tilde{\Theta}\cdot X_1)^{-\Delta-2} (\tilde{\Theta}\cdot X_2)^{-\Delta-2} \Big[ \nonumber\\
&\qquad\quad (\tilde{\Theta}\cdot Z_1)(\tilde{\Theta}\cdot Z_3)(X_1\cdot X_2)(Z_2\cdot Z_4) \nonumber\\
&\qquad - (\tilde{\Theta}\cdot Z_1)(\tilde{\Theta}\cdot Z_3)(X_1\cdot Z_4)(X_2\cdot Z_2) \nonumber\\
&\qquad - (\tilde{\Theta}\cdot Z_1)(\tilde{\Theta}\cdot Z_4)(X_1\cdot X_2)(Z_2\cdot Z_3) \nonumber\\
&\qquad + (\tilde{\Theta}\cdot Z_1)(\tilde{\Theta}\cdot Z_4)(X_1\cdot Z_3)(X_2\cdot Z_2) \nonumber\\[0.5em]
&\qquad - (\tilde{\Theta}\cdot Z_1)(\tilde{\Theta}\cdot Z_3)(X_2\cdot Z_2)(X_1\cdot Z_4) \nonumber\\
&\qquad + (\tilde{\Theta}\cdot Z_1)(\tilde{\Theta}\cdot Z_3)(X_1\cdot X_2)(Z_2\cdot Z_4) \nonumber\\
&\qquad + (\tilde{\Theta}\cdot Z_1)(\tilde{\Theta}\cdot Z_4)(X_2\cdot Z_2)(X_1\cdot Z_3) \nonumber\\
&\qquad - (\tilde{\Theta}\cdot Z_1)(\tilde{\Theta}\cdot Z_4)(X_1\cdot X_2)(Z_2\cdot Z_3) \nonumber\\[0.5em]
&\qquad - (\tilde{\Theta}\cdot Z_2)(\tilde{\Theta}\cdot Z_3)(X_1\cdot X_2)(Z_1\cdot Z_4) \nonumber\\
&\qquad + (\tilde{\Theta}\cdot Z_2)(\tilde{\Theta}\cdot Z_3)(X_1\cdot Z_4)(X_2\cdot Z_1) \nonumber\\
&\qquad + (\tilde{\Theta}\cdot Z_2)(\tilde{\Theta}\cdot Z_4)(X_1\cdot X_2)(Z_1\cdot Z_3) \nonumber\\
&\qquad - (\tilde{\Theta}\cdot Z_2)(\tilde{\Theta}\cdot Z_4)(X_1\cdot Z_3)(X_2\cdot Z_1) \nonumber\\[0.5em]
&\qquad + (\tilde{\Theta}\cdot Z_2)(\tilde{\Theta}\cdot Z_3)(X_2\cdot Z_1)(X_1\cdot Z_4) \nonumber\\
&\qquad - (\tilde{\Theta}\cdot Z_2)(\tilde{\Theta}\cdot Z_3)(X_1\cdot X_2)(Z_1\cdot Z_4) \nonumber\\
&\qquad - (\tilde{\Theta}\cdot Z_2)(\tilde{\Theta}\cdot Z_4)(X_2\cdot Z_1)(X_1\cdot Z_3) \nonumber\\
&\qquad + (\tilde{\Theta}\cdot Z_2)(\tilde{\Theta}\cdot Z_4)(X_1\cdot X_2)(Z_1\cdot Z_3) \Big] \nonumber\\[0.5em]
&= 2(\Delta+1)^2 (\tilde{\Theta}\cdot X_1)^{-\Delta-2} (\tilde{\Theta}\cdot X_2)^{-\Delta-2} \Big[ \nonumber\\
&\qquad\quad (\tilde{\Theta}\cdot Z_1)(\tilde{\Theta}\cdot Z_3) \big( (X_1\cdot X_2)(Z_2\cdot Z_4) - (X_1\cdot Z_4)(X_2\cdot Z_2) \big) \nonumber\\
&\qquad - (\tilde{\Theta}\cdot Z_1)(\tilde{\Theta}\cdot Z_4) \big( (X_1\cdot X_2)(Z_2\cdot Z_3) - (X_1\cdot Z_3)(X_2\cdot Z_2) \big) \nonumber\\
&\qquad - (\tilde{\Theta}\cdot Z_2)(\tilde{\Theta}\cdot Z_3) \big( (X_1\cdot X_2)(Z_1\cdot Z_4) - (X_1\cdot Z_4)(X_2\cdot Z_1) \big) \nonumber\\
&\qquad + (\tilde{\Theta}\cdot Z_2)(\tilde{\Theta}\cdot Z_4) \big( (X_1\cdot X_2)(Z_1\cdot Z_3) - (X_1\cdot Z_3)(X_2\cdot Z_1) \big) \Big]
\end{align}
Now we perform the Schwinger integrals over each of these terms. Starting off with $T_1$, recall that the $\Theta$ and $\eta$ expectation value leaves us with 
\begin{equation}
\langle T_1 \rangle_{\Theta, \eta} = 8 \Big[ (Z_1\cdot Z_3)(Z_2\cdot Z_4) - (Z_1\cdot Z_4)(Z_2\cdot Z_3) \Big] (\tilde{\Theta}\cdot X_1)^{-\Delta-1} (\tilde{\Theta}\cdot X_2)^{-\Delta-1}
\end{equation}
We compute the Gaussian expectation value $\langle \dots \rangle_{\tilde{\Theta}}$ using the identity:
\begin{equation}
A^{-\alpha} = \frac{1}{\Gamma(\alpha)} \int_0^\infty ds \, s^{\alpha-1} e^{-sA}
\end{equation}
Applying this to both denominator factors with $\alpha = \Delta+1$:
\begin{align}
(\tilde{\Theta}\cdot X_1)^{-\Delta-1} &= \frac{1}{\Gamma(\Delta+1)} \int_0^\infty ds \, s^{\Delta} e^{-s(\tilde{\Theta}\cdot X_1)} \\
(\tilde{\Theta}\cdot X_2)^{-\Delta-1} &= \frac{1}{\Gamma(\Delta+1)} \int_0^\infty dt \, t^{\Delta} e^{-t(\tilde{\Theta}\cdot X_2)}
\end{align}
Multiplying these together, the expectation value of the spatial part becomes:
\begin{align}
I_1 &:= \big\langle (\tilde{\Theta}\cdot X_1)^{-\Delta-1} (\tilde{\Theta}\cdot X_2)^{-\Delta-1} \big\rangle_{\tilde{\Theta}} \nonumber\\[0.5em]
&= \frac{1}{\Gamma(\Delta+1)^2} \int_0^\infty ds \int_0^\infty dt \, s^\Delta t^\Delta \Big\langle e^{-s(\tilde{\Theta}\cdot X_1) - t(\tilde{\Theta}\cdot X_2)} \Big\rangle_{\tilde{\Theta}} \nonumber\\[0.5em]
&= \frac{1}{\Gamma(\Delta+1)^2} \int_0^\infty ds \int_0^\infty dt \, s^\Delta t^\Delta \Big\langle e^{\tilde{\Theta}\cdot J} \Big\rangle_{\tilde{\Theta}}
\end{align}
where we have defined the source vector $J = -(sX_1 + tX_2)$.
The Gaussian expectation value of the exponential is $\langle e^{\tilde{\Theta}\cdot J} \rangle_{\tilde{\Theta}} = e^{\frac{1}{2}J^2}$. 
Let us explicitly compute the square of the source vector $J$:
\begin{align}
J^2 &= \big( -(sX_1 + tX_2) \big)^2 \nonumber\\
&= s^2 X_1^2 + t^2 X_2^2 + 2st(X_1\cdot X_2)
\end{align}
Now we enforce the Projective Null Cone (PNC) constraints $X_1^2 = 0$ and $X_2^2 = 0$ which we can use because once the average over all the NN parameters are performed, we have Wick rotated to the Lorentzian embedding space where the notion of a non-trivial PNC makes sense:
\begin{align}
J^2 &= s^2 (0) + t^2 (0) + 2st(X_1\cdot X_2) = 2st(X_1\cdot X_2)
\end{align}
Substituting this back into the expectation value:
\begin{align}
\Big\langle e^{\tilde{\Theta}\cdot J} \Big\rangle_{\tilde{\Theta}} &= e^{\frac{1}{2} (2st(X_1\cdot X_2))} = e^{st(X_1\cdot X_2)}
\end{align}
So the full integral becomes:
\begin{align}
I_1 &= \frac{1}{\Gamma(\Delta+1)^2} \int_0^\infty ds \int_0^\infty dt \, s^\Delta t^\Delta e^{st(X_1\cdot X_2)}
\end{align}
We introduce the variables $u = st$ and $v = s/t$. 
From the definitions, we have $s = \sqrt{uv}$ and $t = \sqrt{\frac{u}{v}}$ and the integral becomes
\begin{align}
I_1 &= \frac{1}{\Gamma(\Delta+1)^2} \int_0^\infty \int_0^\infty \left( \frac{du \, dv}{2v} \right) u^\Delta e^{u(X_1\cdot X_2)} \nonumber\\[0.5em]
&= \frac{1}{2\Gamma(\Delta+1)^2} \left( \int_0^\infty \frac{dv}{v} \right) \left( \int_0^\infty du \, u^\Delta e^{u(X_1\cdot X_2)} \right)
\end{align}
The divergence is entirely in the $\int_0^\infty \frac{dv}{v}$ term. As before, we formally mod out this divergence and we are left with the finite $u$ integral. Since $X_1\cdot X_2 =-\frac{1}{2} (x_1-x_2)^2<0$, we can rewrite the exponent as $-u(-X_1\cdot X_2)$ to get a Gamma function:
\begin{align}
I_{1, \text{finite}} &= \frac{1}{2\Gamma(\Delta+1)^2} \int_0^\infty du \, u^\Delta e^{-u(-X_1\cdot X_2)} \nonumber\\[0.5em]
&= \frac{1}{2\Gamma(\Delta+1)^2} \frac{\Gamma(\Delta+1)}{(-X_1\cdot X_2)^{\Delta+1}} \nonumber\\[0.5em]
&= \frac{1}{2\Gamma(\Delta+1) (-X_1\cdot X_2)^{\Delta+1}}
\end{align}

Re-attaching the kinematic tensor structure we computed earlier, the full expectation value for the $T_1$ component is:
\begin{align}
\langle T_1 \rangle &= 8 \Big[ (Z_1\cdot Z_3)(Z_2\cdot Z_4) - (Z_1\cdot Z_4)(Z_2\cdot Z_3) \Big] \left( \frac{1}{2\Gamma(\Delta+1) (-X_1\cdot X_2)^{\Delta+1}} \right) \nonumber\\[0.5em]
&= \frac{4}{\Gamma(\Delta+1) (-X_1\cdot X_2)^{\Delta+1}} \Big[ (Z_1\cdot Z_3)(Z_2\cdot Z_4) - (Z_1\cdot Z_4)(Z_2\cdot Z_3) \Big]
\end{align}
Similarly, for $T_2$ the $\Theta$ and $\eta$ expectation value leaves us with:
\begin{align}
\langle T_2 \rangle_{\Theta, \eta} &= -4(\Delta+1) \Big[ (\tilde{\Theta}\cdot Z_1) \big( (X_1\cdot Z_3)(Z_2\cdot Z_4) - (X_1\cdot Z_4)(Z_2\cdot Z_3) \big) \nonumber\\
&\qquad\qquad\quad - (\tilde{\Theta}\cdot Z_2) \big( (X_1\cdot Z_3)(Z_1\cdot Z_4) - (X_1\cdot Z_4)(Z_1\cdot Z_3) \big) \Big] (\tilde{\Theta}\cdot X_1)^{-\Delta-2} (\tilde{\Theta}\cdot X_2)^{-\Delta-1}
\end{align}
Applying the Schwinger parameterization to both denominator factors:
\begin{align}
(\tilde{\Theta}\cdot X_1)^{-\Delta-2} &= \frac{1}{\Gamma(\Delta+2)} \int_0^\infty ds \, s^{\Delta+1} e^{-s(\tilde{\Theta}\cdot X_1)} \\
(\tilde{\Theta}\cdot X_2)^{-\Delta-1} &= \frac{1}{\Gamma(\Delta+1)} \int_0^\infty dt \, t^{\Delta} e^{-t(\tilde{\Theta}\cdot X_2)}
\end{align}
We define the fundamental integral for each polarization vector $Z_i$ ($i \in \{1,2\}$):
\begin{align}
I_2(Z_i) &:= \big\langle (\tilde{\Theta}\cdot Z_i) (\tilde{\Theta}\cdot X_1)^{-\Delta-2} (\tilde{\Theta}\cdot X_2)^{-\Delta-1} \big\rangle_{\tilde{\Theta}} \nonumber\\[0.5em]
&= \frac{1}{\Gamma(\Delta+2)\Gamma(\Delta+1)} \int_0^\infty ds \int_0^\infty dt \, s^{\Delta+1} t^\Delta \Big\langle (\tilde{\Theta}\cdot Z_i) e^{\tilde{\Theta}\cdot J} \Big\rangle_{\tilde{\Theta}}
\end{align}
where again the source vector is the same $J = -(sX_1 + tX_2)$.
\begin{align}
\Big\langle (\tilde{\Theta}\cdot Z_i) e^{\tilde{\Theta}\cdot J} \Big\rangle_{\tilde{\Theta}} &= (J\cdot Z_i) e^{\frac{1}{2}J^2} \nonumber\\[0.5em]
J^2 &= s^2 X_1^2 + t^2 X_2^2 + 2st(X_1\cdot X_2) = 2st(X_1\cdot X_2) \nonumber\\
J\cdot Z_i &= -(sX_1 + tX_2) \cdot Z_i = -s(X_1\cdot Z_i) - t(X_2\cdot Z_i) \nonumber\\[0.5em]
\Big\langle (\tilde{\Theta}\cdot Z_i) e^{\tilde{\Theta}\cdot J} \Big\rangle_{\tilde{\Theta}} &= \big( -s(X_1\cdot Z_i) - t(X_2\cdot Z_i) \big) e^{st(X_1\cdot X_2)} \nonumber\\[0.5em]
I_2(Z_i) &= \frac{1}{\Gamma(\Delta+2)\Gamma(\Delta+1)} \int_0^\infty \int_0^\infty ds \, dt \, s^{\Delta+1} t^\Delta \big( -s(X_1\cdot Z_i) - t(X_2\cdot Z_i) \big) e^{st(X_1\cdot X_2)} \nonumber\\[0.5em]
u &= st, \quad v = s/t \implies ds \, dt = \frac{du \, dv}{2v}, \quad s = \sqrt{uv}, \quad t = \sqrt{\frac{u}{v}} \nonumber\\[0.5em]
-s^{\Delta+2} t^\Delta (X_1\cdot Z_i) &= -u^{\Delta+1} v (X_1\cdot Z_i) \nonumber\\
-s^{\Delta+1} t^{\Delta+1} (X_2\cdot Z_i) &= -u^{\Delta+1} (X_2\cdot Z_i) \nonumber\\[0.5em]
I_2(Z_i) &= \frac{1}{2\Gamma(\Delta+2)\Gamma(\Delta+1)} \int_0^\infty \int_0^\infty \frac{du \, dv}{v} \Big[ -u^{\Delta+1} v (X_1\cdot Z_i) - u^{\Delta+1} (X_2\cdot Z_i) \Big] e^{u(X_1\cdot X_2)} \nonumber\\[0.5em]
&= \frac{1}{2\Gamma(\Delta+2)\Gamma(\Delta+1)} \Bigg( -(X_1\cdot Z_i) \int_0^\infty dv \int_0^\infty du \, u^{\Delta+1} e^{u(X_1\cdot X_2)} \nonumber\\
&\qquad\qquad\qquad\qquad\qquad\qquad - (X_2\cdot Z_i) \int_0^\infty \frac{dv}{v} \int_0^\infty du \, u^{\Delta+1} e^{u(X_1\cdot X_2)} \Bigg)
\end{align}

\vspace{1em}

Notice that the first integral in the final expression contains a linear divergence $\int_0^\infty dv$. However, because $Z_1$ and $Z_2$ are polarization vectors specifically defined at $X_1$, kinematic transversality strictly enforces $X_1\cdot Z_1 = 0$ and $X_1\cdot Z_2 = 0$. This gets rid of the linearly divergent term. The remaining divergence is entirely isolated in the $\int_0^\infty \frac{dv}{v}$ term, which we again formally mod out as the redundancy in Schwinger parameter space. We are left with the finite $u$ integral:
\begin{align}
I_{2, \text{finite}}(Z_i) &= \frac{-(X_2\cdot Z_i)}{2\Gamma(\Delta+2)\Gamma(\Delta+1)} \int_0^\infty du \, u^{\Delta+1} e^{-u(-X_1\cdot X_2)} \nonumber\\[0.5em]
&= \frac{-(X_2\cdot Z_i)}{2\Gamma(\Delta+2)\Gamma(\Delta+1)} \frac{\Gamma(\Delta+2)}{(-X_1\cdot X_2)^{\Delta+2}} \nonumber\\[0.5em]
&= \frac{-(X_2\cdot Z_i)}{2\Gamma(\Delta+1) (-X_1\cdot X_2)^{\Delta+2}}
\end{align}
Re-attaching the kinematic tensor structure, we replace the $\tilde{\Theta}$ integrals with $I_{2, \text{finite}}(Z_i)$. The full expectation value for the $T_2$ component is:
\begin{align}
\langle T_2 \rangle &= -4(\Delta+1) \Big[ I_{2, \text{finite}}(Z_1) \big( (X_1\cdot Z_3)(Z_2\cdot Z_4) - (X_1\cdot Z_4)(Z_2\cdot Z_3) \big) \nonumber\\
&\qquad\qquad\qquad - I_{2, \text{finite}}(Z_2) \big( (X_1\cdot Z_3)(Z_1\cdot Z_4) - (X_1\cdot Z_4)(Z_1\cdot Z_3) \big) \Big] \nonumber\\[0.5em]
&= -4(\Delta+1) \Big[ \left( \frac{-(X_2\cdot Z_1)}{2\Gamma(\Delta+1) (-X_1\cdot X_2)^{\Delta+2}} \right) \big( (X_1\cdot Z_3)(Z_2\cdot Z_4) - (X_1\cdot Z_4)(Z_2\cdot Z_3) \big) \nonumber\\
&\qquad\qquad\qquad - \left( \frac{-(X_2\cdot Z_2)}{2\Gamma(\Delta+1) (-X_1\cdot X_2)^{\Delta+2}} \right) \big( (X_1\cdot Z_3)(Z_1\cdot Z_4) - (X_1\cdot Z_4)(Z_1\cdot Z_3) \big) \Big] \nonumber\\[0.5em]
&= \frac{2(\Delta+1)}{\Gamma(\Delta+1) (-X_1\cdot X_2)^{\Delta+2}} \Big[ (X_2\cdot Z_1)(X_1\cdot Z_3)(Z_2\cdot Z_4) - (X_2\cdot Z_1)(X_1\cdot Z_4)(Z_2\cdot Z_3) \nonumber\\
&\qquad\qquad\qquad\qquad\qquad\qquad - (X_2\cdot Z_2)(X_1\cdot Z_3)(Z_1\cdot Z_4) + (X_2\cdot Z_2)(X_1\cdot Z_4)(Z_1\cdot Z_3) \Big]
\end{align}
Using similar techniques we find for $\langle T_3 \rangle$ and $\langle T_4\rangle$:
\begin{align}
\langle T_3 \rangle &= \frac{2(\Delta+1)}{\Gamma(\Delta+1) (-X_1\cdot X_2)^{\Delta+2}} \Big[ (X_1\cdot Z_3)(X_2\cdot Z_1)(Z_4\cdot Z_2) - (X_1\cdot Z_3)(X_2\cdot Z_2)(Z_4\cdot Z_1) \nonumber\\
&\qquad\qquad\qquad\qquad\qquad\qquad - (X_1\cdot Z_4)(X_2\cdot Z_1)(Z_3\cdot Z_2) + (X_1\cdot Z_4)(X_2\cdot Z_2)(Z_3\cdot Z_1) \Big] \\[1em]
\langle T_4 \rangle &= \frac{(\Delta+1)}{\Gamma(\Delta+1) (-X_1\cdot X_2)^{\Delta+2}} \Bigg( \nonumber\\
&\quad\quad \Big[ (Z_1\cdot Z_3) + \frac{(\Delta+2)(X_2\cdot Z_1)(X_1\cdot Z_3)}{(-X_1\cdot X_2)} \Big] \big( (X_1\cdot X_2)(Z_2\cdot Z_4) - (X_1\cdot Z_4)(X_2\cdot Z_2) \big) \nonumber\\
&\quad - \Big[ (Z_1\cdot Z_4) + \frac{(\Delta+2)(X_2\cdot Z_1)(X_1\cdot Z_4)}{(-X_1\cdot X_2)} \Big] \big( (X_1\cdot X_2)(Z_2\cdot Z_3) - (X_1\cdot Z_3)(X_2\cdot Z_2) \big) \nonumber\\
&\quad - \Big[ (Z_2\cdot Z_3) + \frac{(\Delta+2)(X_2\cdot Z_2)(X_1\cdot Z_3)}{(-X_1\cdot X_2)} \Big] \big( (X_1\cdot X_2)(Z_1\cdot Z_4) - (X_1\cdot Z_4)(X_2\cdot Z_1) \big) \nonumber\\
&\quad + \Big[ (Z_2\cdot Z_4) + \frac{(\Delta+2)(X_2\cdot Z_2)(X_1\cdot Z_4)}{(-X_1\cdot X_2)} \Big] \big( (X_1\cdot X_2)(Z_1\cdot Z_3) - (X_1\cdot Z_3)(X_2\cdot Z_1) \big) \Bigg)
\end{align}

Finally, we group the terms by powers of $(-X_1\cdot X_2)$. 
Let us recall the following shorthand definitions \eqref{KDiff}
\begin{align}
S &:= (Z_1\cdot Z_3)(Z_2\cdot Z_4) - (Z_1\cdot Z_4)(Z_2\cdot Z_3) \\
K_{\text{diff}} &:= (Z_1\cdot Z_3)(X_1\cdot Z_4)(X_2\cdot Z_2) + (Z_2\cdot Z_4)(X_1\cdot Z_3)(X_2\cdot Z_1) \nonumber\\
&\quad - (Z_1\cdot Z_4)(X_1\cdot Z_3)(X_2\cdot Z_2) - (Z_2\cdot Z_3)(X_1\cdot Z_4)(X_2\cdot Z_1)
\end{align}
then, the $\mathcal{O}((-X_1\cdot X_2)^{-\Delta-3})$ terms from $\langle T_4 \rangle$ are
all those terms with the coefficient $\frac{2(\Delta+1)(\Delta+2)}{\Gamma(\Delta+1)(-X_1\cdot X_2)^{\Delta+3}}$ which yields a contribution proportional to 
\begin{align}
\Big[ &(X_2\cdot Z_1)(X_1\cdot Z_3)\big(-(X_1\cdot Z_4)(X_2\cdot Z_2)\big) \nonumber\\
&\qquad - (X_2\cdot Z_1)(X_1\cdot Z_4)\big(-(X_1\cdot Z_3)(X_2\cdot Z_2)\big) \nonumber\\
&\qquad - (X_2\cdot Z_2)(X_1\cdot Z_3)\big(-(X_1\cdot Z_4)(X_2\cdot Z_1)\big) \nonumber\\
&\qquad + (X_2\cdot Z_2)(X_1\cdot Z_4)\big(-(X_1\cdot Z_3)(X_2\cdot Z_1)\big) \Big] \nonumber\\[0.5em]
&= - (X_1\cdot Z_3)(X_1\cdot Z_4)(X_2\cdot Z_1)(X_2\cdot Z_2) \nonumber\\
&\quad + (X_1\cdot Z_4)(X_1\cdot Z_3)(X_2\cdot Z_1)(X_2\cdot Z_2) \nonumber\\
&\quad + (X_1\cdot Z_3)(X_1\cdot Z_4)(X_2\cdot Z_2)(X_2\cdot Z_1) \nonumber\\
&\quad - (X_1\cdot Z_4)(X_1\cdot Z_3)(X_2\cdot Z_2)(X_2\cdot Z_1) \nonumber\\[0.5em]
&= 0
\end{align}
This is very similar to what happened for the other technique of finding the same quantity by taking derivatives of $\langle AA \rangle$. For the terms in the next order, we observe directly from the expressions of $\langle T_2 \rangle$ and $\langle T_3 \rangle$ that their contributions are
\begin{align}
\frac{2(\Delta+1)}{\Gamma(\Delta+1) (-X_1\cdot X_2)^{\Delta+2}} K_{\text{diff}}
\end{align}
and 
\begin{align}
\frac{2(\Delta+1)}{\Gamma(\Delta+1) (-X_1\cdot X_2)^{\Delta+2}} K_{\text{diff}}
\end{align}
but there are also contributions coming from $\langle T_4 \rangle$ which after an explicit expansion is found to be 

\begin{align}
\frac{-(\Delta+1)(\Delta+3)}{\Gamma(\Delta+1) (-X_1\cdot X_2)^{\Delta+2}} K_{\text{diff}}
\end{align}
Adding the  $\mathcal{O}((-X_1\cdot X_2)^{-\Delta-2})$ contributions from $T_2$, $T_3$, and $T_4$:
\begin{align}
& \frac{1}{\Gamma(\Delta+1) (-X_1\cdot X_2)^{\Delta+2}} \Big[ 2(\Delta+1) + 2(\Delta+1) - (\Delta+1)(\Delta+3) \Big] K_{\text{diff}} \nonumber\\[0.5em]
&= \frac{\Delta+1}{\Gamma(\Delta+1) (-X_1\cdot X_2)^{\Delta+2}} \Big[ 4 - (\Delta+3) \Big] K_{\text{diff}} \nonumber\\[0.5em]
&= \frac{(\Delta+1)(1-\Delta)}{\Gamma(\Delta+1) (-X_1\cdot X_2)^{\Delta+2}} K_{\text{diff}} \nonumber\\[0.5em]
&= \frac{-(\Delta^2 - 1)}{\Gamma(\Delta+1) (-X_1\cdot X_2)^{\Delta+2}} K_{\text{diff}}
\end{align}
The $\mathcal{O}((-X_1\cdot X_2)^{-\Delta-1})$ components are extracted directly from $\langle T_1 \rangle$ and $\langle T_4 \rangle$ and found to be 
\begin{align}
& \frac{1}{\Gamma(\Delta+1)(-X_1\cdot X_2)^{\Delta+1}} \Big[ 4S - 2(\Delta+1)S \Big] \nonumber\\[0.5em]
&= \frac{-2(\Delta-1)}{\Gamma(\Delta+1)(-X_1\cdot X_2)^{\Delta+1}} \,S
\end{align}
Summing all the contributions yield the complete two-point correlator:
\begin{align}
\langle F(X_1, Z_1, Z_2) F(X_2, Z_3, Z_4) \rangle &= \frac{-2(\Delta-1)}{\Gamma(\Delta+1)(-X_1\cdot X_2)^{\Delta+1}} \,S - \frac{(\Delta^2-1)}{\Gamma(\Delta+1)(-X_1\cdot X_2)^{\Delta+2}} \,K_{\text{diff}}\\
&= \frac{(-1)^{\Delta+1}}{2\Gamma(\Delta+1)} \Big[ -4(\Delta-1)(X_1\cdot X_2)^{-\Delta-1} S + 2(\Delta^2-1)(X_1\cdot X_2)^{-\Delta-2} K_{\text{diff}} \Big]
\end{align}
which is \eqref{eq:full_correlator_generic} upto an overall normalization factor which would also be identical if we kept the overall normalization when computing $\langle A(X_1,Z_1) A(X_2,Z_2)\rangle$ using the Schwinger parameterization in Appendix \ref{2ptappendix}. 

\subsubsection{Expressions in terms of $H(Z_A,Z_B;X_1,X_2)$ and normalization}\label{match}
Here we match the results of the previous appendices to the expected result \eqref{Maxwellexpectedans} by rewriting them in terms of the tensor structure $H(Z_A,Z_B;X_1,X_2)$ introduced in \eqref{newtensorstructure}. Anticipating the expected result \eqref{Maxwellexpectedans} we expand the anti-symmetrized product $$H(Z_1, Z_3 ; X_1, X_2) H(Z_2, Z_4 ; X_1, X_2) - H(Z_1, Z_4 ; X_1, X_2) H(Z_2, Z_3 ; X_1, X_2)$$ using \eqref{newtensorstructure}.
First, multiplying $H(Z_1, Z_3 ; X_1, X_2)$ and $ H(Z_2, Z_4 ; X_1, X_2)$:
\begin{align}
H(Z_1, Z_3 ; X_1, X_2) & H(Z_2, Z_4 ; X_1, X_2) \notag \\ &= 4 \big[ (X_1 \cdot X_2)(Z_1 \cdot Z_3) - (X_1 \cdot Z_3)(X_2 \cdot Z_1) \big] \big[ (X_1 \cdot X_2)(Z_2 \cdot Z_4) - (X_1 \cdot Z_4)(X_2 \cdot Z_2) \big] \notag \\
&= 4(X_1 \cdot X_2)^2 (Z_1 \cdot Z_3)(Z_2 \cdot Z_4) \notag \\
&\quad - 4(X_1 \cdot X_2) \big[ (Z_1 \cdot Z_3)(X_1 \cdot Z_4)(X_2 \cdot Z_2) + (Z_2 \cdot Z_4)(X_1 \cdot Z_3)(X_2 \cdot Z_1) \big] \notag \\
&\quad + 4(X_1 \cdot Z_3)(X_2 \cdot Z_1)(X_1 \cdot Z_4)(X_2 \cdot Z_2)
\end{align}
Similarly, for the swapped term $H(Z_1, Z_4 ; X_1, X_2) H(Z_2, Z_3 ; X_1, X_2)$ we have,
\begin{align}
H(Z_1, Z_4 ; X_1, X_2) & H(Z_2, Z_3 ; X_1, X_2) \notag \\&= 4(X_1 \cdot X_2)^2 (Z_1 \cdot Z_4)(Z_2 \cdot Z_3) \notag \\
&\quad - 4(X_1 \cdot X_2) \big[ (Z_1 \cdot Z_4)(X_1 \cdot Z_3)(X_2 \cdot Z_2) + (Z_2 \cdot Z_3)(X_1 \cdot Z_4)(X_2 \cdot Z_1) \big] \notag \\
&\quad + 4(X_1 \cdot Z_4)(X_2 \cdot Z_1)(X_1 \cdot Z_3)(X_2 \cdot Z_2)
\end{align}
Subtracting the two equations, the quartic $(X \cdot Z)$ terms at the end cancel and grouping the remaining terms, we recognize $K_{\text{diff}}$:
\begin{align}
H(Z_1, Z_3 ; X_1, X_2)& H(Z_2, Z_4 ; X_1, X_2) - H(Z_1, Z_4 ; X_1, X_2) H(Z_2, Z_3 ; X_1, X_2) \\&= 4(X_1 \cdot X_2)^2 \big[ (Z_1 \cdot Z_3)(Z_2 \cdot Z_4) - (Z_1 \cdot Z_4)(Z_2 \cdot Z_3) \big] 
- 4(X_1 \cdot X_2) K_{\text{diff}} \label{intermediatestep}
\end{align}
where $K_{\text{diff}}$ is defined in \eqref{KDiff}. Also, recognizing the term in the brackets multiplied with $-2(X_1 \cdot X_2)^{-3}$ in \eqref{FFMaxwellintermediate} as $K_{\text{diff}}$ defined in \eqref{KDiff}, the above equation can be rewritten in terms of $H(Z_i,Z_j;X_1,X_2)$ using \eqref{intermediatestep} after dividing by an overall $2(X_1 \cdot X_2)^4$ as

\begin{align}\label{FFwith2}
\langle F(X_1, Z_1, Z_2)& F(X_2, Z_3, Z_4) \nonumber \rangle \\
&= \frac{H(Z_1, Z_3 ; X_1, X_2) H(Z_2, Z_4 ; X_1, X_2)-H(Z_1, Z_4 ; X_1, X_2) H(Z_2, Z_3 ; X_1, X_2)}{2(X_1 \cdot X_2)^4}
\end{align}
which is \eqref{Maxwellexpectedans} upto numerical factors.\\

For generic $\Delta$, transposing and multiplying by $-(\Delta-1)(X_1 \cdot X_2)^{-\Delta-3}$ on both sides of \eqref{intermediatestep} we get:
\begin{align}
-4&(\Delta-1)(X_1 \cdot X_2)^{-\Delta-1} \big[ (Z_1 \cdot Z_3)(Z_2 \cdot Z_4) - (Z_1 \cdot Z_4)(Z_2 \cdot Z_3) \big] \notag \\
&= \frac{1-\Delta}{(X_1 \cdot X_2)^{\Delta+3}} \big( H(Z_1, Z_3 ; X_1, X_2) H(Z_2, Z_4 ; X_1, X_2) - H(Z_1, Z_4 ; X_1, X_2) H(Z_2, Z_3 ; X_1, X_2) \big) \notag \\
&\quad - 4(\Delta-1)(X_1 \cdot X_2)^{-\Delta-2} K_{\text{diff}}
\end{align}
Substituting this back into Eq. \eqref{eq:full_correlator_generic} we get
\begin{align}
\langle F(X_1,& Z_1, Z_2) F(X_2, Z_3, Z_4) \rangle \notag\\ &= \frac{1-\Delta}{(X_1 \cdot X_2)^{\Delta+3}} \big( H(Z_1, Z_3 ; X_1, X_2) H(Z_2, Z_4 ; X_1, X_2) - H(Z_1, Z_4 ; X_1, X_2) H(Z_2, Z_3 ; X_1, X_2) \big) \notag \\
&\quad - 4(\Delta-1)(X_1 \cdot X_2)^{-\Delta-2} K_{\text{diff}} + 2(\Delta-1)(\Delta+1)(X_1 \cdot X_2)^{-\Delta-2} K_{\text{diff}} \notag \\
&= \frac{1-\Delta}{(X_1 \cdot X_2)^{\Delta+3}} \big( H(Z_1, Z_3 ; X_1, X_2) H(Z_2, Z_4 ; X_1, X_2) - H(Z_1, Z_4 ; X_1, X_2) H(Z_2, Z_3 ; X_1, X_2) \big) \notag \\
&\quad + 2(\Delta-1)\big[(\Delta+1) - 2\big](X_1 \cdot X_2)^{-\Delta-2} K_{\text{diff}} \notag \\
&= \frac{1-\Delta}{(X_1 \cdot X_2)^{\Delta+3}} \big( H(Z_1, Z_3 ; X_1, X_2) H(Z_2, Z_4 ; X_1, X_2) - H(Z_1, Z_4 ; X_1, X_2) H(Z_2, Z_3 ; X_1, X_2) \big) \notag \\
&\quad + \frac{2(\Delta-1)^2}{(X_1 \cdot X_2)^{\Delta+2}} K_{\text{diff}}
\end{align}
which is \eqref{eq:closed_form}.\\\\
\textbf{Normalization match}\\
To reproduce the Dolan--Osborn normalized Maxwell correlators, we fix the 
overall normalization of the field by choosing
\begin{equation}\label{eq:normalizedfull}
A(X,Z)=\frac{1}{2\sqrt{2}\,\pi}\,
\lim_{\Delta\to 1}\frac{1}{\sqrt{1-\Delta}}\,
(\tilde{\Theta}\cdot X)^{-\Delta-1}
\Big[(\Theta\cdot X)(\eta\cdot Z)-(\eta\cdot X)(\Theta\cdot Z)\Big],
\end{equation}
with all parameters drawn i.i.d. from $\mathcal{N}(0,1)$ and the limit is understood to be taken after evaluating the correlators. 
With this normalization, the two-point function \eqref{eq:AA_final} becomes
\begin{equation}\label{eq:AAnorm}
\langle A(X_1,Z_1)A(X_2,Z_2)\rangle
=\frac{1}{8\pi^2}\,\frac{(-1)^{\Delta+1}}{ (1-\Delta)\Gamma(\Delta+1)}\,
(X_1\cdot X_2)^{-\Delta-1}
\Big[(X_1\cdot X_2)(Z_1\cdot Z_2)-(X_1\cdot Z_2)(X_2\cdot Z_1)\Big],
\end{equation}
where the Schwinger integral prefactor 
$\tfrac{(-1)^{\Delta+1}}{2\Gamma(\Delta+1)}$ has been retained.
Consequently, the field-strength correlator \eqref{eq:closed_form} becomes, 
after the $\Delta\to1$ rescaling in \eqref{eq:normalizedfull},
\begin{align}\label{eq:FFnorm}
\langle F(X_1,Z_1,Z_2)\,&F(X_2,Z_3,Z_4)\rangle
\\&=\frac{H(Z_1,Z_3;X_1,X_2)\,H(Z_2,Z_4;X_1,X_2)
-H(Z_1,Z_4;X_1,X_2)\,H(Z_2,Z_3;X_1,X_2)}
{16\pi^2\,(X_1\cdot X_2)^4}.
\end{align}
Projecting to the Poincar\'e section via \eqref{projections}, with 
$X_1\cdot X_2\to-\tfrac12 x_{12}^2$ and 
$H(Z_A,Z_B;X_1,X_2)\to x_{12}^2\,(z_A\cdot I\cdot z_B)$, the factors of $16$ 
cancel and this becomes
\begin{equation}\label{eq:FFphys}
\langle F_{\mu\nu}(x_1)F_{\sigma\rho}(x_2)\rangle
=\frac{1}{\pi^2 r_{12}^4}
\Big(I_{\mu\sigma}(x_{12})I_{\nu\rho}(x_{12})
-I_{\mu\rho}(x_{12})I_{\nu\sigma}(x_{12})\Big)
\end{equation}
which matches the Dolan--Osborn normalized result \cite{Dolan:2000ut}.

\subsubsection{$\langle \Phi \Phi\rangle$ from truncated ansatz}\label{sec:phiphifromtruncnatedansatz}
It is obvious from the form of the truncated ansatz
\begin{equation}
    A_M(X)=(\Theta \cdot X)^{-1} \eta_M
\end{equation}
that the $\langle \Phi \Phi\rangle$ correlator would have the correct scaling and yield an answer $\sim \frac{1}{(X_1 \cdot X_2)^4}$. To keep track of the proportionality constant we carefully normalize the fields in this section as
\begin{equation}
    A_M = \frac{1}{2 \sqrt{2}\pi}(\Theta \cdot X)^{-1} \eta_M
\end{equation}
The corresponding field strength is
\begin{equation}
    F_{MN} = -\frac{1}{2 \sqrt{2}\pi} (\Theta \cdot X)^{-2} (\Theta_M \eta_N - \Theta_N \eta_M)
\end{equation}                         
Before proceeding to find the $2$-point function $\langle \Phi \Phi\rangle$ as a sanity check with this normalization the $\langle FF\rangle$ correlator in \eqref{FFwith2} gets a contribution of $\frac{1}{8 \pi^2}$ to become 
\begin{align}\label{MaxwellFFwithcoeff}
    \langle F(X_1, Z_1, Z_2) F(X_2, Z_3, Z_4) \rangle 
    &= \frac{H(Z_1, Z_3 ; X_1, X_2) H(Z_2, Z_4 ; X_1, X_2)-H(Z_1, Z_4 ; X_1, X_2) H(Z_2, Z_3 ; X_1, X_2)}{16 \pi^2(X_1 \cdot X_2)^4}
\end{align}
Since after projecting $X_1 \cdot X_2 \to -\frac{1}{2}x_{12}^2$, the factor of $16$ cancels out to yield 
\begin{equation}
    \langle F_{\mu\nu}(x_1) F_{\sigma\rho}(x_2) \rangle =
\frac{1}{\pi^2 r_{12}^4}
\left(
I_{\mu\sigma}(x_{12}) I_{\nu\rho}(x_{12})
-
I_{\mu\rho}(x_{12}) I_{\nu\sigma}(x_{12})
\right)
\end{equation}
which is the expected Maxwell theory result with correct numerical factors \cite{Dolan:2000ut}. The composite scalar $\Phi(X) = \frac{1}{4} F_{MN}F^{MN}$ expands directly to:
\begin{equation}
    \Phi(X)  = \frac{1}{16\pi^2} (\Theta \cdot X)^{-4} \big[ \Theta^2 \eta^2 - (\Theta \cdot \eta)^2 \big]
\end{equation}
Because $\eta$ and $\Theta$ are independent Gaussian random variables in the $D$-dimensional embedding space, we can evaluate the expectation value over them separately. Starting with that over $\eta$ because it appears only in the numerator so that we can apply Wick's theorem directly we get:
\begin{align}
    \Big\langle \Theta^2 \eta^2 - (\Theta \cdot \eta)^2 \Big\rangle_\eta 
    &= \Theta^2 \langle \eta_M \eta^M \rangle_\eta - \Theta_M \Theta_N \langle \eta^M \eta^N \rangle_\eta \nonumber \\[6pt]
    &= \Theta^2 (\delta_M^M) - \Theta_M \Theta_N (\delta^{MN}) \nonumber \\[6pt]
    &= \Theta^2 D - \Theta^2 \nonumber \\[6pt]
    &= \Theta^2 (D - 1) \, .
\end{align}

Substituting this result back into the full expectation value integrates out the $\eta$ variables, leaving an operator defined strictly in terms of $\Theta$:
\begin{equation}
    \langle \Phi(X) \rangle = \frac{D-1}{16\pi^2} \Big\langle (\Theta \cdot X)^{-4} \Theta^2 \Big\rangle_\Theta \, .
\end{equation}

This is essentially the scalar architecture of \cite{Halverson:2024axc} apart from a factor of $\Theta^2$. To evaluate the remaining moment, we again employ Schwinger parameterization to exponentiate the fractional denominator, $(\Theta \cdot X)^{-4} = \frac{1}{\Gamma(4)} \int_0^\infty ds \, s^3 e^{-s(\Theta \cdot X)}$. Defining $J_M = -s X_M$  we get 
\begin{equation}
    \Big\langle (\Theta \cdot X)^{-4} \Theta^2 \Big\rangle_\Theta \nonumber 
    = \frac{1}{6} \int_0^\infty \! ds \, s^3 \Big\langle (\Theta_M \Theta^M) e^{\Theta \cdot J} \Big\rangle_\Theta \
\end{equation}
We compute the expectation value of this tensor polynomial under the shifted Gaussian measure using the generating function $\langle e^{\Theta \cdot J} \rangle = e^{\frac{1}{2}J^2}$. The insertions of $\Theta$ act as parametric derivatives $\partial / \partial J_\mu$ applied to the generating function in the usual way:
\begin{align}
    \Big\langle (\Theta_\mu \Theta^\mu) e^{\Theta \cdot J} \Big\rangle_\Theta 
    &= \frac{\partial}{\partial J_\mu} \frac{\partial}{\partial J^\mu} e^{\frac{1}{2}J^2} \nonumber \\[6pt]
    &= \frac{\partial}{\partial J_\mu} \left( J^\mu e^{\frac{1}{2}J^2} \right) \nonumber \\[6pt]
    &= \left( \frac{\partial J^\mu}{\partial J_\mu} e^{\frac{1}{2}J^2} + J^\mu \frac{\partial}{\partial J_\mu} e^{\frac{1}{2}J^2} \right) \nonumber \\[6pt]
    &= \delta_\mu^\mu e^{\frac{1}{2}J^2} + J^\mu J_\mu e^{\frac{1}{2}J^2} \nonumber \\[6pt]
    &= (D + J^2) e^{\frac{1}{2}J^2} \, .
\end{align}
Putting $
    J^2 = (-s X_\mu)(-s X^\mu) 
    = s^2 X^2 
    = 0  .
$
back into our evaluated moment yields $\langle \Theta^2 e^{\Theta \cdot J} \rangle_\Theta = D$ so that the one point function becomes 
\begin{equation}
    \langle \Phi(X)\rangle = \frac{D(D-1)}{96 \pi^2} \int_0^\infty \! ds \, s^3 \, .
\end{equation}

As a result the 1-point function formally diverges but so does the actual $1$-point function computed in Maxwell-theory and is usually made to vanish via normal ordering. We will ignore such divergent pieces in the following computations as well. We move on to discuss the two point function $\langle \Phi \Phi\rangle$:

\begin{equation}
    \langle \Phi(X_1) \Phi(X_2) \rangle = \frac{1}{256\pi^4} \Big\langle (\Theta \cdot X_1)^{-4} (\Theta \cdot X_2)^{-4} \big[ \Theta^2 \eta^2 - (\Theta \cdot \eta)^2 \big]^2 \Big\rangle
\end{equation}

Again, since the variable $\eta$ appears strictly in the numerator, we evaluate its expectation value first:
\begin{align}
    \Big\langle \big[ \Theta^2 \eta^2 - (\Theta \cdot \eta)^2 \big]^2 \Big\rangle_\eta 
    &= \Big\langle \Theta^4 \eta^4 - 2\Theta^2 \eta^2 (\Theta \cdot \eta)^2 + (\Theta \cdot \eta)^4 \Big\rangle_\eta \nonumber \\[6pt]
    &= \Theta^4 \langle \eta_M \eta^M \eta_N \eta^N \rangle_\eta 
    - 2\Theta^2 \Theta_M \Theta_N \langle \eta_P \eta^P \eta^M \eta^N \rangle_\eta 
    + \Theta_M \Theta_N \Theta_P \Theta_Q \langle \eta^M \eta^N \eta^P \eta^Q \rangle_\eta \nonumber \\[6pt]
    &= \Theta^4 \big( \delta_M^M \delta_N^N + 2\delta_{MN}\delta^{MN} \big) 
    - 2\Theta^2 \Theta_M \Theta_N \big( \delta_P^P \delta^{MN} + 2\delta_P^M \delta^{PN} \big) \nonumber \\
    &\quad + \Theta_M \Theta_N \Theta_P \Theta_Q \big( \delta^{MN}\delta^{PQ} + \delta^{MP}\delta^{NQ} + \delta^{MQ}\delta^{NP} \big) \nonumber \\[6pt]
    &= \Theta^4 \big( D \cdot D + 2D \big) 
    - 2\Theta^2 \Theta_M \Theta_N \big( D\delta^{MN} + 2\delta^{MN} \big) \nonumber \\
    &\quad + \big( (\Theta_M \Theta^M)(\Theta_P \Theta^P) + (\Theta_M \Theta^M)(\Theta_N \Theta^N) + (\Theta_M \Theta^M)(\Theta_N \Theta^N) \big) \nonumber \\[6pt]
    &= \Theta^4 (D^2 + 2D) - 2\Theta^2 (D + 2)(\Theta_M \Theta^M) + \big( \Theta^2 \Theta^2 + \Theta^2 \Theta^2 + \Theta^2 \Theta^2 \big) \nonumber \\[6pt]
    &= \Theta^4 (D^2 + 2D) - 2\Theta^4 (D + 2) + 3\Theta^4 \nonumber \notag \\
    &= \Theta^4 (D^2 - 1) \, .
\end{align}

The remaining expectation value relies entirely on $\Theta$. We evaluate it by employing Schwinger parameterization, $A^{-4} = \frac{1}{\Gamma(4)} \int_0^\infty ds \, s^3 e^{-sA}$, to exponentiate the denominators. This converts the fractional integral into the moment of a Gaussian under an exponential source vector $J_\mu = -(sX_{1\mu} + tX_{2\mu})$:
\begin{align}
    I_\Theta &:= \Big\langle (\Theta \cdot X_1)^{-4} (\Theta \cdot X_2)^{-4} \Theta^4 \Big\rangle_\Theta \nonumber \\[6pt]
    &= \frac{1}{\Gamma(4)^2} \int_0^\infty \! ds \int_0^\infty \! dt \, s^3 t^3 \Big\langle (\Theta_\mu \Theta^\mu)^2 e^{-s(\Theta \cdot X_1) - t(\Theta \cdot X_2)} \Big\rangle_\Theta \nonumber \\[6pt]
    &= \frac{1}{36} \int_0^\infty \! ds \int_0^\infty \! dt \, s^3 t^3 \Big\langle (\Theta_\mu \Theta^\mu \Theta_\nu \Theta^\nu) e^{\Theta \cdot J} \Big\rangle_\Theta \, .
\end{align}

The expectation value of this tensor polynomial under the shifted Gaussian measure is computed using the generating function $\langle e^{\Theta \cdot J} \rangle = e^{\frac{1}{2}J^2}$ and taking derivatives with respect to $J$ similar to the case of $\langle \Phi \rangle$:
\begin{align}
    \Big\langle (\Theta_\mu \Theta^\mu \Theta_\nu \Theta^\nu) e^{\Theta \cdot J} \Big\rangle_\Theta 
    &= \frac{\partial}{\partial J_\mu} \frac{\partial}{\partial J^\mu} \frac{\partial}{\partial J_\nu} \frac{\partial}{\partial J^\nu} e^{\frac{1}{2}J^2} \nonumber \\[6pt]
    &= \frac{\partial}{\partial J_\mu} \frac{\partial}{\partial J^\mu} \frac{\partial}{\partial J_\nu} \left( J^\nu e^{\frac{1}{2}J^2} \right) \nonumber \\[6pt]
    &= \frac{\partial}{\partial J_\mu} \frac{\partial}{\partial J^\mu} \Big[ (D + J^2) e^{\frac{1}{2}J^2} \Big] \nonumber \\[6pt]
    &= \frac{\partial}{\partial J_\mu} \Big[ (D + 2 + J^2) J_\mu e^{\frac{1}{2}J^2} \Big] \nonumber \\[6pt]
    &= \Big[ D(D+2) + 2(D+2)J^2 + (J_\mu J^\mu)^2 \Big] e^{\frac{1}{2}J^2} \, .
\end{align}
Again because the insertion points $X_1$ and $X_2$ lie on the physical null cone ($X_1^2 = 0$ and $X_2^2 = 0$), the square of the source vector simplifies to:
\begin{align}
    J^2 &= J_\mu J^\mu = (sX_{1\mu} + tX_{2\mu})(sX_1^\mu + tX_2^\mu) \nonumber \\
    &= s^2 X_1^2 + t^2 X_2^2 + 2st(X_1 \cdot X_2) \nonumber \\
    &= 2st(X_1 \cdot X_2) \, ,
\end{align}
We substitute $J^2$ back into the generating polynomial and perform the change of integration variables $u = st$ and $v = s/t$. This introduces the measure $ds \, dt = \frac{du \, dv}{2v}$:
\begin{align}
    I_\Theta &= \frac{1}{36} \int_0^\infty \frac{dv}{2v} \int_0^\infty \! du \, u^3 \Big[ D(D+2) + 4(D+2)(X_1 \cdot X_2) u + 4(X_1 \cdot X_2)^2 u^2 \Big] e^{(X_1 \cdot X_2) u} \nonumber \\[6pt]
    &\cong \frac{1}{72} \Bigg( D(D+2) \int_0^\infty \! du \, u^3 e^{(X_1 \cdot X_2) u} + 4(D+2)(X_1 \cdot X_2) \int_0^\infty \! du \, u^4 e^{(X_1 \cdot X_2) u}  \notag \\  
    &\quad\quad\quad\quad\quad\quad +~  4(X_1 \cdot X_2)^2 \int_0^\infty \! du \, u^5 e^{(X_1 \cdot X_2) u} \Bigg) \nonumber \\[6pt]
    &= \frac{1}{72} \Bigg( D(D+2)\frac{\Gamma(4)}{(X_1 \cdot X_2)^4} - 4(D+2)(X_1 \cdot X_2)\frac{\Gamma(5)}{(X_1 \cdot X_2)^5} + 4(X_1 \cdot X_2)^2\frac{\Gamma(6)}{(X_1 \cdot X_2)^6} \Bigg) \nonumber \\[6pt]
    &= \frac{1}{72(X_1 \cdot X_2)^4} \Big( 6D(D+2) - 96(D+2) + 480 \Big) \nonumber \\[6pt]
    &= \frac{(D-6)(D-8)}{12(X_1 \cdot X_2)^4} \, .
\end{align}
where in the 2nd step we have dropped the divergent $v$ integral due to reasons discussed in Appendix \ref{FFfromFF}. So the two point correlator is 
\begin{equation}
    \langle \Phi(X_1) \Phi(X_2) \rangle = \frac{(D^2 - 1)(D-6)(D-8)}{3072\pi^4 (X_1 \cdot X_2)^4} \, .
\end{equation}
For $d=4$ i.e. $D=6$ the above result vanishes unlike in pure Maxwell theory \cite{Dolan:2000ut}. If the higher point correlators also vanished then this would describe a TQFT, but that is unlikely to happen.

\subsubsection{$\langle \Phi \Phi \rangle$ from primary $F$}\label{sec:phiphifromprimaryF}

In this section, we compute the two-point correlator $\langle \Phi(X_1) \Phi(X_2) \rangle$ using the truncated neural network ansatz but a primary $F$. The non-primary $A$ and $F$ as in the previous subsection are:
\begin{equation}
\begin{aligned}
    A_M &= \frac{1}{2\sqrt{2}\pi} (\Theta \cdot X)^{-1} \eta_M \, , \\
    F_{MN} &= -\frac{1}{2\sqrt{2}\pi} (\Theta \cdot X)^{-2} (\Theta_M \eta_N - \Theta_N \eta_M) \, .
\end{aligned}
\end{equation}
Its contraction with the position vector yields a non-zero longitudinal vector $V_N$:
\begin{equation}
    V_N := X^M F_{MN} = -\frac{1}{2\sqrt{2}\pi} (\Theta \cdot X)^{-2} \big[ (\Theta \cdot X)\eta_N - (\eta \cdot X)\Theta_N \big] \neq 0 \, .
\end{equation}

For a primary $F$ we must extract a strictly transverse representative. On the null cone where $X^2 = 0$ there is no projector entirely out of $X$ to achieve this \footnote{The most general rank-(1,1) tensor constructed solely from $X$ and the metric is $P^M_{\;\;N} = a \delta^M_{\;\;N} + b X^M X_N$. Transversality and the PNC condition $X^2=0$ implies $a=0$. The remaining tensor $P^M_{\;\;N} = b X^M X_N$ yields $(P^2)^M_{\;\;N} = b^2 X^M X^2 X_N = 0$ which on imposing $P^2 = P$, implies $P=0$.}. But one can introduce a preferred embedding space vector $\bar{X}^M$ satisfying $X \cdot \bar{X} = -1$ and $\bar{X}^2 = 0$ which gives a transverse representative:
\begin{equation}
    F^{\perp}_{MN} = F_{MN} + \bar{X}_M V_N - \bar{X}_N V_M \, .
\end{equation}
One can check that transversality is automatically restored:
\begin{equation}
    X^M F^{\perp}_{MN} = X^M F_{MN} + (X \cdot \bar{X})V_N - (X \cdot V)\bar{X}_N = V_N - V_N - 0 = 0 \, .
\end{equation}
So we have traded the non-transversality of $F$ for a preferred vector $\bar{X}$. This is equivalent to restricting to the original $F$ but raising with $K^{MN}$ of Costa et al. \cite{Costa:2011mg}. Since we are interested in a $(0,2)$ tensor we define
\begin{equation}
    K_M^{\;\;A} := \delta_M^A + X_M \bar{X}^A + \bar{X}_M X^A \, .
\end{equation}
Using this we can define a strictly transverse field strength via $\hat{F}_{MN} = K_M^{\;\;A} K_N^{\;\;B} F_{AB}$. Expanding the projector we find
\begin{align}
    \hat{F}_{MN} &= (\delta_M^A + X_M \bar{X}^A + \bar{X}_M X^A)(\delta_N^B + X_N \bar{X}^B + \bar{X}_N X^B) F_{AB} \nonumber \\
    &= F_{MN} + \bar{X}_M (X^A F_{AN}) + X_M (\bar{X}^A F_{AN}) + \bar{X}_N (X^B F_{MB}) + X_N (\bar{X}^B F_{MB}) \nonumber \\
    &\quad + X_M \bar{X}_N (\bar{X}^A X^B F_{AB}) + \bar{X}_M X_N (X^A \bar{X}^B F_{AB}) \nonumber \\
    &= F_{MN} + \bar{X}_M V_N - \bar{X}_N V_M + X_M (\bar{X}^A F_{AN}) - X_N (\bar{X}^B F_{BM}) \nonumber \\
    &\quad + X_M \bar{X}_N (\bar{X} \cdot V) - \bar{X}_M X_N (\bar{X} \cdot V) \nonumber \\
    &= F^{\perp}_{MN} + X_M \big( \bar{X}^A F_{AN} + \bar{X}_N (\bar{X} \cdot V) \big) - X_N \big( \bar{X}^B F_{BM} + \bar{X}_M (\bar{X} \cdot V) \big) \, .
\end{align}
Thus, $\hat{F}_{MN}$ equals $F^{\perp}_{MN}$ modulo pure gauge terms proportional to $X_M$ and $X_N$: Constructing the scalar primary by tracing the original $F$ with the modified metric $K^{AB}$ is therefore mathematically equivalent to tracing the explicitly projected tensor $\hat{F}_{MN}$ with the standard flat embedding metric $\eta^{MN}$. Importantly, this does not affect the correlators of $F$ as we now show:\\
\textbf{1-point function}\\
$$\begin{aligned}
\langle F ^\perp_{MN} \rangle &= \big\langle F_{MN} + \bar{X}_M V_N - \bar{X}_N V_M \big\rangle \\
&= \langle F_{MN} \rangle + \bar{X}_M \langle V_N \rangle - \bar{X}_N \langle V_M \rangle \\[6pt]
\text{where} \quad \langle V_N \rangle &= \langle X^A F_{AN} \rangle = X^A \langle F_{AN} \rangle \\
&= X^A \left\langle -\frac{1}{2\sqrt{2}\pi} (\Theta \cdot X)^{-2} (\Theta_A \eta_N - \Theta_N \eta_A) \right\rangle \\[8pt]
&= -\frac{1}{2\sqrt{2}\pi} \Big[ \big\langle (\Theta \cdot X)^{-2} (\Theta \cdot X) \big\rangle_\Theta \langle \eta_N \rangle_\eta - \big\langle (\Theta \cdot X)^{-2} \Theta_N \big\rangle_\Theta X^A \langle \eta_A \rangle_\eta \Big] \\[8pt]
&= 0 \quad \big(\text{since } \langle \eta_N \rangle_\eta = 0 \text{ and } \langle \eta_A \rangle_\eta = 0 \big) \\[12pt]
\Rightarrow \langle F^\perp_{MN} \rangle &= \langle F_{MN} \rangle + \bar{X}_M (0) - \bar{X}_N (0)=  \langle F_{MN} \rangle
\end{aligned}$$
\textbf{2-point function}
\begin{align}
\langle F^\perp_{MN} F^\perp_{PQ} \rangle &= \big\langle \left( F_{MN} + \bar{X}_M V_N - \bar{X}_N V_M \right) \left( F_{PQ} + \bar{X}_P V_Q - \bar{X}_Q V_P \right) \big\rangle \notag \\
&= \langle F_{MN} F_{PQ} \rangle ]\notag \\
&\quad + \bar{X}_P \langle F_{MN} V_Q \rangle - \bar{X}_Q \langle F_{MN} V_P \rangle \notag \\
&\quad + \bar{X}_M \langle V_N F_{PQ} \rangle - \bar{X}_N \langle V_M F_{PQ} \rangle \notag \\
&\quad + \bar{X}_M \bar{X}_P \langle V_N V_Q \rangle - \bar{X}_M \bar{X}_Q \langle V_N V_P \rangle - \bar{X}_N \bar{X}_P \langle V_M V_Q \rangle + \bar{X}_N \bar{X}_Q \langle V_M V_P \rangle \label{FcaretFcaret}
\end{align}
Now we show why each of these terms except the first one vanishes: Recall that $\langle F_{MN}(X_1) F_{PQ}(X_2) \rangle$ is proportional to $(H_{MP}H_{NQ} - H_{MQ}H_{NP})$ on physical slices (cf. \ref{match}), so:

$$\begin{aligned}
X_1^M \langle F_{MN}(X_1) F_{PQ}(X_2) \rangle &\propto X_1^M \big( H_{MP} H_{NQ} - H_{MQ} H_{NP} \big) \\
&= (X_1^M H_{MP}) H_{NQ} - (X_1^M H_{MQ}) H_{NP} \\
&= 0
\end{aligned}$$
where we used 
$$\begin{aligned}
X_1^M H_{MN}(X_1, X_2) &= X_1^M \big( -2 (X_1 \cdot X_2) \eta_{MN} + 2 X_{2M} X_{1N} \big) \\
&= -2 (X_1 \cdot X_2) X_{1N} + 2 (X_1 \cdot X_2) X_{1N} \\
&= 0
\end{aligned}$$
Therefore, $\langle V_N(X_1) F_{PQ}(X_2) \rangle = 0$. Also note that
$$\langle V_N(X_1) V_Q(X_2) \rangle = X_1^M X_2^P \langle F_{MN}(X_1) F_{PQ}(X_2) \rangle$$ so by $\langle V_N(X_1) F_{PQ}(X_2) \rangle = 0$ the $\langle VV \rangle$ correlator also vanishes. Therefore \eqref{FcaretFcaret} implies 
$$\langle \hat{F}_{MN} \hat{F}_{PQ} \rangle= \langle F_{MN}F_{PQ} \rangle$$
Returning back to our expression for $\hat{F}_{MN}$, we can distribute the effective metrics onto the NN parameter vectors, defining $\hat{\Theta}_M := K_M^{\;\;A} \Theta_A$ and $\hat{\eta}_N := K_N^{\;\;B} \eta_B$:
\begin{equation}
\begin{aligned}
    \hat{\Theta}_M &= \Theta_M + X_M(\bar{X} \cdot \Theta) + \bar{X}_M(X \cdot \Theta) \, , \\
    \hat{\eta}_N &= \eta_N + X_N(\bar{X} \cdot \eta) + \bar{X}_N(X \cdot \eta) \, .
\end{aligned}
\end{equation}
Substituting these into the field strength we get 
\begin{align}
    \hat{F}_{MN} &:= K_M^{\;\;A} K_N^{\;\;B} F_{AB} \nonumber \\[6pt]
    &= K_M^{\;\;A} K_N^{\;\;B} \left( -\frac{1}{2\sqrt{2}\pi} (\Theta \cdot X)^{-2} (\Theta_A \eta_B - \Theta_B \eta_A) \right) \nonumber \\[6pt]
    &= -\frac{1}{2\sqrt{2}\pi} (\Theta \cdot X)^{-2} \Big( K_M^{\;\;A} K_N^{\;\;B} \Theta_A \eta_B - K_M^{\;\;A} K_N^{\;\;B} \Theta_B \eta_A \Big) \nonumber \\[6pt]
    &= -\frac{1}{2\sqrt{2}\pi} (\Theta \cdot X)^{-2} \Big( (K_M^{\;\;A} \Theta_A)(K_N^{\;\;B} \eta_B) - (K_N^{\;\;B} \Theta_B)(K_M^{\;\;A} \eta_A) \Big) \nonumber \\[6pt]
    &= -\frac{1}{2\sqrt{2}\pi} (\Theta \cdot X)^{-2} \big( \hat{\Theta}_M \hat{\eta}_N - \hat{\Theta}_N \hat{\eta}_M \big) \, .
\end{align}
so that we can construct $\Phi(X)$ as follows
\begin{align}\label{eq:PhiThetaeta}
    \Phi(X) &:= \frac{1}{4} \hat{F}_{MN} \hat{F}^{MN} \nonumber \\[6pt]
    &= \frac{1}{4} \left( -\frac{1}{2\sqrt{2}\pi} (\Theta \cdot X)^{-2} (\hat{\Theta}_M \hat{\eta}_N - \hat{\Theta}_N \hat{\eta}_M) \right) \left( -\frac{1}{2\sqrt{2}\pi} (\Theta \cdot X)^{-2} (\hat{\Theta}^M \hat{\eta}^N - \hat{\Theta}^N \hat{\eta}^M) \right) \nonumber \\[6pt]
    &= \frac{1}{32\pi^2} (\Theta \cdot X)^{-4} \Big( \hat{\Theta}_M \hat{\eta}_N \hat{\Theta}^M \hat{\eta}^N - \hat{\Theta}_M \hat{\eta}_N \hat{\Theta}^N \hat{\eta}^M - \hat{\Theta}_N \hat{\eta}_M \hat{\Theta}^M \hat{\eta}^N + \hat{\Theta}_N \hat{\eta}_M \hat{\Theta}^N \hat{\eta}^M \Big) \nonumber \\[6pt]
    &= \frac{1}{32\pi^2} (\Theta \cdot X)^{-4} \Big( \hat{\Theta}^2 \hat{\eta}^2 - (\hat{\Theta} \cdot \hat{\eta})^2 - (\hat{\Theta} \cdot \hat{\eta})^2 + \hat{\eta}^2 \hat{\Theta}^2 \Big) \nonumber \\[6pt]
    &= \frac{1}{16\pi^2} (\Theta \cdot X)^{-4} \Big[ \hat{\Theta}^2 \hat{\eta}^2 - (\hat{\Theta} \cdot \hat{\eta})^2 \Big] \,.
\end{align}
To express this completely in terms of the fundamental variables, we expand the inner products utilizing the geometric constraints $X_i^2 = 0$, $\bar{X}_i^2 = 0$, and $X_i \cdot \bar{X}_i = -1$:
\begin{align}
    \hat{\Theta}^2 &= \big( \Theta + X_i(\bar{X}_i \cdot \Theta) + \bar{X}_i(X_i \cdot \Theta) \big)^2 \nonumber \\
    &= \Theta^2 + 2(X_i \cdot \Theta)(\bar{X}_i \cdot \Theta) + 2(\bar{X}_i \cdot \Theta)(X_i \cdot \Theta) - 2(X_i \cdot \Theta)(\bar{X}_i \cdot \Theta) \nonumber \\
    &= \Theta^2 + 2(X_i \cdot \Theta)(\bar{X}_i \cdot \Theta) \label{eq:ThetaHatSq} \, .
\end{align}
Similarly, the square of $\hat{\eta}$ simplifies to:
\begin{equation}
    \hat{\eta}^2 = \eta^2 + 2(X_i \cdot \eta)(\bar{X}_i \cdot \eta) \label{eq:EtaHatSq} \, .
\end{equation}
Expanding the cross-term and canceling the pairwise redundancies yields:
\begin{align}
    \hat{\Theta} \cdot \hat{\eta} &= \big( \Theta + X_i(\bar{X}_i \cdot \Theta) + \bar{X}_i(X_i \cdot \Theta) \big) \cdot \big( \eta + X_i(\bar{X}_i \cdot \eta) + \bar{X}_i(X_i \cdot \eta) \big) \nonumber \\
    &= (\Theta \cdot \eta) + (X_i \cdot \Theta)(\bar{X}_i \cdot \eta) + (\bar{X}_i \cdot \Theta)(X_i \cdot \eta) \label{eq:ThetaEtaCross} \, .
\end{align}
Substituting equations \eqref{eq:ThetaHatSq}, \eqref{eq:EtaHatSq}, and \eqref{eq:ThetaEtaCross} back into \eqref{eq:PhiThetaeta} yields:
\begin{align}
    \Phi(X_i) &= \frac{1}{16\pi^2} (\Theta \cdot X_i)^{-4} \Big[ \hat{\Theta}^2 \hat{\eta}^2 - (\hat{\Theta} \cdot \hat{\eta})^2 \Big] \nonumber \\[6pt]
    &= \frac{1}{16\pi^2} (\Theta \cdot X_i)^{-4} \Bigg( \Big[ \Theta^2 + 2(X_i \cdot \Theta)(\bar{X}_i \cdot \Theta) \Big] \Big[ \eta^2 + 2(X_i \cdot \eta)(\bar{X}_i \cdot \eta) \Big] \nonumber \\
    &\qquad\qquad\qquad\qquad - \Big[ (\Theta \cdot \eta) + (X_i \cdot \Theta)(\bar{X}_i \cdot \eta) + (\bar{X}_i \cdot \Theta)(X_i \cdot \eta) \Big]^2 \Bigg) \nonumber \\[6pt]
    &= \frac{1}{16\pi^2} (\Theta \cdot X_i)^{-4} Q(X_i) \, ,
\end{align}
where the polynomial $Q(X_i)$ is defined as:
\begin{equation}
\begin{aligned}
    Q(X_i) &:= \Big[ \Theta^2 + 2(X_i \cdot \Theta)(\bar{X}_i \cdot \Theta) \Big] \Big[ \eta^2 + 2(X_i \cdot \eta)(\bar{X}_i \cdot \eta) \Big] \\
    &\quad - \Big[ (\Theta \cdot \eta) + (X_i \cdot \Theta)(\bar{X}_i \cdot \eta) + (\bar{X}_i \cdot \Theta)(X_i \cdot \eta) \Big]^2 \, .
\end{aligned}
\end{equation}
To compute the two-point function $\langle \Phi(X_1) \Phi(X_2) \rangle$, we choose 
\begin{equation}
    \bar{X}_1 = \frac{X_2}{\alpha} \quad \text{and} \quad \bar{X}_2 = \frac{X_1}{\alpha} \, ,
\end{equation}
where $\alpha = -X_1 \cdot X_2$. Under this choice, the expressions for the effective metrics at both points become identical: recall $K_i^{AB} = \delta^{AB} + X_i^A \bar{X}_i^B + \bar{X}_i^A X_i^B$ at each insertion point, so that substituting $\bar{X}_1 = X_2/\alpha$ yields:
\begin{equation}
    K_1^{AB} = \delta^{AB} + X_1^A \left(\frac{X_2^B}{\alpha}\right) + \left(\frac{X_2^A}{\alpha}\right) X_1^B = \delta^{AB} + \frac{1}{\alpha}\big(X_1^A X_2^B + X_2^A X_1^B\big) \, .
\end{equation}
Similarly, at point 2, substituting $\bar{X}_2 = X_1/\alpha$ gives:
\begin{equation}
    K_2^{AB} = \delta^{AB} + X_2^A \left(\frac{X_1^B}{\alpha}\right) + \left(\frac{X_1^A}{\alpha}\right) X_2^B = \delta^{AB} + \frac{1}{\alpha}\big(X_2^A X_1^B + X_1^A X_2^B\big) \, .
\end{equation}
Therefore we have $K_1^{AB} = K_2^{AB} := K^{AB}$. Now, we can rewrite the constituent brackets of $Q(X_i)$ directly in terms of the effective metric $K_i^{AB}$. For instance, factoring out the parameters $\Theta$ from the first bracket yields:
\begin{equation}
    \Theta^2 + 2(X_i \cdot \Theta)(\bar{X}_i \cdot \Theta) = \Theta_A \delta^{AB} \Theta_B + \Theta_A \big( X_i^A \bar{X}_i^B + \bar{X}_i^A X_i^B \big) \Theta_B = \Theta_A K_i^{AB} \Theta_B := (\Theta K_i \Theta) \, .
\end{equation}
Similarly, the second bracket becomes $(\eta K_i \eta)$. Finally, the cross-term inside the squared bracket factors as:
\begin{equation}
    (\Theta \cdot \eta) + (X_i \cdot \Theta)(\bar{X}_i \cdot \eta) + (\bar{X}_i \cdot \Theta)(X_i \cdot \eta) = \Theta_A \big( \delta^{AB} + X_i^A \bar{X}_i^B + \bar{X}_i^A X_i^B \big) \eta_B := (\Theta K_i \eta) \, .
\end{equation}
Substituting these traced forms back into $Q(X_i)$, and utilizing the fact that $K_1 = K_2 = K$, the polynomial at either insertion point simplifies to
\begin{equation}
    Q = (\Theta K \Theta)(\eta K \eta) - (\Theta K \eta)^2 \, .
\end{equation}
where we have omitted the indices for simplicity. 
The full two-point expectation value becomes
\begin{equation}
    \langle \Phi(X_1) \Phi(X_2) \rangle = \frac{1}{256\pi^4} \left\langle \frac{Q^2}{(\Theta \cdot X_1)^4 (\Theta \cdot X_2)^4} \right\rangle_{\Theta, \eta} \, .
\end{equation}

We evaluate the expectation value over $\eta$ by expanding $Q^2$ and applying Wick's theorem for $\eta$ (where $\langle \eta_M \eta_N \rangle = \delta_{MN}$) and idempotence of $K$
\begin{align}
    \langle Q^2 \rangle_\eta &= \Big\langle \big[ (\Theta K \Theta)(\eta K \eta) - (\Theta K \eta)^2 \big]^2 \Big\rangle_\eta \nonumber \\[6pt]
    &= (\Theta K \Theta)^2 \langle (\eta K \eta)^2 \rangle_\eta - 2(\Theta K \Theta) \langle (\eta K \eta)(\Theta K \eta)^2 \rangle_\eta + \langle (\Theta K \eta)^4 \rangle_\eta \nonumber \\[6pt]
    &= (\Theta K \Theta)^2 \Big( \text{Tr}(K)^2 + 2\text{Tr}(K^2) \Big) - 2(\Theta K \Theta) \Big( \text{Tr}(K)(\Theta K \Theta) + 2(\Theta K K K \Theta) \Big) + 3(\Theta K \Theta)^2 \nonumber \\[6pt]
    &= (\Theta K \Theta)^2 \big( d^2 + 2d \big) - 2(\Theta K \Theta) \big( d(\Theta K \Theta) + 2(\Theta K \Theta) \big) + 3(\Theta K \Theta)^2 \nonumber \\[6pt]
    &= (\Theta K \Theta)^2 (d^2 + 2d - 2d - 4 + 3) \nonumber \\[6pt]
    &= (\Theta K \Theta)^2 (d^2 - 1) \, .
\end{align}

We substitute the expectation value over $\eta$ back into the full correlator and evaluate the remaining expectation value over $\Theta$ utilizing Schwinger parameterization. By introducing the source vector $J=-(sX_1+tX_2)$ as usual and noting that $K^{AB} J_B = 0$ we see that all $J$-dependent cross-terms vanish as follows
\begin{align}
    \langle \Phi(X_1) \Phi(X_2) \rangle &= \frac{d^2 - 1}{256\pi^4} \left\langle \frac{(\Theta_A K^{AB} \Theta_B)^2}{(\Theta \cdot X_1)^4 (\Theta \cdot X_2)^4} \right\rangle \nonumber \\[6pt]
    &= \frac{d^2 - 1}{256\pi^4 \, \Gamma(4)^2} \int_0^\infty \! ds \int_0^\infty \! dt \, s^3 t^3 \left\langle (\Theta_A K^{AB} \Theta_B)^2 e^{-s(\Theta \cdot X_1) - t(\Theta \cdot X_2)} \right\rangle \nonumber \\[6pt]
    &= \frac{d^2 - 1}{9216\pi^4} \int_0^\infty \! ds \int_0^\infty \! dt \, s^3 t^3 \left\langle (\Theta_A K^{AB} \Theta_B)^2 e^{\Theta \cdot J} \right\rangle \quad  \nonumber \\[6pt]
    &= \frac{d^2 - 1}{9216\pi^4} \int_0^\infty \! ds \int_0^\infty \! dt \, s^3 t^3 \, e^{\frac{1}{2}J^2} \Big\langle \big( (\Theta + J)_A K^{AB} (\Theta + J)_B \big)^2 \Big\rangle \nonumber \\[6pt]
    &= \frac{d^2 - 1}{9216\pi^4} \int_0^\infty \! ds \int_0^\infty \! dt \, s^3 t^3 \, e^{st(X_1 \cdot X_2)} \Big\langle \big( \Theta_A K^{AB} \Theta_B + 2 \Theta_A K^{AB} J_B + J_A K^{AB} J_B \big)^2 \Big\rangle \nonumber \\[6pt]
    &= \frac{d^2 - 1}{9216\pi^4} \int_0^\infty \! ds \int_0^\infty \! dt \, s^3 t^3 \, e^{st(X_1 \cdot X_2)} \Big\langle \big( \Theta_A K^{AB} \Theta_B + 0 + 0 \big)^2 \Big\rangle \nonumber \\[6pt]
    &= \frac{d^2 - 1}{9216\pi^4} \int_0^\infty \! ds \int_0^\infty \! dt \, s^3 t^3 \, e^{st(X_1 \cdot X_2)} \langle (\Theta_A K^{AB} \Theta_B) (\Theta_C K^{CD} \Theta_D) \rangle \, .
\end{align}
The integrand contains
\begin{align}
    \langle (\Theta_A K^{AB} \Theta_B) (\Theta_C K^{CD} \Theta_D) \rangle_0 &= K^{AB} K^{CD} \langle \Theta_A \Theta_B \Theta_C \Theta_D \rangle_0 \nonumber \\[6pt]
    &= K^{AB} K^{CD} \big( \delta_{AB} \delta_{CD} + \delta_{AC} \delta_{BD} + \delta_{AD} \delta_{BC} \big) \nonumber \\[6pt]
    &= K^A_{\;\;A} K^C_{\;\;C} + K^{AB} K_{AB} + K^{AB} K_{BA} \nonumber \\[6pt]
    &= K^A_{\;\;A} K^C_{\;\;C} + 2 K^{AB} K_{AB} \, ,
\end{align}
where the last line follows from the symmetry $K^{AB} = K^{BA}$. The trace of $K$ is 
\begin{align}
    K^A_{\;\;A} &= \delta^A_{\;\;A} + \frac{1}{\alpha} \big( X_1^A X_{2A} + X_2^A X_{1A} \big) \nonumber \\
    &= (d + 2) + \frac{1}{\alpha} \big( (-\alpha) + (-\alpha) \big) \nonumber \\
    &= d + 2 - 2 \nonumber \\
    &= d \, .
\end{align}
Furthermore, because $K$ is an idempotent projector, it satisfies $K^{AB} K_{BC} = K^A_{\;\;C}$, which immediately implies that its full contraction is equal to its trace: $K^{AB} K_{AB} = K^A_{\;\;A} = d$. Substituting these scalar values back into the Wick contraction gives $
    K^A_{\;\;A} K^C_{\;\;C} + 2K^{AB} K_{AB} = (d)(d) + 2(d) = d(d+2) $
Substituting this back we get
\begin{align}
    \langle \Phi(X_1) \Phi(X_2) \rangle &= \frac{d^2 - 1}{256\pi^4 \, \Gamma(4)^2} \int_0^\infty \! ds \int_0^\infty \! dt \, s^3 t^3 \left\langle (\Theta K \Theta)^2 e^{\Theta \cdot J} \right\rangle_\Theta \nonumber \\[6pt]
    &= \frac{d^2 - 1}{256\pi^4 (36)} \int_0^\infty \! ds \int_0^\infty \! dt \, s^3 t^3 \big( d(d+2) \big) e^{st(X_1 \cdot X_2)} \nonumber \\[6pt]
    &= \frac{(d^2 - 1)d(d+2)}{9216\pi^4} \int_0^\infty \frac{dv}{2v} \int_0^\infty \! du \, u^3 e^{-u\alpha} \nonumber \\[6pt]
    &\cong \frac{d(d - 1)(d + 1)(d + 2)}{18432\pi^4} \left( \frac{\Gamma(4)}{\alpha^4} \right) \nonumber \\[6pt]
    &= \frac{d(d - 1)(d + 1)(d + 2)}{3072\pi^4 \, \alpha^4} \, .
\end{align}
Evaluating this explicitly at $d=4$, we find:
\begin{equation}
    \langle \Phi(X_1) \Phi(X_2) \rangle = \frac{15}{128\pi^4 \, \alpha^4} \, .
\end{equation}
Translating from the embedding dot product back to the physical spacetime distance using $\alpha = -X_1 \cdot X_2 = r_{12}^2 / 2 \implies \alpha^4 = r_{12}^8 / 16$, the final result is 
\begin{equation}
    \langle \Phi(x_1) \Phi(x_2) \rangle = \frac{15}{128\pi^4 \left(\frac{r_{12}^8}{16}\right)} = \frac{15}{8\pi^4 \, r_{12}^8} \, .
\end{equation}
which is off from the true Maxwell result \cite{Dolan:2000ut}
\begin{equation}\label{eq:trueMaxwell2ptPhiPhi}
     \langle \Phi(x_1) \Phi(x_2) \rangle = \frac{3}{\pi^4 \, r_{12}^8} \, .
\end{equation}
by a factor of $\frac{5}{8}$. Thus although the $\langle FF \rangle$ of our NN architecture matches exactly that of the Maxwell CFT, this doesn't hold for the two-point function $\langle \Phi \Phi \rangle$ even after making $F$ a primary. The discrepancy appears from the missing $(d+1)(d+2)$ terms in the correct Maxwell result (see e.g. \eqref{eq:ddimPhiPhimaxwell}). These factors are precisely an artifact of the Non-Gaussianities which will be suppressed in the large-N technique discussed in the next section.

\subsubsection{Correlators from ``large-$N$ methods''}\label{sec:largeN}

While the neural network architecture proposed in Section \ref{Maxwellsection} successfully reproduces the fundamental 2-point correlator of the free Maxwell field strength $F_{MN}$, as shown in earlier sections, computing higher-point functions or correlators of composite operators directly from the single-channel architecture can introduce non-Gaussian artifacts arising from the architecture's non-linear dependence on the NN parameters (which was necessary to obtain the correct CFT tensor structures). However, to recover generalized free field behavior (and thereby genuine Maxwell theory), we can employ a large $N$ limit technique as also discussed for the case of scalars in \cite{Halverson:2024axc}. The essential idea of applying NNGP in this case is that if we take a countably infinite number of i.i.d. copies of our architecture and sum them up, then a version of the central limit theorem guarantees that the output statistics converge to those of a Gaussian random field as $N \to \infty$, irrespective of the non-Gaussianities present in any individual channel. Hence, we upgrade the single-channel ansatz $A(X,Z)$ to an ensemble of $N$ independent, identically distributed (i.i.d.) channels:
\begin{equation}\label{eq:largeNansatzappendix}
    A^{(N)}_M(X) = \frac{1}{\sqrt{N}} \sum_{c=1}^N A^{(c_M)}_M(X)
\end{equation}
where $M$ labels different fields, and $c_M$ labels different members of the Gaussian ensemble. Our discussion here is independent of the single channel architecture for $A(X,Z)$ that is being used as long as it produces the correct field strength $1$-point and $2$-point correlators. Each channel is initialized with independent Gaussian parameters centered at zero, $\langle A^{(c_M)} \rangle = 0$, and the channels are uncorrelated such that $\langle A^{(c_1)}(X_1) A^{(c_2)}(X_2) \rangle = \delta^{c_1 c_2} \langle A(X_1) A(X_2) \rangle$. 

When computing the 4-point correlator of this field, we explicitly expand the multi-channel sum. The expectation value of four independent zero-mean variables vanishes unless the channel indices pair up in either self-pairs ($c_1 = c_2 \neq c_3 = c_4$) or all-equal sets ($c_1 = c_2 = c_3 = c_4$):
\begin{align}
    \langle A^{(N)}_1 A^{(N)}_2 A^{(N)}_3 A^{(N)}_4 \rangle &= \frac{1}{N^2} \sum_{c_1, c_2, c_3, c_4 = 1}^N \langle A^{(c_1)}_1 A^{(c_2)}_2 A^{(c_3)}_3 A^{(c_4)}_4 \rangle \notag \\
    &= \frac{1}{N^2} \Bigg( \sum_{c=1}^N \langle A^{(c)}_1 A^{(c)}_2 A^{(c)}_3 A^{(c)}_4 \rangle \notag \\
    &\qquad \qquad + \sum_{c \neq d}^N \langle A^{(c)}_1 A^{(c)}_2 \rangle \langle A^{(d)}_3 A^{(d)}_4 \rangle \notag \\
    &\qquad \qquad + \sum_{c \neq d}^N \langle A^{(c)}_1 A^{(c)}_3 \rangle \langle A^{(d)}_2 A^{(d)}_4 \rangle \notag \\
    &\qquad \qquad + \sum_{c \neq d}^N \langle A^{(c)}_1 A^{(c)}_4 \rangle \langle A^{(d)}_2 A^{(d)}_3 \rangle \Bigg)
\end{align}
Because there are $N$ terms in the first sum and $N(N-1)$ terms in each of the pairwise sums, we collect the coefficients:
\begin{align}
    \langle A^{(N)}_1 A^{(N)}_2 A^{(N)}_3 A^{(N)}_4 \rangle &= \frac{1}{N} \langle A_1 A_2 A_3 A_4 \rangle \notag \\
    &\quad + \frac{N-1}{N} \Big( \langle A_1 A_2 \rangle \langle A_3 A_4 \rangle + \langle A_1 A_3 \rangle \langle A_2 A_4 \rangle + \langle A_1 A_4 \rangle \langle A_2 A_3 \rangle \Big)
\end{align}
In the case that all fields are identical or exactly two of them are equal to each other, the first term contributes for finite $N$ as shown in Appendix \ref{AAAAresults}. But the large-$N$ limit enforces Wick factorization:
\begin{align}
    \lim_{N \to \infty} \langle A^{(N)}_1 A^{(N)}_2 A^{(N)}_3 A^{(N)}_4 \rangle &= \langle A_1 A_2 \rangle \langle A_3 A_4 \rangle + \langle A_1 A_3 \rangle \langle A_2 A_4 \rangle + \langle A_1 A_4 \rangle \langle A_2 A_3 \rangle
\end{align}
Because coordinate derivatives commute with the statistical expectation values over the network parameters, the field strength tensor $F^{(N)}$, being linear in $A$ defined via $F_{MN}^{(N)} = \partial_M A_N^{(N)} - \partial_N A_M^{(N)} = \frac{1}{\sqrt{N}} \sum_{c=1}^N F_{MN}^{(c)}$, inherits identical large-$N$ statistics:
\begin{align}
    \langle F^{(N)}_1 F^{(N)}_2 F^{(N)}_3 F^{(N)}_4 \rangle &= \frac{1}{N} \langle F_1 F_2 F_3 F_4 \rangle_{\text{single}} \notag \\
    &\quad + \frac{N-1}{N} \Big( \langle F_1 F_2 \rangle \langle F_3 F_4 \rangle + \langle F_1 F_3 \rangle \langle F_2 F_4 \rangle + \langle F_1 F_4 \rangle \langle F_2 F_3 \rangle \Big)
\end{align}
Consequently, the correlators of $F_{MN}$ also factorize:
\begin{align}
    \lim_{N \to \infty} \langle F^{(N)}_1 F^{(N)}_2 F^{(N)}_3 F^{(N)}_4 \rangle &= \langle F_1 F_2 \rangle \langle F_3 F_4 \rangle + \langle F_1 F_3 \rangle \langle F_2 F_4 \rangle + \langle F_1 F_4 \rangle \langle F_2 F_3 \rangle
\end{align}

The scalar primary $\Phi:=\frac{1}{4}F^2$, being quadratic in $F$, does not obey Gaussian statistics, but we can recover the correct correlators due to the Gaussianity of $F^{(N)}$. We use normal ordering to subtract the singular coincident self-contractions:
\begin{equation}
    \Phi^{(N)}(X) = \frac{1}{4} :F^{(N)}_{MN}(X) F^{(N) MN}(X):
\end{equation}
More precisely, using $F^{(N)} = \frac{1}{\sqrt{N}} \sum_{c=1}^N F^{(c)}$, the scalar $\Phi^{(N)}$ expands as a double sum:
\begin{align}
    \Phi^{(N)}(X) &= \frac{1}{4N} \sum_{c=1}^N \sum_{d=1}^N :F^{(c)}_{MN}(X) F^{(d) MN}(X): \notag \\
    &:= \frac{1}{4N} \sum_{c,d=1}^N \Big( F^{(c)}_{MN}(X) F^{(d) MN}(X) - \delta^{cd} \langle F_{MN}(X) F^{MN}(X) \rangle \Big)
\end{align}
Therefore, for the two-point function:
\begin{align}
    \langle \Phi^{(N)}(X_1) \Phi^{(N)}(X_2) \rangle &= \frac{1}{16 N^2} \sum_{c,d,e,f=1}^N \langle :F^{(c)}_{M_1 N_1}(X_1) F^{(d) M_1 N_1}(X_1): \notag \\
    &\qquad \qquad \qquad \qquad \times :F^{(e)}_{M_2 N_2}(X_2) F^{(f) M_2 N_2}(X_2): \rangle \notag \\
    &= \frac{1}{16 N^2} \sum_{c,d,e,f=1}^N  \langle F^{(c)}_{M_1 N_1}(X_1) F^{(e)}_{M_2 N_2}(X_2) \rangle \langle F^{(d) M_1 N_1}(X_1) F^{(f) M_2 N_2}(X_2) \rangle \notag \\
    &\quad + \langle F^{(c)}_{M_1 N_1}(X_1) F^{(f) M_2 N_2}(X_2) \rangle \langle F^{(d) M_1 N_1}(X_1) F^{(e)}_{M_2 N_2}(X_2) \rangle \notag \\
    &\quad + \text{connected piece} \notag \\
    &=\frac{1}{16 N^2} \sum_{c,d=1}^N \Big( \langle F^{(c)}_{M_1 N_1}(X_1) F^{(c)}_{M_2 N_2}(X_2) \rangle \langle F^{(d) M_1 N_1}(X_1) F^{(d) M_2 N_2}(X_2) \rangle \notag \\
    &\qquad \qquad \qquad + \langle F^{(c)}_{M_1 N_1}(X_1) F^{(d) M_2 N_2}(X_2) \rangle \langle F^{(d) M_1 N_1}(X_1) F^{(c)}_{M_2 N_2}(X_2) \rangle \Big) \notag \\
    &\quad + \mathcal{O}\left(\frac{1}{N}\right) \notag \\
    &= \frac{1}{16} \Big( \langle F_{M_1 N_1}(X_1) F_{M_2 N_2}(X_2) \rangle \langle F^{M_1 N_1}(X_1) F^{M_2 N_2}(X_2) \rangle \notag \\
    &\qquad \quad + \langle F_{M_1 N_1}(X_1) F^{M_2 N_2}(X_2) \rangle \langle F^{M_1 N_1}(X_1) F_{M_2 N_2}(X_2) \rangle \Big) \notag \\
    &\quad + \mathcal{O}\left(\frac{1}{N}\right) \notag \\
    &= \frac{1}{8} \langle F_{M_1 N_1}(X_1) F_{M_2 N_2}(X_2) \rangle \langle F^{M_1 N_1}(X_1) F^{M_2 N_2}(X_2) \rangle + \mathcal{O}\left(\frac{1}{N}\right)
\end{align}

In the above, the normal-ordering subtraction cancels the divergent intra-coordinate pairings (the terms proportional to $\delta^{cd} \delta^{ef}$), yielding the first line.
The disconnected components require the channel indices across the two spatial coordinates to match. The first trace requires $c=e$ and $d=f$, which generates $N^2$ terms in the sum. The second trace requires $c=f$ and $d=e$, which generates another $N^2$ terms, which cancels from the $N^2$ appearing from the normalization in \eqref{eq:largeNansatzappendix}. The single connected configuration where all four channels match ($c=d=e=f$) contains only $N$ terms, and is dropped in the $N \to \infty$ limit. Since we already have the correct $\langle FF \rangle$ correlator, it is intuitively obvious that the result of $\langle \Phi \Phi\rangle$ using this method is bound to match that of Maxwell. To wit, from \eqref{MaxwellFF} in Section \ref{Maxwellsection}, the 2-point correlator in components is
\begin{align}
    \langle F_{M_1 N_1}(X_1) F_{M_2 N_2}(X_2) \rangle &= \frac{1}{16\pi^2(X_1 \cdot X_2)^4} \big( H_{M_1 M_2}(X_1, X_2) H_{N_1 N_2}(X_1, X_2) \notag \\ & \quad\quad\quad\quad\quad- H_{M_1 N_2}(X_1, X_2) H_{N_1 M_2}(X_1, X_2) \big)
\end{align}
where $H_{MN}(X_i, X_j) = -2 (X_i \cdot X_j) \delta_{MN} + 2 X_{jM} X_{iN}$.
Squaring this we get
\begin{align}\label{eq:someeqn}
    \big( H_{M_1 M_2} H_{N_1 N_2} - H_{M_1 N_2} H_{N_1 M_2} \big)^2 &= H_{M_1 M_2} H^{M_1 M_2} H_{N_1 N_2} H^{N_1 N_2} \notag \\
    &\quad - H_{M_1 M_2} H^{M_1 N_2} H_{N_1 N_2} H^{N_1 M_2} \notag \\
    &\quad - H_{M_1 N_2} H^{M_1 M_2} H_{N_1 M_2} H^{N_1 N_2} \notag \\
    &\quad + H_{M_1 N_2} H^{M_1 N_2} H_{N_1 M_2} H^{N_1 M_2} \notag \\
    &= 2 (H_{MA} H^{MA})^2 - 2 (H_{MA} H_N^{\phantom{N}A}) (H^{MB} H^N_{\phantom{N}B})
\end{align}

The first term in the above equation:
\begin{align}
    H_{MA} H^{MA} &= \big( -2 (X_1 \cdot X_2) \delta_{MA} + 2 X_{2M} X_{1A} \big) \big( -2 (X_1 \cdot X_2) \delta^{MA} + 2 X_2^M X_1^A \big) \notag \\
    &= 4 (X_1 \cdot X_2)^2 \delta_{MA} \delta^{MA} - 4 (X_1 \cdot X_2) X_{2M} X_1^M \notag \\
    &\quad - 4 (X_1 \cdot X_2) X_2^A X_{1A} + 4 (X_{2M} X_2^M)(X_{1A} X_1^A) \notag \\
    &= 4 (X_1 \cdot X_2)^2 (d+2) - 4 (X_1 \cdot X_2)^2 - 4 (X_1 \cdot X_2)^2 + 0 \notag \\
    &= 4d (X_1 \cdot X_2)^2
\end{align}
The second term in \eqref{eq:someeqn} involves:
\begin{align}
    H_{MA} H_N^{\phantom{N}A} &= \big( -2(X_1 \cdot X_2) \delta_{MA} + 2 X_{2M} X_{1A} \big) \big( -2(X_1 \cdot X_2) \delta_N^A + 2 X_{2N} X_1^A \big) \notag \\
    &= 4(X_1 \cdot X_2)^2 \delta_{MN} - 4(X_1 \cdot X_2) X_{2N} X_{1M} - 4(X_1 \cdot X_2) X_{2M} X_{1N} + 4 X_{2M} X_{2N} (X_1 \cdot X_1) \notag \\
    &= 4(X_1 \cdot X_2)^2 \delta_{MN} - 4(X_1 \cdot X_2) (X_{1M} X_{2N} + X_{2M} X_{1N})
\end{align}
We then square this result to get:
\begin{align}
    (H_{MA} H_N^{\phantom{N}A}) (H^{MB} H^N_{\phantom{N}B}) &= \big[ 4(X_1 \cdot X_2)^2 \delta_{MN} - 4(X_1 \cdot X_2) (X_{1M} X_{2N} + X_{2M} X_{1N}) \big] \notag \\
    &\qquad \times \big[ 4(X_1 \cdot X_2)^2 \delta^{MN} - 4(X_1 \cdot X_2) (X_1^M X_2^N + X_2^M X_1^N) \big] \notag \\
    &= 16(X_1 \cdot X_2)^4 (d+2) - 16(X_1 \cdot X_2)^3 (4 X_1 \cdot X_2) + 16(X_1 \cdot X_2)^2 \big( 2 (X_1 \cdot X_2)^2 \big) \notag \\
    &= 16(X_1 \cdot X_2)^4 (d+2) - 64 (X_1 \cdot X_2)^4 + 32 (X_1 \cdot X_2)^4 \notag \\
    &= 16d (X_1 \cdot X_2)^4
\end{align}
Putting these results into \eqref{eq:someeqn} we get:
\begin{align}
    2 (H_{MA} H^{MA})^2 - 2 (H_{MA} H_N^{\phantom{N}A}) (H^{MB} H^N_{\phantom{N}B}) &= 2 \big( 4d (X_1 \cdot X_2)^2 \big)^2 - 2 \big( 16d (X_1 \cdot X_2)^4 \big) \notag \\
    &= 32d^2 (X_1 \cdot X_2)^4 - 32d (X_1 \cdot X_2)^4 \notag \\
    &= 32d(d-1) (X_1 \cdot X_2)^4
\end{align}
which in turn yields the two-point correlator: 
\begin{align}
    \langle \Phi(X_1) \Phi(X_2) \rangle &= \frac{1}{8} \left( \frac{1}{16\pi^2(X_1 \cdot X_2)^4} \right)^2 \times 32d(d-1) (X_1 \cdot X_2)^4 \notag \\
    &= \frac{d(d-1)}{64\pi^4 (X_1 \cdot X_2)^4}
\end{align}
Projecting to physical spacetime using $(X_1 \cdot X_2) =-\frac{1}{2} x_{12}^2$, the correlator is 
\begin{equation}\label{eq:ddimPhiPhimaxwell}
    \langle \Phi(x_1) \Phi(x_2) \rangle = \frac{d(d-1)}{4\pi^4\,x_{12}^8}
\end{equation}
For $d=4$ this exactly reproduces the Maxwell CFT result \eqref{eq:trueMaxwell2ptPhiPhi}.\\\\
Finally, we discuss the case of the three-point function $\langle \Phi(X_1) \Phi(X_2) \Phi(X_3) \rangle$
\begin{equation}
    \langle \Phi(X_1) \Phi(X_2) \Phi(X_3) \rangle = \frac{1}{4^3} \langle :F_{M_1 N_1}(X_1) F^{M_1 N_1}(X_1): :F_{M_2 N_2}(X_2) F^{M_2 N_2}(X_2): :F_{M_3 N_3}(X_3) F^{M_3 N_3}(X_3): \rangle
\end{equation}

Again, since $F_{M_iN_i}(X_i)$ are Gaussian, we can apply Wick's theorem to find $8$ equal terms which simplify to 
\begin{align}
    \langle \Phi(X_1) \Phi(X_2) \Phi(X_3) \rangle = \frac{1}{8} \langle F_{M_1 N_1}(X_1) F_{M_2 N_2}(X_2) \rangle \langle F^{M_2 N_2}(X_2) F_{M_3 N_3}(X_3) \rangle \langle F^{M_3 N_3}(X_3) F^{M_1 N_1}(X_1) \rangle
\end{align}

Using the Maxwell $\langle FF\rangle $ correlator from \eqref{MaxwellFFwithcoeff},
the numerator $\mathcal{N}$ of the 3-point correlator can be expanded to
\begin{align}
    \mathcal{N} &\propto \big( H_{M_1 M_2}(X_1, X_2) H_{N_1 N_2}(X_1, X_2) - H_{M_1 N_2}(X_1, X_2) H_{N_1 M_2}(X_1, X_2) \big) \notag \\
    &\quad \times \big( H^{M_2 M_3}(X_2, X_3) H^{N_2 N_3}(X_2, X_3) - H^{M_2 N_3}(X_2, X_3) H^{N_2 M_3}(X_2, X_3) \big) \notag \\
    &\quad \times \big( H_{M_3}^{\phantom{M_3} M_1}(X_3, X_1) H_{N_3}^{\phantom{N_3} N_1}(X_3, X_1) - H_{M_3}^{\phantom{M_3} N_1}(X_3, X_1) H_{N_3}^{\phantom{N_3} M_1}(X_3, X_1) \big) \notag \\[1em]
    &= 4 \big( H_{M_1 M_2}(X_1, X_2) H^{M_2 M_3}(X_2, X_3) H_{M_3}^{\phantom{M_3} M_1}(X_3, X_1) \big)^2 \notag \\
    &\quad - 4 \big( H_{M_1 M_2}(X_1, X_2) H^{M_2 M_3}(X_2, X_3) H_{M_3}^{\phantom{M_3} N_1}(X_3, X_1) \notag \\
    &\qquad \qquad \times H_{N_1 N_2}(X_1, X_2) H^{N_2 N_3}(X_2, X_3) H_{N_3}^{\phantom{N_3} M_1}(X_3, X_1) \big)\label{numnum}
\end{align}

The proportionality constant is independent of spacetime dimension $d$.
The terms in \eqref{numnum} contain

\begin{align}
    &H_{M_1 M_2}(X_1, X_2) H^{M_2 M_3}(X_2, X_3) H_{M_3}^{\phantom{M_3} N_1}(X_3, X_1) \notag \\
    &= \big( -2 (X_1 \cdot X_2) \eta_{M_1 M_2} + 2 X_{2M_1} X_{1M_2} \big) \big( -2 (X_2 \cdot X_3) \eta^{M_2 M_3} + 2 X_3^{M_2} X_2^{M_3} \big) \notag \\
    &\qquad \times \big( -2 (X_1 \cdot X_3) \eta_{M_3}^{\phantom{M_3} N_1} + 2 X_{1M_3} X_3^{N_1} \big) \notag \\[1em]
    &= -8 (X_1 \cdot X_2)(X_2 \cdot X_3)(X_1 \cdot X_3) \eta_{M_1}^{\phantom{M_1} N_1}  + 8 (X_1 \cdot X_2)(X_2 \cdot X_3) X_{1M_1} X_3^{N_1} \notag \\
    &\quad + 8 (X_1 \cdot X_2)(X_1 \cdot X_3) X_{3M_1} X_2^{N_1}  - 8 (X_1 \cdot X_2) (X_2 \cdot X_1) X_{3M_1} X_3^{N_1} \notag \\
    &\quad + 8 (X_2 \cdot X_3)(X_1 \cdot X_3) X_{2M_1} X_1^{N_1}  - 8 (X_2 \cdot X_3) X_{2M_1} (X_1 \cdot X_1) X_3^{N_1} \notag \\
    &\quad - 8 (X_1 \cdot X_3) X_{2M_1} (X_1 \cdot X_3) X_2^{N_1} + 8 X_{2M_1} (X_1 \cdot X_3) (X_2 \cdot X_1) X_3^{N_1} \notag \\[1em]
    &= -8 (X_1 \cdot X_2)(X_2 \cdot X_3)(X_1 \cdot X_3) \eta_{M_1}^{\phantom{M_1} N_1} + 8 (X_1 \cdot X_2)(X_2 \cdot X_3) X_{1M_1} X_3^{N_1} \notag \\
    &\quad + 8 (X_1 \cdot X_2)(X_1 \cdot X_3) X_{3M_1} X_2^{N_1} - 8 (X_1 \cdot X_2)^2 X_{3M_1} X_3^{N_1} \notag \\
    &\quad + 8 (X_2 \cdot X_3)(X_1 \cdot X_3) X_{2M_1} X_1^{N_1} - 8 (X_1 \cdot X_3)^2 X_{2M_1} X_2^{N_1} + 8 (X_1 \cdot X_2)(X_1 \cdot X_3) X_{2M_1} X_3^{N_1} \label{eq:H_sequence_expanded}
\end{align}
The first term in \eqref{numnum} is just the square of the above after setting $N_1=M_1$ which simplifies to
\begin{align}
    H_{M_1 M_2}(X_1, X_2) &H^{M_2 M_3}(X_2, X_3) H_{M_3}^{\phantom{M_3} M_1}(X_3, X_1) \notag \\
    &= -8 (X_1 \cdot X_2) (X_2 \cdot X_3) (X_1 \cdot X_3) D \notag \\
    &\quad + 8 (X_1 \cdot X_2) (X_1 \cdot X_3) (X_2 \cdot X_3) \notag \\
    &\quad + 8 (X_2 \cdot X_3) (X_1 \cdot X_3) (X_1 \cdot X_2) - 0 \notag \\
    &\quad + 8 (X_1 \cdot X_2) (X_2 \cdot X_3) (X_1 \cdot X_3) - 0 \notag \\
    &\quad + 8 (X_1 \cdot X_2) (X_1 \cdot X_3) (X_2 \cdot X_3) \notag \\
    &= 8 (X_1 \cdot X_2) (X_2 \cdot X_3) (X_1 \cdot X_3) \big( -D + 1 + 1 + 1 + 1 \big) \notag \\
    &= 8 (X_1 \cdot X_2) (X_2 \cdot X_3) (X_1 \cdot X_3) \big( 4 - D \big)\label{eq:first_piece}
\end{align}
where $\delta^M {}_M=D$ and $X_i^2=0$ has been used. The second term in \eqref{numnum} can be simplified using \eqref{eq:H_sequence_expanded} and the PNC constraints as:
\begin{align}
    &H_{M_1 M_2}(X_1, X_2) H^{M_2 M_3}(X_2, X_3) H_{M_3}^{\phantom{M_3} N_1}(X_3, X_1) H_{N_1 N_2}(X_1, X_2) H^{N_2 N_3}(X_2, X_3) H_{N_3}^{\phantom{N_3} M_1}(X_3, X_1) \notag \\[1em]
    &= \begin{aligned}[t]
        &\Big( -8 (X_1 \cdot X_2) (X_2 \cdot X_3) (X_1 \cdot X_3) \eta_{M_1}^{\phantom{M_1} N_1} + 8 (X_1 \cdot X_2)(X_2 \cdot X_3) X_{1M_1} X_3^{N_1} \\
        &\quad + 8 (X_1 \cdot X_2)(X_1 \cdot X_3) X_{3M_1} X_2^{N_1} - 8 (X_1 \cdot X_2)^2 X_{3M_1} X_3^{N_1} \\
        &\quad + 8 (X_2 \cdot X_3)(X_1 \cdot X_3) X_{2M_1} X_1^{N_1} - 8 (X_1 \cdot X_3)^2 X_{2M_1} X_2^{N_1} \\
        &\quad + 8 (X_1 \cdot X_2)(X_1 \cdot X_3) X_{2M_1} X_3^{N_1} \Big)
       \end{aligned} \notag \\
    &\quad \times \begin{aligned}[t]
        &\Big( -8 (X_1 \cdot X_2) (X_2 \cdot X_3) (X_1 \cdot X_3) \eta_{N_1}^{\phantom{N_1} M_1} + 8 (X_1 \cdot X_2)(X_2 \cdot X_3) X_{1N_1} X_3^{M_1} \\
        &\quad + 8 (X_1 \cdot X_2)(X_1 \cdot X_3) X_{3N_1} X_2^{M_1} - 8 (X_1 \cdot X_2)^2 X_{3N_1} X_3^{M_1} \\
        &\quad + 8 (X_2 \cdot X_3)(X_1 \cdot X_3) X_{2N_1} X_1^{M_1} - 8 (X_1 \cdot X_3)^2 X_{2N_1} X_2^{M_1} \\
        &\quad + 8 (X_1 \cdot X_2)(X_1 \cdot X_3) X_{2N_1} X_3^{M_1} \Big)
       \end{aligned} \notag \\[1em]
    &= 64 (X_1 \cdot X_2)^2 (X_2 \cdot X_3)^2 (X_1 \cdot X_3)^2 D \notag \\
    &\quad - 16 (X_1 \cdot X_2) (X_2 \cdot X_3) (X_1 \cdot X_3) \big( 32 (X_1 \cdot X_2) (X_2 \cdot X_3) (X_1 \cdot X_3) \big) \notag \\
    &\quad + \big( 256 - 256 + 256 - 256 + 256 + 256 \big) (X_1 \cdot X_2)^2 (X_2 \cdot X_3)^2 (X_1 \cdot X_3)^2 \notag \\[1em]
    &= 64 \big( D - 2 \big) (X_1 \cdot X_2)^2 (X_2 \cdot X_3)^2 (X_1 \cdot X_3)^2 \label{eq:2ndpiece}
\end{align}

Combining \eqref{eq:first_piece} and \eqref{eq:2ndpiece} we finally obtain from \eqref{numnum} and $d=D-2$

\begin{align}
    \mathcal{N} &\propto 
     4 \big[ -8 (X_1 \cdot X_2) (X_2 \cdot X_3) (X_1 \cdot X_3) (d - 2) \big]^2 \notag \\
    &\quad - 4 \big[ 64 (X_1 \cdot X_2)^2 (X_2 \cdot X_3)^2 (X_1 \cdot X_3)^2 (d) \big] \notag \\[1em]
    &= 256 (X_1 \cdot X_2)^2 (X_2 \cdot X_3)^2 (X_1 \cdot X_3)^2 \big[ (d - 2)^2 - d \big] \notag \\
    &= 256 (X_1 \cdot X_2)^2 (X_2 \cdot X_3)^2 (X_1 \cdot X_3)^2 (d - 1)(d - 4)
\end{align}
which vanishes for $d=4$. Recalling that the proportionality above is dimension independent and the denominator is non-zero, we conclude that the $3$-point function $\langle \Phi \Phi \Phi \rangle$ vanishes for $d=4$ as expected in Maxwell theory \cite{Dolan:2000ut,El-Showk:2011xbs}.

\bibliographystyle{jhep}
\bibliography{main}
\end{document}